%% file: meta_article_main.tex
\documentclass[11pt,letterpaper]{article}
\usepackage[margin=1in]{geometry}
\usepackage{amsmath,amssymb}
\usepackage{graphicx}
\graphicspath{{figures/}}
\usepackage{bm}
\usepackage{booktabs}
\usepackage{xcolor}
\usepackage[colorlinks=true,linkcolor=blue,citecolor=blue,urlcolor=blue]{hyperref}
\usepackage[numbers,sort&compress]{natbib}
\usepackage{microtype}
\usepackage{placeins}
\usepackage[font=small,labelfont=bf]{caption}
\usepackage{longtable}

\newcommand{\aB}{a_{\mathrm{B}}^{*}}
\newcommand{\Ha}{\mathrm{Ha}^{*}}
\newcommand{\rs}{r_{s}}
\newcommand{\rc}{r_{c}}
\newcommand{\kF}{k_{\mathrm{F}}}

\newcommand{\nIRM}{n_{\mathrm{IRM}}}
\newcommand{\nc}{n_{c}}
\newcommand{\nph}{n_{1/2}}
\newcommand{\nimpurity}{\bar{n}_{i}}

\newcommand{\br}{\mathbf{r}}
\newcommand{\bq}{\mathbf{q}}
\newcommand{\bG}{\mathbf{G}}

\newcommand{\bR}{\mathbf{R}}

\newenvironment{ruledtabular}{}{}

\newcommand{\pacs}[1]{}
\newcommand{\keywords}[1]{}

\providecommand{\rhob}{\bar{\rho}}
\providecommand{\ee}{e^2/\epsilon\ell}
\providecommand{\av}[1]{\langle #1 \rangle}
\providecommand{\Sc}{S_c}
\providecommand{\Sst}{S_{\mathrm{st}}}

\title{\textbf{Disorder, Interactions, and the Fate of the\\
Two-Dimensional Electron System:\\
Wigner Crystal, Metal-Insulator Transition, and Amorphous Solid}}
\author{Sankar Das Sarma\\[6pt]
\small Condensed Matter Theory Center and Joint Quantum Institute,\\
\small Department of Physics, University of Maryland,\\
\small College Park, Maryland 20742, USA\\
\small \href{mailto:dassarma@umd.edu}{dassarma@umd.edu}}
\date{\today}

\begin{document}

\maketitle

\begin{abstract}
We present a unified theoretical study of the interplay between Coulomb
disorder and Coulomb interactions in two-dimensional electron systems,
covering the Wigner crystallization transition, the metal-insulator
transition, and the crossover from a Wigner crystal to an amorphous
localized solid in both zero magnetic field and the lowest Landau level.
Using Hartree-Fock theory, variational Monte Carlo, and percolation and
strong-localization arguments, we show that quenched charged disorder
destroys true long-range Wigner crystalline order in the thermodynamic
limit at any nonzero charged-impurity density, within the
Imry-Ma-Larkin framework; the finite-system density image is that of an
amorphous solid with crystalline coherence over at most a few lattice
spacings.  For the two-dimensional metal--insulator transition we show, in a
controlled nonlinear-Thomas--Fermi model of long-range Coulomb
disorder at setback $d$, that the half-depletion density of the
semiclassical landscape and the Ioffe--Regel--Mott density of
Born--RPA transport track each other to within a factor of order
unity over the physically relevant range of impurity density and
setback, with a slow $n_i^{-1/10} d^{-1/5}$ spread; the agreement
reflects the common energy scale $E_{\mathrm{dis}} \sim E_F$ at the
transition rather than a theorem, and modest independent short-range
transport channels can shift the ratio by factors of order two.  The
article is a theoretical study under well-defined approximations; it
takes the recent scanning tunneling microscopy imaging experiments on
two-dimensional electron systems and the longstanding
disorder-versus-interaction debate on the two-dimensional
metal--insulator transition as motivation, and leaves to the reader
the comparison between its calculated signatures and the published
experimental analyses.
\end{abstract}

\tableofcontents
\newpage

\input{meta_sec1_intro}
\FloatBarrier
\input{meta_sec2_hf_letter}
\FloatBarrier
\input{meta_sec3_vmc}

\FloatBarrier
\input{meta_sec4_lll}

\FloatBarrier
\input{meta_sec5_tf_irm}
\FloatBarrier
\input{meta_sec6_pan}
\FloatBarrier
\input{meta_sec7_conclusion}
\FloatBarrier

\clearpage
\appendix
\input{meta_appendix}

\bibliographystyle{unsrtnat}
\bibliography{meta_bibliography}

\end{document}

%% file: meta_sec1_intro.tex

\section{Introduction}
\label{sec:intro}

A two-dimensional (2D) electron system crystallizes when the Coulomb
energy sufficiently dominates the kinetic energy, that is, when the
dimensionless density parameter
$\rs = 1/(\sqrt{\pi n}\,\aB)$ exceeds a critical value $\rc$.  Quantum
Monte Carlo places $\rc$ in the range of a few tens, with the specific
value depending on the spin sector and the trial-wave-function family
used~\cite{Tanatar1989,Attaccalite2002,Drummond2009,Azadi2024};
Drummond and Needs~\cite{Drummond2009} report the clean
paramagnetic-fluid to antiferromagnetic-crystal transition at
$\rc = 31(1)$, with the ferromagnetic crystal becoming stable at a
different density.  The spin sector relevant to a given calculation
therefore matters when quoting an $\rc$.
This number is a property of the interacting electrons alone: it is
the density below which the interaction-driven spontaneous breaking
of continuous translational symmetry becomes energetically favorable.
Throughout this paper $a$ denotes the average inter-particle
separation, which is also, up to a numerical factor of order unity,
the lattice constant of the pristine triangular Wigner crystal at
the same density.
Hartree--Fock (HF) theory, which neglects correlation, places the
crystallization transition at a lower $\rs$ than the quantum
many-body result; a commensurate HF search puts the transition at
$\rs \simeq 2.65$ in the fully spin-polarized gas~\cite{Bernu2008,%
Bernu2011,Trail2003}.  In the lowest Landau level, where kinetic
energy is quenched and the entire problem is the projected Coulomb
interaction, HF produces a Wigner crystal at every partial filling
$\nu < 1$~\cite{FPA1979,YoshiokaFukuyama1979,YoshiokaLee1983,%
MakiZotos1983}.

The physical question that this paper addresses is what happens to
this interaction-driven crystallization in the presence of quenched
Coulomb disorder.  The motivation is twofold.  First, recent scanning
tunneling microscopy (STM) experiments on gate-tunable monolayer and
bilayer MoSe$_2$~\cite{Xiang2025,Ge2026,Xiang2024,Li2021} and on
Bernal bilayer graphene in a strong perpendicular
field~\cite{Tsui2024} report real-space density images that are
interpreted as Wigner solids.  In the zero-field experiments these
images are obtained at electron densities corresponding to
$\rs \lesssim 20$, well inside the clean quantum liquid regime, and
the solid-like appearance becomes more robust as the short-range
disorder density is increased~\cite{Ge2026}.  The experimental
groups~\cite{Ge2026,Xiang2024} and parts of the theoretical
literature~\cite{Chui1995,Joy2025,Shayegan2022} discuss the
disorder-stabilized Wigner-solid picture.  The present paper
addresses, within a specific set of microscopic approximations, an
alternative structural picture~\cite{Babbar2026a}: that at
$\rs < \rc$ a correlated electron state can be organized by a
quenched impurity landscape into an \emph{amorphous solid} whose
density pattern resembles a solid locally while lacking true
long-range crystalline order.  Which picture is the correct
description of a given imaged sample is a question for the
experimental analysis; the role of the present paper is to work out
what the amorphous-solid picture predicts quantitatively within
Hartree--Fock, variational Monte Carlo, and lowest-Landau-level
Hartree--Fock, and to set out the structural observables that
distinguish it from a disorder-broken Wigner crystal.

Second, the metal--insulator transition (MIT) in two dimensions has
been debated for more than three decades.  Scaling theory predicts
that all states in a 2D disordered system are localized at any
disorder, which would preclude a true metallic phase at $T = 0$.
Experimentally~\cite{Kravchenko1994}, however, 2D electron systems in
Si MOSFETs and in GaAs/AlGaAs heterostructures exhibit a
low-temperature resistivity that decreases with decreasing
temperature on the metallic side of a critical carrier density $n_c$,
and increases on the insulating side.  Two mutually exclusive-looking
theoretical pictures have been proposed for the transition: (i)
classical percolation through a disorder-induced puddle
landscape~\cite{Efros1989,EPB1993,Fogler2004,DasSarma2005}, where
$n_c$ is identified as the density at which the connected region of
the self-consistent electron density first spans the sample; and
(ii) a quantum Ioffe--Regel--Mott (IRM) strong-localization crossover,
where $n_c$ is identified as the density at which $k_F \ell_{\rm MFP}
\sim 1$~\cite{IoffeRegel1960,Mott1972,DasSarma2013,DasSarma2014}.
The empirical observation is that the two definitions give the same
$n_c$ to within a numerical factor of order unity, over a range of
disorder strengths and setback distances.  Why they should coincide
is a quantitative question with no obvious a priori answer.

The two topics---whether a 2D electron solid can be interaction-driven
or impurity-driven, and what controls the critical density of the 2D
MIT---are facets of the same underlying problem: how do Coulomb
interactions and Coulomb disorder compete to organize the electrons
in two dimensions, and which of them wins at which density, and at
which disorder strength.  This paper studies that competition with a
battery of complementary microscopic methods: self-consistent
Hartree--Fock theory of a periodic 16-electron supercell at zero
field; variational Monte Carlo (VMC) with $N = 16$ spin-polarized
electrons at $\rs = 10$, $20$, $30$, $50$ in three disorder models;
Hartree--Fock on a torus in the lowest Landau level; and nonlinear
Thomas--Fermi (TF) screening with explicit computation of the
percolation threshold and the Ioffe--Regel--Mott density.  The
results are compared with the quantum HF and classical molecular
dynamics (MD) study of Das Sarma and Pan~\cite{DSPan2026}, which uses
a different disorder model and arrives at a consistent picture.

The main organizing variable throughout the paper is the dimensionless
disorder-to-electron density ratio
\begin{equation}
\eta_{\mathrm{dis}} \equiv \nimpurity/n,
\label{eq:ratio}
\end{equation}
where $\nimpurity$ is the density of charged impurities seen by the
2D electron gas (including, for Coulomb impurities at setback $d$,
the appropriate screening function; the explicit form appears in
Sec.~\ref{sec:tf_irm}).  The reason for this choice is already clear
on general grounds and is derived in detail in Secs.~\ref{sec:hf_letter}
and \ref{sec:vmc}: at $\rs < \rc$ the clean system is a correlated
liquid, and whether its density image looks solid is controlled by
the ratio of the disorder energy to the Coulomb repulsion scale
$e^2/a$, not by $\rs$ alone; and at $\rs > \rc$ the clean HF system
is a crystal whose coherence is cut off by the random pinning at a
scale set by $\eta_{\mathrm{dis}}$.  In both regimes
$\eta_{\mathrm{dis}}$ decides the outcome.

\emph{A note on notation.}  Because the material of Secs.~\ref{sec:hf_letter}--\ref{sec:tf_irm}
was developed in independent calculations with different natural
notational conventions, the symbol for the dimensionless impurity
density varies modestly between sections: $n_i/n$ with no bar in the
HF and VMC analyses (Secs.~\ref{sec:hf_letter}--\ref{sec:vmc}),
$N_i/N$ for the finite LLL torus (Sec.~\ref{sec:lll}), $\bar{n}_i$
with bar in the TF/MIT analysis (Sec.~\ref{sec:tf_irm}) where the
spatial-averaged impurity density enters the continuum-Gaussian
construction, and $c = \bar{n}_i/n_0$ in the comparison with
Ref.~\cite{DSPan2026} (Sec.~\ref{sec:pan_qhf}).  These are the same
physical quantity.  Separately, the dimensionless disorder parameter
$\eta_{\mathrm{dis}} \equiv \bar{n}_i/n$ of
Eq.~(\ref{eq:ratio}) coincides with the quantity $\eta = W/E_F$ used
in Sec.~\ref{sec:pan_qhf} only through its definition in each
section; the two $\eta$'s are not the same variable.  The difference
in notations in different sections is consistent with the different
problems and models being considered in each section although they
all bear upon the issue of central conceptual interest: competition
between disorder and interaction in 2D electron layers.

A theorem shapes everything that follows.  The Imry--Ma--Larkin
argument~\cite{ImryMa1975,Larkin1970,Chitra2002} states that a
system with a continuous order parameter cannot possess true
long-range order in two dimensions in the presence of any nonzero
random field: the energy that an arbitrarily weak random field gains
by locally following the order-parameter texture, over a region of
linear size $R$, grows faster with $R$ than the elastic cost of the
texture.  For the 2D Wigner crystal, which is characterized by the
spontaneous breaking of continuous translational symmetry, this
theorem forbids true crystalline Bragg peaks at any nonzero charged
disorder.  What remains is an \emph{amorphous solid}: a state with
no long-range translational order, but with crystalline coherence
over a finite length $R_c$ (the Larkin length) that is set by the
competition between elastic stiffness and random-force variance.

The Larkin theorem has two regimes that must be distinguished.  In
the weak-pinning (collective) regime $R_c \gg a$, the random field
couples linearly to the elastic displacement of a well-formed
crystal and the Imry--Ma--Larkin estimate for $R_c$ is sharp.  In
the strong-pinning regime $R_c \sim a$, the coupling is to the
individual-pinning scale and the elastic description of a crystal
breaks down; the state is then more naturally described as a glass
of impurity-trapped electrons than as an elastically pinned Wigner
crystal.  The amorphous solid of
Secs.~\ref{sec:hf_letter}--\ref{sec:lll} sits predominantly in the
strong-pinning regime, where the two descriptions agree on the
qualitative conclusion (no long-range Bragg order, coherence over at
most a few lattice spacings) but on different microscopic grounds.
Our finite-system calculations ($N=16$ and $N=36$ in Sec.~\ref{sec:hf_letter};
$N_\phi \leq 200$ in Sec.~\ref{sec:lll}) resolve coherence lengths
at the lattice spacing and up but cannot in principle distinguish
$R_c = a$ from $R_c = 3a$, and we are careful below not to claim
otherwise.

A second, physical observation sharpens the picture.  Consider a
2D electron system of density $n$ in a landscape of quenched Coulomb
impurities of density $\nimpurity$.  At fixed impurity density,
lowering $n$ increases $\rs$ and also increases
$\eta_{\mathrm{dis}} = \nimpurity/n$.  In the limit
$n \ll \nimpurity$ the state is organized by the impurity landscape
rather than by mutual Coulomb repulsion: there are more pinning
centers than electrons, and the electrons occupy a subset of the
attractive wells (fewer electrons than wells, so not every well is
filled).  In the opposite limit $\nimpurity \ll n$ the state is a
weakly disordered correlated liquid.  Across the $(\rs, \eta_{\mathrm{dis}})$ plane sampled by the
microscopic calculations of
Secs.~\ref{sec:hf_letter}--\ref{sec:lll}, weak-disorder changes to
the structural measures we use are small or statistically
unresolved, and clear suppression of the connected crystalline
correlation sets in once $\eta_{\mathrm{dis}}$ becomes appreciable;
no window of disorder-enhanced crystalline coherence is seen within
those measures over the sampled range.  We stop short
of inferring from this a universal impossibility of
disorder-stabilized local order: the microscopic result is a
statement about the specific disorder models, parameter ranges, and
structural diagnostics of this paper, and the Imry--Ma--Larkin
theorem below bears on long-range order rather than on finite-domain
energetics.  The physically relevant parameters for the appearance of
a solid-looking density are $\eta_{\mathrm{dis}}$ and the ratio of
the disorder energy to the Coulomb scale $e^2/a$, and the question
with a sharp answer within the present approximations is which of
these two energies organizes the electrons.  We emphasize in this
context that even liquids have strong short-range order on the scale
of the interparticle separation $a$, and thus short-range order by
itself does not imply local solidity; it only implies the existence
of a strong repulsive interaction.

Each section considers a slightly different problem and model.  The
remainder of the paper substantiates this picture through explicit
calculation.  Section~\ref{sec:hf_letter} develops the
zero-field Hartree--Fock theory of a spin-polarized 2D electron gas
with randomly placed Coulomb impurities of both signs, establishes a
density-image-level diagnostic of ``amorphous solid versus Wigner
crystal,'' and maps out the $(\rs, \nimpurity/n)$ structural
crossover diagram.  Section~\ref{sec:vmc} performs variational
Monte Carlo calculations with $N = 16$ spin-polarized electrons at
$\rs = 10, 20, 30, 50$ in three disorder models (point-like
impurities, impurity-strength scan, and a layer of modulation-type
donors), and shows that at no density, in any of the three models,
in any of three trial-wave-function families, does charged disorder
increase the connected crystalline correlation.
Section~\ref{sec:lll} applies the same programme to the lowest
Landau level, in the presence of a strong perpendicular magnetic
field, where kinetic energy is quenched and the clean system
is already a Wigner crystal: in this case the impurity-driven loss
of crystalline coherence is seen in a controlled setting whose
dimensionless control parameter is $\eta_{\mathrm{dis}}$ alone.
Section~\ref{sec:tf_irm} studies the problem of long-range Coulomb
disorder at setback $d$ with nonlinear Thomas--Fermi screening,
compares the half-depletion (``percolation'') density of the
semiclassical landscape with the Ioffe--Regel--Mott density of
Born--RPA transport, and connects the result to the
longstanding debate on the 2D metal--insulator transition.
Section~\ref{sec:pan_qhf} compares with the independent quantum
Hartree--Fock and classical molecular dynamics study of Das Sarma and
Pan~\cite{DSPan2026}, which identifies the Larkin coherence length
with weak-pinning scaling $R_c \propto W^{-1}$ in the HF amplitude
scan and $R_c \propto \nimpurity^{-1/2}$ in the classical
concentration scan, in agreement with the picture built up here.
Section~\ref{sec:discussion} discusses the implications for the
MoSe$_2$ STM imaging experiments and for the 2D MIT in Si and GaAs,
and Sec.~\ref{sec:conclusion} summarizes.  Much of the work with most
of the results presented in Secs.~\ref{sec:hf_letter}--\ref{sec:tf_irm}
is new.

%% file: meta_sec2_hf_letter.tex

\section{Zero-Field Hartree--Fock Theory of a Two-Dimensional Electron Gas
with Charged Impurities}
\label{sec:hf_letter}

We begin with the mean-field (Hartree--Fock) treatment of the
simplest version of the problem: a spin-polarized 2D electron gas in
a landscape of randomly placed Coulomb impurities of both signs.
Hartree--Fock, which neglects correlation, places the clean
crystallization at a lower $\rs$ than the quantum many-body result
(a commensurate HF search gives $\rs \simeq 2.65$ in the fully
spin-polarized gas, Sec.~\ref{sec:intro}); this is the virtue of the
method here, not its defect, because any disorder-driven enhancement
of crystallization would appear most readily within HF.  The main HF
result below is that no such enhancement is found.

\input{body_hf}


\input{body_hf_sm}

%% file: body_hf.tex
\subsection{Model}

We consider $N$ spin-polarized electrons of band mass $m^*$ in a medium of dielectric constant $\epsilon$, with areal density $n=1/[\pi(r_s a_B^*)^2]$ and $a_B^*=\hbar^2\epsilon/(m^*e^2)$. Lengths are measured in units of $r_s a_B^*$, $\mathbf{x}=\mathbf{r}/(r_s a_B^*)$, and energies in the effective Hartree $H^*=e^2/(\epsilon a_B^*)$. The Hamiltonian is
\begin{align}
H={}&-\frac{1}{2r_s^2}\sum_i \nabla_{\mathbf{x}_i}^2
+\frac{1}{r_s}\sum_{i<j}\frac{1}{|\mathbf{x}_i-\mathbf{x}_j|}\nonumber\\
&-\frac{Z}{r_s}\sum_{i,\alpha}\frac{s_\alpha}{|\mathbf{x}_i-\mathbf{X}_\alpha|}
+H_{\rm bg}.
\label{hf:eq:H}
\end{align}
The first term is the kinetic energy, which scales as $r_s^{-2}$. The second is the electron-electron Coulomb repulsion, which scales as $r_s^{-1}$. The third couples each electron to quenched point charges $s_\alpha Ze$ ($s_\alpha=\pm1$) at random positions $\mathbf{X}_\alpha$ in the electron plane through the bare Coulomb interaction. The impurities come in equal numbers of each sign, so their net charge vanishes. $H_{\rm bg}$ is a uniform positive background that neutralizes the electrons and removes the $\mathbf{q}=0$ components of all Coulomb terms.

The dimensionless parameters are $r_s$, $Z$, and $n_i/n$, where $n_i$ is the total impurity density. We take $Z=1$ unless stated otherwise. The electron-impurity and electron-electron terms carry the same prefactor $1/r_s$. The ratio of these two bare energy scales is therefore fixed by $Z$ and $n_i/n$ alone; the response of the electrons to the impurities still depends on $r_s$ through the kinetic energy. For short-range disorder of fixed amplitude $W$, by contrast, the ratio to the Coulomb scale $E_C\sim H^*/r_s$ grows as $r_s$, or falls as $r_s^{-1}$ if $W$ is fixed relative to $E_F$.

\subsection{Method}

We solve Eq.~(\ref{hf:eq:H}) in unrestricted HF theory in a periodic triangular supercell containing $N_c=16$ electrons, spanning $4\times4$ primitive cells of a triangular lattice with spacing $a=(2\pi/\sqrt3)^{1/2}r_s a_B^*$. Orbitals are expanded in 163 plane waves ($|\mathbf{G}|\le1.67\,G_1$, with $G_1=4\pi/\sqrt3 a$). The Fock matrix contains four terms: the kinetic energy; the Hartree potential $2\pi\rho_{\mathbf{d}}/(Ar_s|\mathbf{d}|)$, where $\rho_{\mathbf{d}}$ are the density Fourier coefficients per cell; the exchange potential; and the impurity potential
\begin{equation}
U_{\mathbf{d}}=-\frac{2\pi Z}{A\,r_s|\mathbf{d}|}\sum_\alpha s_\alpha e^{-i\mathbf{d}\cdot\mathbf{X}_\alpha}.
\label{hf:eq:U}
\end{equation}
Here $A$ is the supercell area and $\mathbf{d}$ runs over supercell reciprocal vectors. The exchange Coulomb kernel is averaged over each Bloch-mesh cell, with the singular cell integrated exactly. The plane-wave cutoff regularizes the in-plane impurity potential at short distance. Electronic screening of the impurities is generated self-consistently; impurity setback and gate screening are not included. Occupations are Fermi--Dirac at $k_BT=0.005E_F$, reached by annealing; every state has a density-matrix residual $10^{-6}$.

For each liquid-side point we start from three seeds: the uniform state, the self-consistent Hartree state, and a lattice-seeded state. For configuration A at $r_s=4$ the seeds are the clean crystal, the Hartree state, and, where available, the solution at the next lower impurity density. We retain the lowest free energy. Direct-only Hartree with identical impurities isolates exchange. The impurity densities $n_i/n=1/8$, $1/4$, $1/2$, and 1 correspond to 2, 4, 8, and 16 impurities per supercell, drawn as nested subsets of each random configuration. We study five configurations at $r_s=2.4$, labeled A--E, and three at $r_s=2.0$. Configuration A also provides a crystalline comparison at $r_s=4$.

\subsection{Clean reference}

The uniform-state energy on a finite Bloch mesh fluctuates with the discrete sampling of the Fermi sea; its error reaches $6\times10^{-4}H^*$ on a $4\times4$ mesh. We therefore compare lattice-seeded solutions with the exact uniform HF energy $E_u=r_s^{-2}-8/(3\pi r_s)$.

\begin{figure}[!tbp]
\includegraphics[width=0.85\columnwidth]{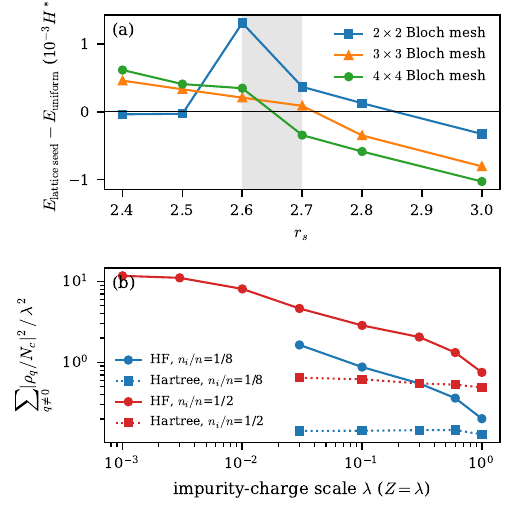}
\caption{(a) Lattice-seeded clean energy minus the exact uniform HF energy for three Bloch meshes; the band marks the estimated crossing. (b) Density power per $\lambda^2$ at $r_s=2.4$ with the charges of configuration A scaled by $\lambda$.}
\label{hf:fig:conv}
\end{figure}

Figure~\ref{hf:fig:conv}(a) shows the comparison. On the $4\times4$ mesh the crystal lies $E_u$ by $3.4\times10^{-4}$, $5.9\times10^{-4}$, and $1.0\times10^{-3}H^*$ at $r_s=2.7$, $2.8$, and $3.0$. At $r_s=2.5$ and $2.6$ the lattice seed yields only a weak modulation (density contrast $C\le0.006$), and at $r_s=2.4$ it relaxes to uniform density. These differences are comparable to the remaining mesh error, so we regard $r_c\simeq2.65$ as an estimated crossing within this search; our conclusions do not depend on its precise value. It agrees with the boundary $r_s\approx2.7$ which the fully polarized unrestricted HF ground state becomes an incommensurate triangular crystal with about 11\% more sites than electrons~\cite{Bernu2008}. With spin freedom the commensurate crystal is favored for $r_s>1.22$~\cite{Bernu2011}. Our commensurate search excludes incommensurate states~\cite{Trail2003,Bernu2008}. The uniform state is our model of the electron liquid, and the results are the response of this reference. We study the liquid side at $r_s=2.0$ and $2.4$.

The disordered calculations use a $2\times2$ mesh. Repeating configuration A at $r_s=2.4$ and $n_i/n=1/8$ on a $4\times4$ mesh raises the registered fraction (defined) from 0.63 to 0.94 and the local bond order from 0.52 to 0.61; for this case the coarse mesh understates crystalline order.

\begin{figure*}[!htbp]
\includegraphics[width=0.9\textwidth]{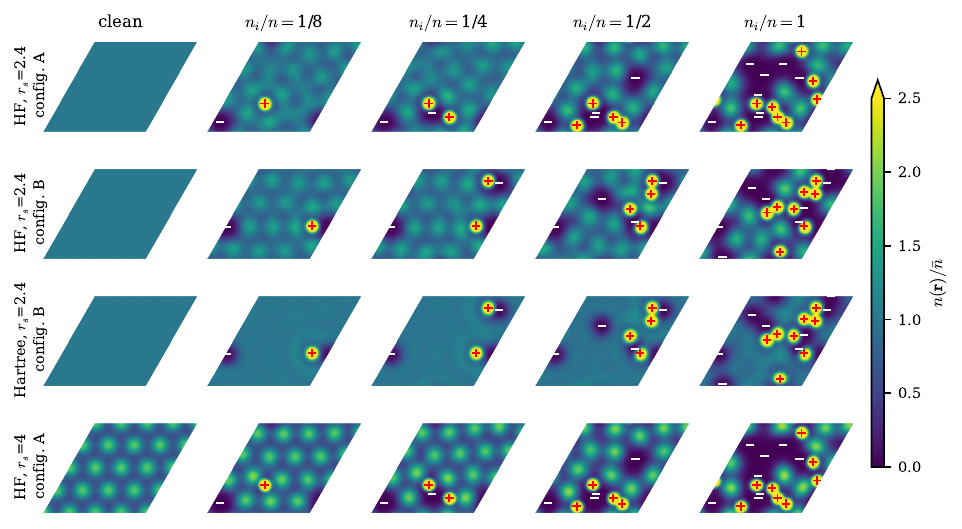}
\caption{Self-consistent densities $n(\mathbf{r})/\bar n$ in one 16-electron supercell for increasing $n_i/n$. Red $+$ and white $-$ mark attractive and repulsive impurities. Rows as labeled; $r_s=4$ is the clean-crystal reference.}
\label{hf:fig:maps}
\end{figure*}

\subsection{Exchange-enhanced, nonlinear response}

To isolate the response mechanism we scale all impurity charges by $\lambda$ at fixed positions [Fig.~\ref{hf:fig:conv}(b)]. The Hartree response reproduces the linear response of the uniform reference to within 10\% at $\lambda=0.03$, using an independently computed static susceptibility. The HF response reaches its linear plateau only for $\lambda\lesssim0.003$. On that plateau the total density power at $n_i/n=1/2$ exceeds the Hartree value by a factor of 18, which corresponds to an rms amplitude ratio of about 4.2. For the spin-polarized gas $2k_F/G_1=(2\sqrt3/\pi)^{1/2}\simeq1.05$, so the $2k_F$ scale of this response lies next to the first reciprocal shell of the Wigner lattice. The corresponding real-space Friedel period is $\pi/k_F\simeq0.825a$.

At $\lambda=1$ the HF power per $\lambda^2$ has fallen 16-fold its plateau. Impurities with $Z=1$ act far outside the linear regime, so the density patterns are nonlinear responses.

\subsection{Structural measures}

We locate maxima of $n(\mathbf{r})$ exceeding $\bar n$. A maximum within $0.5a$ of a positive impurity is labeled \emph{bound}; the others are \emph{free}. For every maximum we compute the charge $Q_j$ in its Voronoi cell and compare it with the charge $e\bar nA_j$ that the uniform density would place in the same cell. For the free maxima we compute the mean local bond order $\langle|\psi_6|\rangle$, where $|\psi_6|=|z^{-1}\sum_je^{6i\theta_j}|$ is taken over the $z$ Delaunay neighbors. For all maxima we compute the registered fraction: the fraction of pristine lattice sites lying within $0.25a$ of a maximum after optimal translation.

Each state is compared with a null model matched to its selection rules. The bound maxima are kept in place. The free maxima are placed at random subject to the two conditions that hold in the data: they stay at least $0.5a$ from every positive impurity, and they are separated from all other maxima by at least the minimum separation observed among the free maxima of that state. The null configurations pass through the identical triangulation and registration pipeline. We report the percentile of each observed quantity among 300 null samples. A percentile of 100 means that the observed value exceeds every sample, i.e., an estimated tail probability $1/300$. Percentiles measure unusualness relative to each state's own null, not absolute order.

\begin{figure}[!tbp]
\includegraphics[width=0.85\columnwidth]{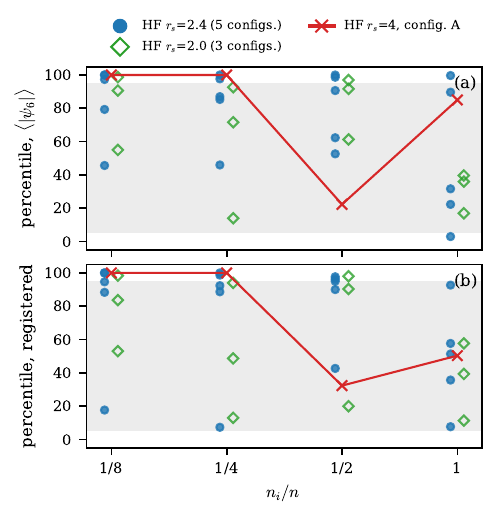}
\caption{Percentile of (a) the mean local bond order of free maxima and (b) the registered fraction, each among 300 samples of a null model matched to the individual state. Shading: 5--95 percentile range. Line: $r_s=4$ crystal.}
\label{hf:fig:null}
\end{figure}

\subsection{Density maxima}

Figure~\ref{hf:fig:maps} shows representative densities. On the liquid side, every HF state with impurities contains 14--17 maxima for 16 electrons. The Hartree densities with the same impurities contain 11--21 maxima, with no fixed relation to the electron number. The Voronoi cells of free maxima hold 0.94--0.98 of their uniform-density charge for $n_i/n\le1/4$ and 0.78--0.86 at $n_i/n=1$. Free maxima therefore redistribute charge within their cells rather than accumulating it. Bound maxima hold 1.2--2.1 times the uniform-density charge. These are features of the density; we do not address orbital localization, and the supercell admits Bloch states.

\subsection{Order versus impurity density}

Figure~\ref{hf:fig:null} gives the percentiles. At $r_s=2.4$ the median percentiles of the local bond order and the registered fraction across the five configurations are 97 and 95 at $n_i/n=1/8$; 87 and 92 at $1/4$; 91 and 95 at $1/2$; and 32 and 51 at $n_i/n=1$. Two configurations at $n_i/n=1/8$ reach a registered fraction of 0.94 with $\langle|\psi_6|\rangle=0.72$ and 0.81, exceeding every null sample: crystal-like order across the cell. Results at $r_s=2.0$ are similar. The sixfold-coordinated fraction of free maxima drops from 0.68 at $n_i/n=1/8$ to 0.26 at $n_i/n=1$.

The clean crystal at $r_s=4$ with configuration A sits at the 100th percentile in both measures for $n_i/n\le1/4$. It falls to the 22nd and 32nd percentiles at $n_i/n=1/2$ and lies at the 85th and 50th at $n_i/n=1$. At $n_i/n=1/2$ the liquid-side states thus remain in the upper tail of their nulls while this crystal-side state does not.

\subsection{Origin of the dilute-limit order}

In a 16-electron periodic cell, the impurities at $n_i/n=1/8$ are a single $\pm$ pair repeated with period $4a$, an arrangement commensurate with the triangular lattice. The Hartree control at $n_i/n=1/8$ has registered-fraction percentiles 15--85 with median 81: elevated, but the HF median of 95. This is consistent with a commensurate imprint that exists without exchange and that exchange strengthens, both through the enhanced response and by organizing the density into maxima close in number to the electrons. A 36-electron supercell separates the two contributions. There the local bond order of free maxima exceeds at least 98\% of the matched null samples for $n_i/n\le2/9$ in every configuration and falls into the null range at $n_i/n=1$, while the cell-wide registered fraction stays at 0.3--0.5. The local order is robust; the cell-wide registry of the smaller cell includes a commensurate contribution.

\subsection{Short-range disorder}

We provide in  the calculated results for HF theory in the presence of short-range Gaussian random disorder of rms strength $W$, computed in the same cell and tested against the same null model. There the disorder does not produce apparent local crystalline order, in contrast to the Coulomb case: on the liquid side the median bond-order percentile stays between 34 and 87 for $W/E_F=0.5$--10. It drives both the Wigner crystal (large $r_s$) and the liquid (small $r_s$) into the amorphous-solid regime through disorder-induced crossovers. The local crystalline order found here is thus tied to dilute charged impurities.

\subsection{Strong disorder}

As $n_i/n$ increases, attractive centers acquire maxima, repulsive centers expel density, and random $\pm$ charges generate long-wavelength potential variations ($v(q)\propto1/q$). At $n_i/n=1$ the typical order of liquid-side states falls toward the null level, with substantial configuration dependence, and the density becomes an amorphous pattern that follows the impurity configuration. Section~\ref{sec:hf_sm} collects the 16-electron results in a diagram in the $r_s$--$n_i/n$ plane.

\subsection{Structural observables relevant to STM imaging}

Within the specific HF model and parameter range reported above, three structural observations connect to the experimental literature on disordered 2D electron solids imaged by scanning tunneling microscopy~\cite{Xiang2025,Ge2026}; we present them as structural signatures of the amorphous-solid picture, leaving comparison with the published image analyses to the experimental literature.

(i) Density maxima at the interelectron spacing arise from a uniform reference state under charged disorder at every impurity density studied. Such maxima alone do not discriminate between a liquid response, a pinned crystal, and an impurity-templated state. Crystalline information resides in order beyond nearest neighbors, tested against a matched null model and compared with the mapped defect positions.

(ii) Charged impurities up to $n_i/n=1/2$ produce density patterns that are typically more ordered than the matched null in a state that is uniform when clean. In the 16-electron cell some configurations reach crystal-like registry across the cell; in the 36-electron cell the enhanced order is local. This order depends on the impurity configuration and likely includes a contribution from the commensurate periodic cell. Short-range disorder produces nonperiodic density puddles instead, so the two defect populations characterized separately in Ref.~\cite{Ge2026} are expected to affect density images differently. The experimental defect arrangement is neither periodic nor commensurate. Its effect on the extent of such order is a separate question that our calculation does not settle.

(iii) The amorphous regime at $n_i/n\simeq1$ is the structural counterpart of the strong-disorder picture of Babbar \textit{et al.}~\cite{Babbar2026a}, in which the low-density solid is controlled by the random impurity potential rather than by a clean crystalline order parameter. Their identification rests on transport, which we do not compute. Random charged impurities, modeled here, are the dominant scatterers in doped 2D semiconductors~\cite{Ando1982,Huang2024}. A quantitative comparison with Ref.~\cite{Ge2026} ($r_s\approx18$) requires the defect density per carrier, impurity setback, and gate screening.

%% file: body_hf_sm.tex
\subsection{Diagram in the $r_s$--$n_i/n$ plane}
\label{sec:hf_sm}
\label{hfsm:sec:map}

\textit{Data.}---The diagram uses the 16-electron supercell ($4\times4$ triangular), 163 plane waves, a $2\times2$ Bloch mesh, and $k_BT=0.005E_F$, with charged impurities of valence $Z=1$ in equal numbers of each sign. All points use one nested impurity configuration (configuration A of the preceding subsection). The grid is $r_s=2.0,\,2.4,\,3,\,4,\,8,\,16$ and $n_i/n=0,\,1/8,\,1/4,\,1/2,\,1,\,2,\,10$, i.e., 0, 2, 4, 8, 16, 32, and 160 impurities per supercell. For each point, HF solutions are started from the clean state (uniform for $r_s<r_c$, crystalline otherwise), the Hartree state with the same impurities, and, where available, a third seed: the solution at the next lower impurity density, or a lattice-seeded state for $r_s=2.4$ and for $r_s=8$ and 16 at $n_i/n\le1/4$. The lowest free energy is retained. All states satisfy a density-matrix residual below $10^{-6}$, except $r_s=16$ at $n_i/n=1/8$, which is not fully converged.

\textit{Measures.}---Two quantities classify each state.
(i) The first-shell coherence length $\xi_B$ is the first $e^{-1}$ crossing of the demodulated first-shell density correlation, evaluated on the density clipped at $2\bar n$. The clipping removes the peak height of maxima at attractive centers, which is set by the short-distance regularization. In this cell $\xi_B$ is bounded above by half the cell side, $2a$, and below by the value $0.64a$ produced by a featureless spectrum.
(ii) The bound fraction $f_b$ is the fraction of density maxima lying within $0.5a$ of an attractive impurity.

\textit{Classification.}---A state is labeled \emph{WS} if $\xi_B>a$, i.e., if first-shell density-wave coherence extends beyond one pristine lattice spacing within the cell. It is labeled \emph{AS} (amorphous solid) if $\xi_B\le a$ and $f_b\ge0.4$, i.e., if the density is organized around the impurities, with a large fraction of maxima at attractive centers. States with $\xi_B\le a$ and $f_b<0.4$ are labeled \emph{FL} for $r_s<r_c$, where the clean reference is uniform and the density is the static response of the liquid. The same states are labeled \emph{crossover} for $r_s>r_c$. The labels refer to structural regimes in a finite cell with quenched disorder; they do not identify thermodynamic phases or broken symmetries. With random pinning no state has long-range crystalline order, and ``WS'' denotes coherence exceeding the accessible range of the cell. The two thresholds separate the data cleanly: $\xi_B$ is either $\le0.81a$ or $\ge1.31a$, and $f_b$ is either $\le0.38$ or $\ge0.50$.

\begin{figure}[h]
\includegraphics[width=0.62\textwidth]{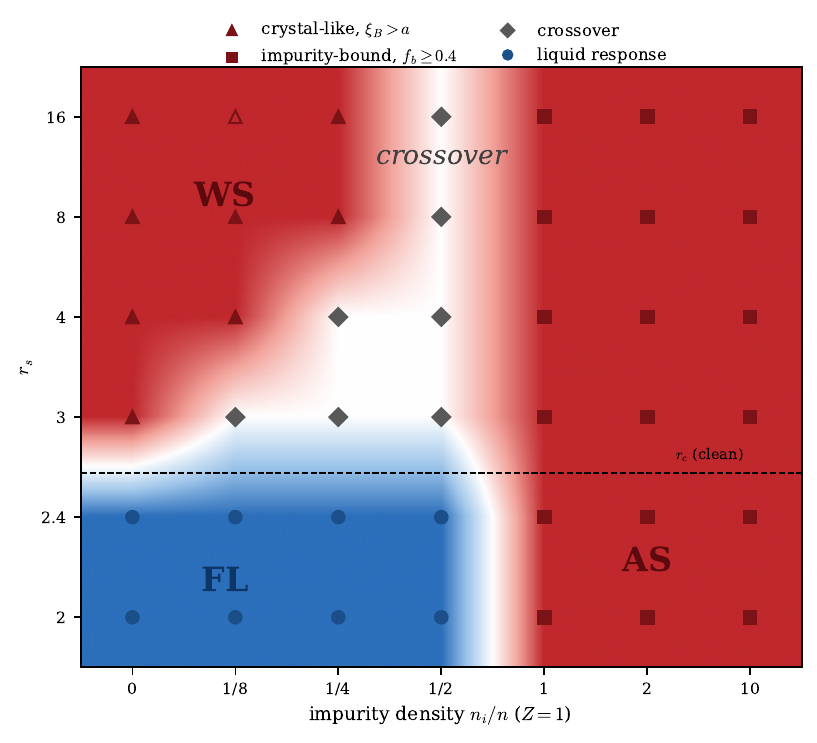}
\caption{Structural regimes of the 16-electron HF calculation in the $r_s$--$n_i/n$ plane at $Z=1$. Symbols mark computed states: triangles, $\xi_B>a$ (WS); squares, $f_b\ge0.4$ (AS); circles, FL; diamonds, crossover. The open triangle is not fully converged. The background color interpolates between point labels (blue, FL; white, crossover; red, WS and AS) and is a guide to the eye. The dashed line marks the estimated clean crossing $r_c\simeq2.65$ of the commensurate search. Both axes list the sampled values.}
\label{hfsm:fig:map}
\end{figure}

\begin{table}[h]
\caption{Coherence length $\xi_B/a$ (clipped density; $>2$ denotes the cell ceiling) and bound fraction $f_b$ for each point of Fig.~\ref{hfsm:fig:map}. Dashes: uniform clean state.}
\label{hfsm:tab:map}
\begin{ruledtabular}
\begin{tabular}{c|ccccccc|ccccccc}
 & \multicolumn{7}{c|}{$\xi_B/a$ at $n_i/n=$} & \multicolumn{7}{c}{$f_b$ at $n_i/n=$}\\
$r_s$ & 0 & 1/8 & 1/4 & 1/2 & 1 & 2 & 10 & 0 & 1/8 & 1/4 & 1/2 & 1 & 2 & 10\\ \hline
2.0 & -- & 0.62 & 0.62 & 0.62 & 0.64 & 0.62 & 0.67 & 0 & 0.06 & 0.12 & 0.25 & 0.50 & 0.94 & 1.00\\
2.4 & -- & 0.69 & 0.63 & 0.62 & 0.64 & 0.63 & 0.67 & 0 & 0.06 & 0.12 & 0.25 & 0.50 & 0.94 & 1.00\\
3 & $>2$ & 0.70 & 0.67 & 0.63 & 0.64 & 0.62 & 0.67 & 0 & 0.06 & 0.12 & 0.25 & 0.50 & 0.94 & 1.00\\
4 & $>2$ & $>2$ & 0.81 & 0.64 & 0.63 & 0.66 & 0.67 & 0 & 0.06 & 0.12 & 0.25 & 0.50 & 1.00 & 1.00\\
8 & $>2$ & $>2$ & $>2$ & 0.70 & 0.62 & 0.65 & 0.67 & 0 & 0.12 & 0.19 & 0.38 & 0.75 & 1.00 & 1.00\\
16 & $>2$ & 1.31 & $>2$ & 0.65 & 0.65 & 0.65 & 0.68 & 0 & 0.07 & 0.13 & 0.27 & 0.75 & 1.00 & 1.00\\
\end{tabular}
\end{ruledtabular}
\end{table}

\textit{Features.}---Four regimes appear. For $r_s<r_c$ and $n_i/n\le1/2$ the density is the static response of the uniform reference (FL): it is strongly modulated and locally ordered (Fig.~\ref{hf:fig:null}), but first-shell coherence does not extend beyond one lattice spacing and few maxima sit on impurities. For $n_i/n\ge1$ the density is organized by the impurities (AS) at every $r_s$ studied, whether the clean reference is uniform or crystalline: half or more of the maxima sit on attractive centers and coherence is at the resolution floor. Coherence beyond the lattice spacing (WS) occupies the clean crystal and the dilute-impurity region above $r_c$, and this region widens with $r_s$: at $r_s=3$ it is limited to the clean system, and at $r_s=8$ and 16 it extends to $n_i/n=1/4$. Between WS and AS on the crystal side lies a crossover region in which first-shell coherence is lost while fewer than 40\% of the maxima are bound to impurities. The layout parallels the regimes discussed in Ref.~\cite{Babbar2026a}: a weak-disorder crystalline regime at large $r_s$, a disorder-dominated regime at large impurity density, and a liquid at weak interaction and weak disorder.

\clearpage
\subsection{36-electron supercell}
\label{hfsm:sec:36}

\textit{Setup.}---The supercell spans $6\times6$ primitive cells and contains $N_c=36$ electrons. The plane-wave cutoff is the same as in the preceding subsection ($|\mathbf{G}|\le1.67G_1$, 367 plane waves). We sample the Brillouin zone at $\Gamma$ only, which for this cell gives a $k$-point density comparable to the $2\times2$ mesh of the 16-electron cell. With this sampling the clean lattice-seeded state relaxes to uniform density at $r_s=2.0$ and $2.2$ ($C<10^{-4}$). At $r_s=2.4$ it keeps a weak commensurate modulation ($C=0.003$). Impurity configurations in this cell are drawn independently of those in the 16-electron cell and are labeled A$'$--C$'$. The liquid-side calculations use $r_s=2.0$ (configuration A$'$) and $r_s=2.2$ (configurations A$'$--C$'$). The clean crystal at $r_s=4$ (configuration A$'$) serves as the crystalline reference. Integer, charge-neutral impurity numbers give $n_i/n=1/9$, $2/9$, $1/2$, and 1 (4, 8, 18, and 36 impurities). Seeds, selection by free energy, and the matched null model (bound maxima fixed; free maxima at least $0.5a$ from attractive impurities and separated by the observed minimum spacing) are as in the preceding subsection. HF states are compared with 200 null samples, Hartree states with 100, and the $r_s=4$ states with 150.

\begin{figure}[h]
\includegraphics[width=0.92\textwidth]{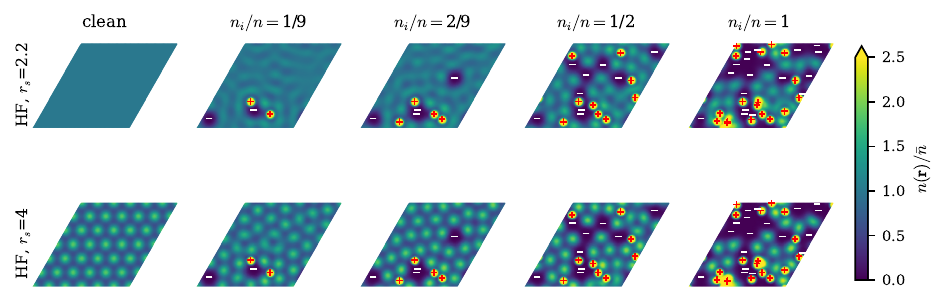}
\caption{Self-consistent HF densities in the 36-electron supercell for configuration A$'$ at $r_s=2.2$ (clean liquid) and $r_s=4$ (clean crystal). Red $+$ and white $-$ mark attractive and repulsive impurities. Color scale capped at $2.5\bar n$.}
\label{hfsm:fig:maps36}
\end{figure}

\begin{figure}[h]
\includegraphics[width=0.48\textwidth]{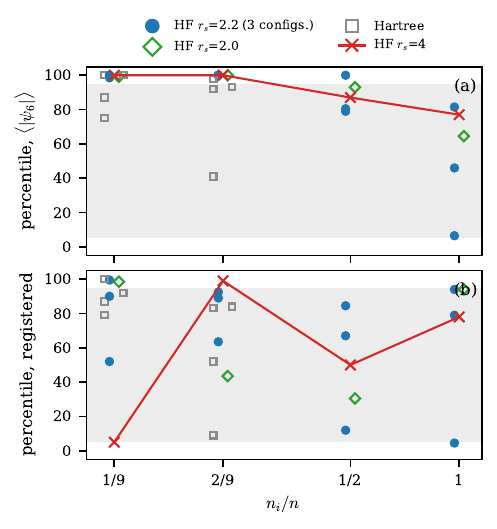}
\caption{36-electron supercell: percentile of (a) the mean local bond order of free maxima and (b) the registered fraction within the matched null model. Shading: 5--95 percentile range. Filled circles, HF at $r_s=2.2$ (three configurations); open diamonds, HF at $r_s=2.0$; open squares, Hartree with the same impurities at $n_i/n=1/9$ and $2/9$; line, HF at $r_s=4$ (configuration A$'$).}
\label{hfsm:fig:null36}
\end{figure}

\begin{table}[h]
\caption{36-electron HF results. $M$: number of density maxima; $M_b$: maxima bound to attractive impurities; $\langle|\psi_6|\rangle$: mean local bond order of free maxima, with percentile within the matched null in parentheses; ``reg.'': registered fraction, with percentile; $Q/e\bar nA$: mean ratio of the Voronoi charge of free maxima to the uniform-density charge of the same cell.}
\label{hfsm:tab:36}
\begin{ruledtabular}
\begin{tabular}{llcccccc}
$r_s$, config. & $n_i/n$ & $M$ & $M_b$ & $\langle|\psi_6|\rangle$ (pct.) & reg. (pct.) & sixfold & $Q/e\bar nA$\\ \hline
2.2, A$'$ & 1/9 & 37 & 2 & 0.42 (98.5) & 0.50 (99.5) & 0.26 & 0.97\\
       & 2/9 & 37 & 4 & 0.52 (100) & 0.33 (63.5) & 0.61 & 0.95\\
       & 1/2 & 36 & 9 & 0.40 (79) & 0.25 (12) & 0.44 & 0.90\\
       & 1   & 31 & 16 & 0.36 (46) & 0.33 (94) & 0.33 & 0.84\\
2.2, B$'$ & 1/9 & 40 & 2 & 0.45 (100) & 0.31 (52) & 0.47 & 0.98\\
       & 2/9 & 38 & 4 & 0.49 (100) & 0.42 (92.5) & 0.59 & 0.96\\
       & 1/2 & 36 & 9 & 0.40 (80.5) & 0.39 (84.5) & 0.52 & 0.92\\
       & 1   & 36 & 17 & 0.32 (6.5) & 0.36 (79) & 0.42 & 0.89\\
2.2, C$'$ & 1/9 & 38 & 2 & 0.51 (100) & 0.36 (90) & 0.39 & 0.98\\
       & 2/9 & 37 & 4 & 0.49 (100) & 0.42 (89) & 0.52 & 0.95\\
       & 1/2 & 41 & 9 & 0.54 (100) & 0.33 (67) & 0.44 & 0.91\\
       & 1   & 35 & 19 & 0.40 (81.5) & 0.22 (4.5) & 0.44 & 0.82\\
2.0, A$'$ & 1/9 & 36 & 2 & 0.43 (99) & 0.42 (98.5) & 0.32 & 0.98\\
       & 2/9 & 36 & 4 & 0.52 (100) & 0.31 (43.5) & 0.53 & 0.95\\
       & 1/2 & 36 & 9 & 0.40 (93) & 0.25 (30.5) & 0.41 & 0.91\\
       & 1   & 31 & 16 & 0.38 (64.5) & 0.33 (94) & 0.40 & 0.84\\
4, A$'$ & 1/9 & 36 & 2 & 0.69 (100) & 0.22 (5) & 0.71 & --\\
       & 2/9 & 36 & 4 & 0.58 (100) & 0.58 (99) & 0.59 & --\\
       & 1/2 & 35 & 9 & 0.41 (87) & 0.31 (50) & 0.38 & --\\
       & 1   & 31 & 17 & 0.39 (77) & 0.28 (78) & 0.36 & --\\
\end{tabular}
\end{ruledtabular}
\end{table}

\textit{Results.}---The 36-electron calculations reproduce the trends of the 16-electron cell (Figs.~\ref{hfsm:fig:maps36}, \ref{hfsm:fig:null36}; Table~\ref{hfsm:tab:36}). On the liquid side the HF density contains 31--41 maxima for 36 electrons. At dilute impurity density ($n_i/n=1/9$ and $2/9$) the local bond order of free maxima exceeds at least 98\% of the matched null samples in every configuration, with $\langle|\psi_6|\rangle=0.42$--$0.52$. It decreases at $n_i/n=1/2$ (percentiles 79--100) and falls into the null range at $n_i/n=1$ (percentiles 6.5--81.5), where 16--19 of the maxima sit on attractive centers. The $r_s=4$ crystal shows the same approach to the null at $n_i/n\ge1/2$. The Hartree densities with the same impurities contain 31--38 maxima. Their mean local bond order at $n_i/n=1/9$ and $2/9$ is 0.41, against 0.48 for HF, and their bond-order percentiles span 41--100.

Two features differ from the 16-electron cell. The registered fraction at dilute impurity density is 0.31--0.50, whereas two 16-electron configurations reached 0.94; its percentiles range from 43.5 to 99.5. Cell-wide registry is therefore weaker in the larger cell, while local order persists. This indicates a commensurate contribution to the cell-wide registry of the 16-electron cell, where the impurities at $n_i/n=1/8$ form a single $\pm$ pair repeated with period $4a$. The Voronoi charge of free maxima is 0.95--0.98 of the uniform-density charge of the same cell at dilute impurity density and 0.82--0.89 at $n_i/n=1$, as in the 16-electron cell: free maxima redistribute charge within their cells, and charge accumulates at attractive centers.

\textit{Additional 16-electron data.}---At $r_s=2.0$ (three configurations) the median percentiles of local bond order and registered fraction in the 16-electron cell are 91 and 84 at $n_i/n=1/8$, 72 and 49 at $1/4$, 92 and 90 at $1/2$, and 36 and 39 at $n_i/n=1$, following the $r_s=2.4$ trend of the preceding subsection.

\clearpage
\subsection{Short-range Gaussian disorder}
\label{hfsm:sec:sr}

\textit{Model and method.}---We replace the charged-impurity term of Eq.~(\ref{hf:eq:H}) by a Gaussian white-noise potential $V(\mathbf{r})$. It is generated on a $32\times32$ grid spanning the supercell (grid spacing $a/8$) from independent unit-normal variables, with the mean removed and the rms normalized to $W$, and it enters the Fock matrix through its Fourier coefficients on the supercell reciprocal vectors of the basis. The potential is zero-range on the scale of the lattice spacing, with the grid spacing as its short-distance cutoff. All other elements are those of the charged-impurity calculations: the 16-electron cell, 163 plane waves, the $2\times2$ Bloch mesh, $k_BT=0.005E_F$, three seeds (the clean state, the Hartree state with the same potential, and a lattice-seeded state), and selection by free energy. We use $W/E_F=0.5$, 1, 2, 5, and 10, with three disorder realizations at $r_s=2.4$ and one at $r_s=2.0$ and $r_s=4$. All states satisfy a density-matrix residual below $10^{-6}$. Because $W$ is fixed in units of $E_F$, its ratio to the Coulomb scale falls as $r_s^{-1}$, unlike the charged-impurity case, where the ratio of bare energy scales is independent of $r_s$. The structural measures and the matched null model are those of the preceding subsection; with no discrete impurity centers, all density maxima are free, and the null model randomizes all of them subject to the observed minimum separation (300 samples).

\begin{figure}[h]
\includegraphics[width=0.95\textwidth]{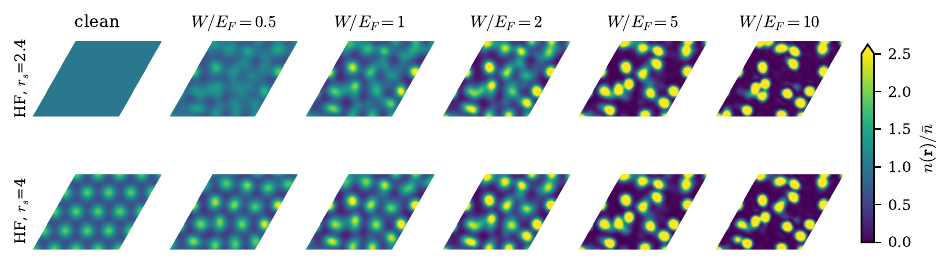}
\caption{Self-consistent HF densities with short-range Gaussian disorder of rms strength $W$ (one realization) on the liquid side, $r_s=2.4$, and on the crystal side, $r_s=4$. Color scale capped at $2.5\bar n$.}
\label{hfsm:fig:srmaps}
\end{figure}

\begin{figure}[h]
\includegraphics[width=0.48\textwidth]{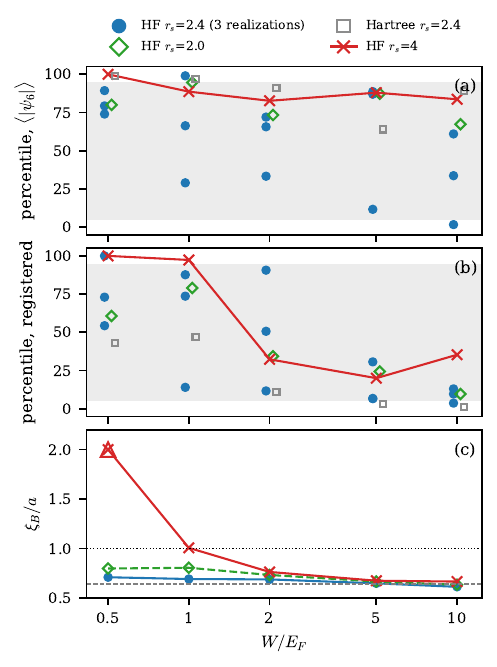}
\caption{Short-range disorder: percentile of (a) the mean local bond order and (b) the registered fraction within the matched null model, and (c) the first-shell coherence length $\xi_B$ of the clipped density. Shading in (a,b): 5--95 percentile range. Filled circles, HF at $r_s=2.4$ (three realizations; median in (c)); open diamonds, HF at $r_s=2.0$; open squares, Hartree at $r_s=2.4$; line, HF at $r_s=4$. In (c) the open triangle marks the cell ceiling $\xi_B>2a$, the dotted line $\xi_B=a$, and the dashed line the resolution floor $0.64a$.}
\label{hfsm:fig:srnull}
\end{figure}

\textit{Results.}---On the liquid side (Figs.~\ref{hfsm:fig:srmaps} and \ref{hfsm:fig:srnull}), short-range disorder turns the uniform density into an increasingly inhomogeneous, nonperiodic landscape of density puddles: the density contrast at $r_s=2.4$ grows from $C=0.02$--0.05 at $W/E_F=0.5$ to 2.7--3.0 at $W/E_F=10$. The median percentile of the local bond order stays within the matched-null range at every disorder strength; individual realizations lie between the 2nd and 99th percentiles without a systematic trend. The median over the three realizations at $r_s=2.4$ is 79, 66, 66, 87, and 34 for $W/E_F=0.5$, 1, 2, 5, and 10, and the values at $r_s=2.0$ are 80, 95, 73, 87, and 67. The coherence length stays at 0.61--0.85$a$, and the registered fraction falls from percentiles 54--100 at $W/E_F=0.5$ to 4--13 at $W/E_F=10$, below the null: the density is increasingly selected by the disorder rather than by lattice registry. The Hartree control with the same potential gives bond-order percentiles 64--99, comparable to HF: exchange does not enhance local order for this disorder.

On the crystal side ($r_s=4$) weak disorder preserves the crystal: at $W/E_F=0.5$ the coherence length exceeds the $2a$ ceiling and every lattice site is registered. At $W/E_F=1$ the coherence length is $1.01a$, at $W/E_F=2$ it falls to $0.76a$, and at $W/E_F=5$ and 10 it is $0.67a$, equal to the liquid-side values. The registered-fraction percentile drops from 100 to 20--35 over the same range, and at $W/E_F=10$ the density contrast ($C=2.8$) matches the liquid side. Increasing short-range disorder thus drives both the Wigner crystal at large $r_s$ and the Fermi liquid at small $r_s$ into a common disorder-selected amorphous-solid (AS) regime through disorder-induced crossovers; for $r_s=4$ the crossover lies near $W/E_F\simeq1$--2. These are structural crossovers, not thermodynamic phase transitions.

\textit{Comparison with charged impurities.}---The two disorder types agree at strong disorder: in both, liquid-side and crystal-side states converge to an amorphous density selected by the disorder. They differ on the liquid side at weak disorder. Dilute charged impurities produce local bond order above the matched null: median percentile 97 at $n_i/n=1/8$ in the 16-electron cell (Fig.~\ref{hf:fig:null}) and at least 98 in every 36-electron configuration at $n_i/n\le2/9$ (Sec.~\ref{hfsm:sec:36}). Short-range Gaussian disorder, analyzed identically, produces no systematic local order beyond the null at any strength. A plausible reason is that a dilute set of Coulomb centers acts at discrete points and leaves impurity-free regions, where the exchange-enhanced response of the liquid, peaked near $2k_F\simeq G_1$, organizes the density locally, whereas a zero-range Gaussian potential acts everywhere, with no preferred wave vector, and imposes its own nonperiodic pattern. In these calculations, apparent local crystalline order in the density of a liquid arises with dilute charged disorder and not with short-range disorder.

%% file: meta_sec3_vmc.tex

\section{Variational Monte Carlo of the 2{D} Electron System
with Charged Impurities}
\label{sec:vmc}

The Hartree--Fock analysis of Sec.~\ref{sec:hf_letter} is a
mean-field calculation.  To test whether its conclusions survive at
the fully correlated many-body level, we turn to variational
Monte Carlo (VMC) with Jastrow--Slater trial wave
functions~\cite{Foulkes2001}, in which the quantum crystallization
density of the clean 2D electron gas is in the range of a few tens
and depends on the spin sector: Drummond and Needs~\cite{Drummond2009}
report the paramagnetic-to-antiferromagnetic-crystal transition at
$\rc = 31(1)$, with the ferromagnetic crystal becoming stable at a
different density; the fully spin-polarized gas studied here enters
the Wigner regime in the same general range.
The HF estimate $\rc^{\mathrm{HF}} \simeq 2.65$ lies well below the
VMC/DMC estimate; a disorder effect that is visible in HF below
$\rc^{\mathrm{HF}}$ should also be visible in VMC below the VMC
$\rc$ if the mechanism is real.

\input{body_vmc}

%% file: body_vmc.tex
\subsection{Models, trial wave functions, diagnostics, and energy scales}
\label{vmc:sec:model}

\subsubsection{Hamiltonian}
We consider $N=16$ fully spin-polarized electrons in a rhombic periodic cell that accommodates a $4\times4$ triangular lattice of spacing $a=1.90463\,\rs\aB$, interacting through the 2D Coulomb interaction with a neutralizing background, in the presence of $N_i$ quenched point charges $Z_I$ (in units of $e$) located in a plane a distance $d$ from the electron plane at positions $\mathbf{R}_I$. In effective atomic units ($\aB=\hbar^2\kappa/me^2$, $\Ha=e^2/\kappa\aB$),
\begin{equation}
\begin{split}
H={}&-\frac12\sum_i\nabla_i^2+\sum_{i<j}\frac{1}{|\br_i-\br_j|}\\
&-\sum_{i,I}\frac{Z_I}{\sqrt{|\br_i-\mathbf{R}_I|^2+d^2}}+E_{\rm bg},
\end{split}
\label{vmc:eq:H}
\end{equation}
with all periodic images included through 2D Ewald sums for both the electron--electron and the electron--impurity terms, and $E_{\rm bg}$ the background terms, which are constants. The impurities are frozen; there is no disorder averaging, and each realization is a distinct Hamiltonian whose ground state is calculated separately. This is the appropriate description of an STM experiment, which images one realization.

\begin{figure}[!htbp]
\includegraphics[width=\columnwidth]{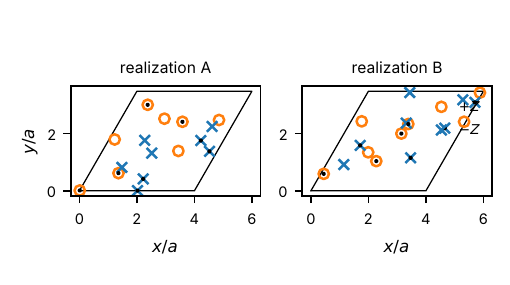}
\caption{The two random impurity landscapes used throughout, in units of the lattice spacing $a$ of the $4\times4$ triangular reference lattice. Circles: $Z_I=+Z$ (attractive); crosses: $Z_I=-Z$ (repulsive). The dots mark the 8-impurity subset nested in each 16-impurity set. In the donor-layer model all 16 sites of landscape A carry $Z_I=+1$. The same fractional positions are used at every $\rs$.}
\label{vmc:fig:landscapes}
\end{figure}

Three disorder models are studied, each at $\rs=10$, 20, 30, and 50.
\begin{enumerate}
\item[(i)] \emph{Strong point-like disorder}: $N_i=8$ or $16$, half with $Z_I=+1$ and half with $Z_I=-1$ (a neutral set), at $d=\aB$, for two independent random realizations A and B of the positions, with the 8-impurity set nested in the 16-impurity set. The same fractional positions are used at every density (Fig.~\ref{vmc:fig:landscapes}), so that the impurity spacing scales with the electron spacing and $n/n_i$ is fixed at $2$ ($N_i=8$) or $1$ ($N_i=16$).
\item[(ii)] \emph{Impurity-strength scan}: the 16-impurity landscape A (and B as a check at $\rs=20$) with $Z_I=\pm Z$, $Z=1/50$, $1/20$, $1/10$, $1/5$, and $1$, at $d=\aB$.
\item[(iii)] \emph{Donor layer}: $N_i=16$ impurities of charge $Z_I=+1$ at the 16 positions of landscape A, i.e., a neutral electron--donor system with $n/n_i=1$ as in a modulation-doped structure, at setbacks $d=1$, 5, 10, and $20\,\aB$.
\end{enumerate}
Model (i) is the strong-coupling limit in which each attractive impurity is a hydrogenic trap; model (ii) connects that limit continuously to the clean liquid; model (iii) spans, through the setback, the range from point-like traps to the smooth long-range potential fluctuations in which puddles rather than traps are the natural objects. Models (i) and (ii) coincide at $Z=1$; models (ii) and (iii) have the same impurity positions.

\subsubsection{Energy and length scales}
\label{vmc:sec:scales}
Three energies compete: the Fermi energy of the polarized gas $E_F=2/\rs^2$ ($0.020$, $0.0050$, $0.0022$, $0.0008\,\Ha$ at $\rs=10$, 20, 30, 50), the nearest-neighbor Coulomb energy $e^2/a=0.525/\rs$ ($0.0525$, $0.0263$, $0.0175$, $0.0105\,\Ha$), and the impurity binding energy. For a point charge $Z$ at setback $d\lesssim\aB$ the 2D hydrogenic binding energy is of order $2Z^2\,\Ha$ with orbit radius $\aB/2Z$; for $Z=1$ at $d=\aB$ the VMC gives a bound orbit of rms radius $1.5$--$1.8\aB$, i.e., a binding energy of order $0.5\,\Ha$, ten to fifty times $e^2/a$. This is the trapping regime: whatever the electrons do collectively, an attractive impurity captures one of them. For $Z=1/50$ the binding energy is $\sim10^{-3}\,\Ha$ and the orbit radius $\sim25\aB$, comparable to $a$; this is the perturbative regime. The crossover, where $2Z^2\sim e^2/a$, is at $Z\approx0.1$--$0.2$ for $\rs=20$--$30$, and that is where the physics of model (ii) changes.

For the donor layer the relevant comparison is different. A single donor at setback $d\gg\aB$ binds classically, with well depth $e^2/d$ ($0.2$, $0.1$, $0.05\,\Ha$ at $d=5$, 10, $20\aB$) and harmonic localization length $\ell=(d^3\aB)^{1/4}$ ($3.3$, $5.6$, $9.5\aB$). Both are again large against $e^2/a$ at every $\rs$ studied, so the setback alone never switches off single-donor trapping. What decides whether the landscape consists of individual traps or of smooth puddles is the ratio $d/a$ of the setback to the donor spacing (which equals the electron spacing, $n_i=n$): for $d\gtrsim a/2$ neighboring donor potentials overlap into a smooth landscape whose corrugation is of order $e^2/a$, while for $d\ll a$ each donor is its own trap. At fixed $d=10\aB$, $d/a=0.53$, $0.26$, $0.18$, $0.11$ at $\rs=10$, 20, 30, 50; at $d=20\aB$ the same numbers are doubled and at $d=5\aB$ halved; at $d=\aB$ every donor is a point trap at every density. Lowering the density at fixed disorder therefore takes the system from the smooth-potential regime to the trapping regime, which is the physical situation in a gate-tunable device, and raising the setback at fixed density does the reverse.

These numbers are the content of the paper in compressed form. At $\rs<r_c$ the electrons are a liquid, and whether they look solid in a density image is decided by the ratio of the disorder energy to $e^2/a$ and by the ratio of the disorder length to $a$, not by $\rs$.

\subsubsection{Trial wave functions}
We use conventional Slater--Jastrow VMC \cite{Foulkes2001}. Three trial families are optimized independently for every Hamiltonian:
\begin{align}
\Psi_{\rm PW}&=\det[e^{i\mathbf{k}_m\cdot\br_j}]\,e^{-U(\{\br\})}\,e^{W(\{\br\})},\\
\Psi_{\rm PG}&=\det[\phi_\beta(\br_j-\mathbf{s}_m)]\,e^{-U}\,e^{W},\quad \mathbf{s}_m\ \text{on the lattice},\\
\Psi_{\rm FG}&=\det[\phi_\beta(\br_j-\mathbf{c}_m)]\,e^{-U}\,e^{W},\quad \mathbf{c}_m\ \text{relaxed},
\end{align}
where the plane-wave (PW) determinant uses the 16 lowest closed-shell momenta of the cell, $\phi_\beta$ is a periodized Gaussian of optimized width $\beta^{-1/2}$, $U$ is a two-body Jastrow factor of the standard Pad\'e form with an optimized parameter, and $W=\ell\sum_{i,I}\kappa_I\,\mathrm{sgn}(Z_I)\exp[-(\sqrt{|\br_i-\mathbf{R}_I|^2+d^2}-d)/\ell]$ is a one-body impurity factor with independently optimized amplitudes $\kappa_\pm$ and range $\ell$, which lets electrons bind to attractive impurities and avoid repulsive ones. In the pinned-Gaussian family (PG) the orbital centers $\mathbf{s}_m$ are the sites of the pristine triangular lattice; in the free-center family (FG) the centers $\mathbf{c}_m$ are obtained by relaxing 16 classical point charges in the electron--electron Ewald potential plus the impurity potential smeared over the Gaussian orbital, starting from the lattice and from random positions and keeping the lowest-energy arrangement. The FG family exists precisely so that the calculation can find a disorder-adapted crystalline arrangement if one is energetically preferred: a locally crystalline state whose electrons sit at positions selected by the disorder is representable in FG but not in PW or PG. For the clean system PW and PG are used. All parameters are optimized by stochastic reconfiguration for every Hamiltonian; production runs use 24 Markov chains and 1200 blocks (28\,800 stored configurations per state), with errors from chain-to-chain and reblocked statistics. The lowest-energy family is used for the figures; all families are reported in the tables, where the lowest-energy one is marked with an asterisk. Computational details are in Appendix~\ref{sec:SM_vmc}.

VMC with these trial functions is not exact, and we do not claim quantitative accuracy for energies. The structural conclusions reported below are comparisons between the clean and disordered states of the same model in the same trial family, carried out in whichever families are physically applicable for each disorder model (both PW and PG for Model (i) at $Z=1$; all three families for Model (ii)'s charge scan where free-center optimization is defined; FG for the donor layer).  Within these subsets the structural inferences are consistent across the families compared, and that consistency, rather than any exact-ground-state claim, is the ground of the discussion that follows.

\subsubsection{Diagnostics: what a density image measures and what a crystal requires}
\label{vmc:sec:diag}
The structure factor $S(\bq)=\av{|\rho_\bq|^2}/N$, $\rho_\bq=\sum_j e^{i\bq\cdot\br_j}$, evaluated at the reciprocal vectors of the periodic cell, decomposes exactly as
\begin{equation}
S(\bq)=\underbrace{\frac{|\av{\rho_\bq}|^2}{N}}_{\Sst(\bq)}+\underbrace{\frac{\av{|\rho_\bq|^2}-|\av{\rho_\bq}|^2}{N}}_{\Sc(\bq)}.
\label{vmc:eq:decomp}
\end{equation}
The static part $\Sst$ is the Fourier power of the one-body density $n(\br)=\av{\sum_j\delta(\br-\br_j)}$; it is nonzero only when something external (here the impurities, or the trial lattice in PG/FG) selects positions. A time-averaged density image, which is what STM records in the simplest limit (Sec.~\ref{vmc:sec:stm}), measures $n(\br)$ and hence $\Sst$ and nothing else. The connected part $\Sc$ measures the correlation of density fluctuations; a clean crystal of $N$ electrons in a periodic cell, which has uniform $n(\br)$ because nothing selects its position, has all of its crystalline signal in $\Sc$, and a crystal pinned by disorder keeps its crystalline correlations in $\Sc$ while acquiring a static part. The central point of this paper is that disorder moves weight from $\Sc$ to $\Sst$. We quote $S$, $\Sst$, and $\Sc$ at the first star of reciprocal vectors $\{\bG_1\}$ of the pristine triangular lattice, which is the crystal wave vector and also where the liquid's $S(\bq)$ peaks, and we show full maps; we also record the maximum of $\Sc(\bq)$ over all wave vectors, which would detect a crystal of any orientation.

Real-space order is quantified by the pair correlation $g(r)$ and by the sixfold bond-orientational order $\psi_{6,i}=\frac16\sum_{j\in\mathcal N_6(i)}e^{6i\theta_{ij}}$ over the six nearest neighbors of electron $i$; we report the local amplitude $\av{|\psi_{6,i}|}$ and the global amplitude $\av{|\frac1N\sum_i\psi_{6,i}|}$, and for model (i) also the local amplitude of the unbound electrons alone, with the impurity-bound electrons removed from the neighbor lists. The clean liquid at $\rs=10$--20 has $\av{|\psi_6|}\simeq0.37$--$0.38$ and the clean $\rs=50$ crystal has $0.43$; the local amplitude is far from zero in the liquid because nearest-neighbor shells are always roughly sixfold at these densities. This is the quantitative form of the statement that every correlated liquid has crystalline coherence on the scale of $a$. Pinning is quantified by the mean number of electrons within a given distance of an attractive impurity, compared with the value that a uniform density would give (for eight attractive impurities, $72/\rs^2$ within $3\aB$ and $1.8$ within $a/4$). For the donor layer we also report the fraction of the cell with $n(\br)>\bar n/2$ and whether that region, or its complement, spans the periodic cell; on $N=16$ this is a statement about morphology, not a percolation calculation.

\subsubsection{What $\Sc$ measures and what it does not}
\label{vmc:sec:Sc_caveat}
A caveat to Eq.~(\ref{vmc:eq:decomp}) deserves to be stated plainly, because it bears on how the tables below should be read.  The connected structure factor $\Sc(\bq)$ measures the correlation of \emph{fluctuations} of $\rho_\bq$ about its mean, not positional order as such.  Positional order can reside wholly, or partly, or not at all in $\Sc$ depending on the trial state and the ensemble averaging.  A trial wave function whose electrons are tied to definite lattice sites, as in the lattice-Gaussian (PG) family, places the Bragg weight in $\Sst$ and leaves $\Sc$ to measure the residual fluctuations about the lattice; a trial wave function whose density is translation-invariant, as in the plane-wave (PW) family, places all crystalline signal that is there in $\Sc$ and has $\Sst=0$ identically.  The paper's own clean benchmarks at $\rs=50$ in Table~\ref{vmc:tab:z1} illustrate this: the clean PG state has $(S,\Sst,\Sc) \simeq (3.46, 2.28, 1.18)$, while the clean PW trial has $\Sc\simeq1.66$, so a ranking by $\Sc$ alone would favor the delocalized PW reference over the lattice-pinned PG trial that is at lower energy and is the physical reference crystal at this density.  The lesson is that $\Sc$ is not an orientation-independent crystallinity measure, and a decrease in $\Sc$ does not by itself establish the loss of positional order.  We therefore quote $S$ and $\Sst$ alongside $\Sc$, and rely on real-space diagnostics -- the sixfold bond amplitude $\av{|\psi_6|}$, the impurity-registered fraction of density maxima, the pair correlation $g(r)$, and the first-shell density coherence length -- as the primary structural inferences in the discussion below.  $\Sc$ enters as a fluctuation diagnostic that complements these, not as a standalone crystallinity measure.

\subsubsection{Definitions}
To keep the terminology fixed: a \emph{Wigner crystal} (WC) is interaction-driven periodic order of the electron density, with Bragg peaks; it exists only in the clean limit and is characterized by positional order that is coherent across the sample (reflected in $S(\bG)\propto N$, with the weight distributed between $\Sst$ and $\Sc$ according to the ensemble convention). A \emph{pinned Wigner crystal} is a WC whose translational coherence has been cut off by disorder at a length $\xi\gg a$ with a density of defects that is small inside domains; it retains crystalline correlations over many shells. An \emph{amorphous solid} (AS) is a state with a static, inhomogeneous density whose peaks are organized by the disorder, with crystalline correlations extending at most to the nearest-neighbor shell. ``Wigner solid'' (WS) is a term in common use in the imaging-experiment literature; in the HF classification of Sec.~\ref{sec:hf_letter} it was used as a finite-cell structural label for states with first-shell density-wave coherence extending beyond one lattice spacing in the cell, and the same shorthand is used here, with the operational structural meaning carried by the WC/pinned-WC/AS distinction above.  Part of the purpose of the structural diagnostics introduced here is to tell, within the present approximations, which of the above pictures a given calculated state resembles.

\subsection{Strong point-like disorder at $\rs=10$, 20, 30, 50}
\label{vmc:sec:strong}

\begin{figure*}[!htbp]
\includegraphics[width=\textwidth]{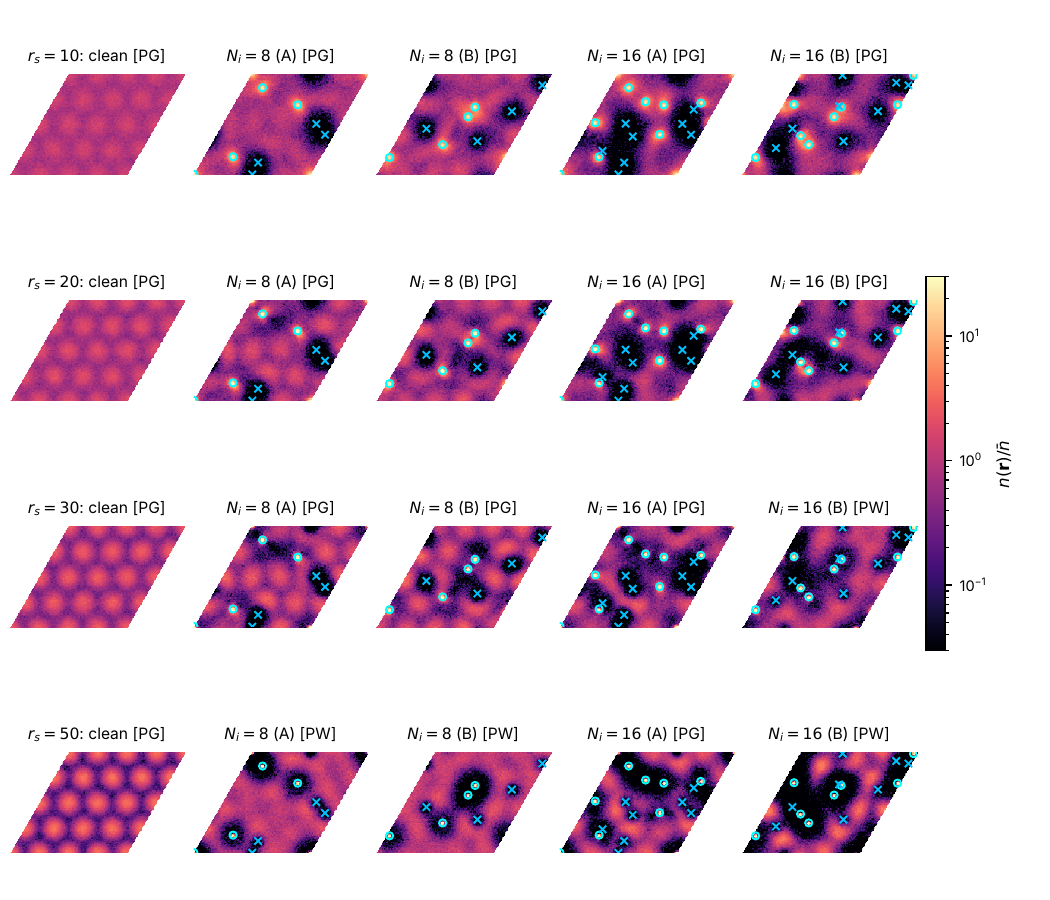}
\caption{One-body density $n(\br)/\bar n$ for model (i) ($Z=\pm1$, $d=\aB$), logarithmic color scale, lowest-energy trial family in each panel. Rows: $\rs=10$, 20, 30, 50; columns: clean, 8 impurities (A, B), 16 impurities (A, B). Circles: attractive impurities; crosses: repulsive ones. The lattice modulation in the clean PG panels is the imprint of the trial orbitals at $\rs=10$ and 20, an incipient crystal at $\rs=30$, and the clean crystal at $\rs=50$. In every disordered panel the bright spots sit on the attractive impurities and the dark holes on the repulsive ones, at every $\rs$.}
\label{vmc:fig:z1dens}
\end{figure*}

\begin{figure*}[!htbp]
\includegraphics[width=\textwidth]{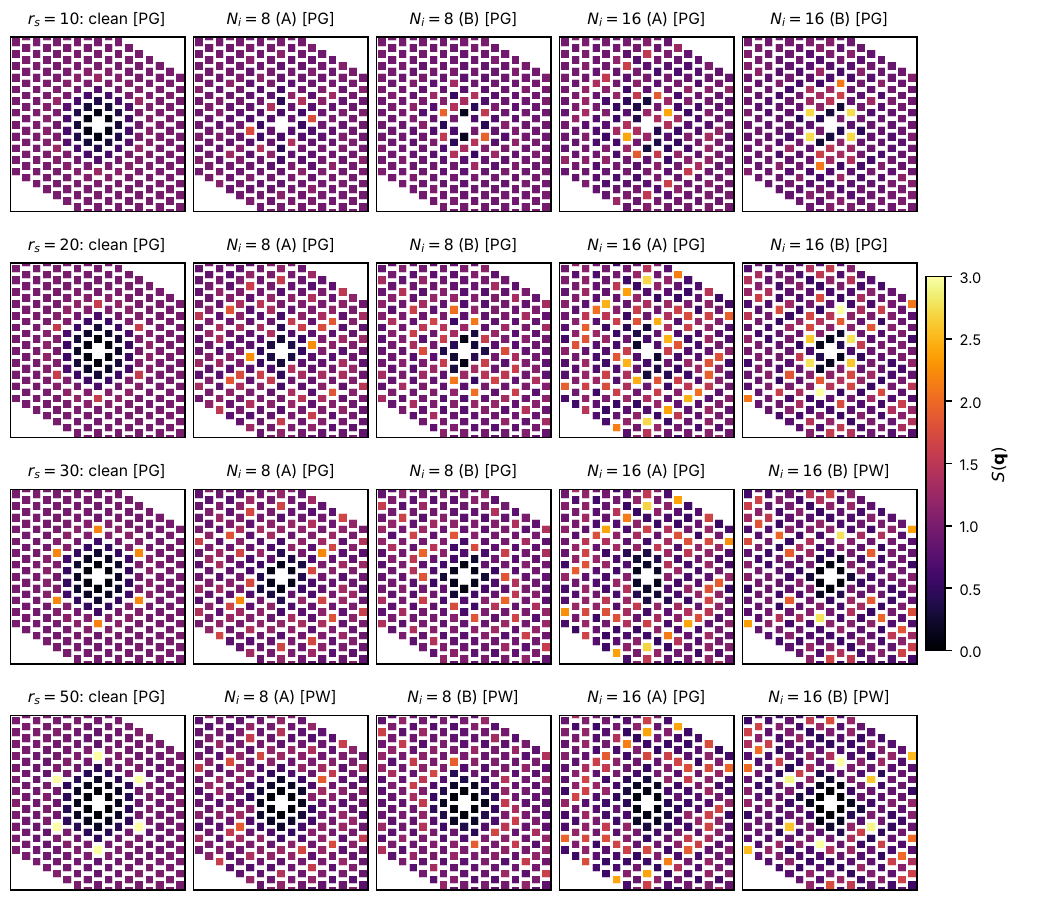}
\caption{Total structure factor $S(\bq)$ at the reciprocal vectors of the cell for the states of Fig.~\ref{vmc:fig:z1dens} (model (i), $Z=\pm1$, $d=\aB$; color scale 0--3, saturating). The clean $\rs=30$ and $50$ PG states show the six first-star spots of the (incipient) crystal. In the disordered states the strong spots are at irregular wave vectors set by the impurity geometry and are static (Fig.~\ref{vmc:fig:z1conn}).}
\label{vmc:fig:z1tot}
\end{figure*}

\begin{figure*}[!htbp]
\includegraphics[width=\textwidth]{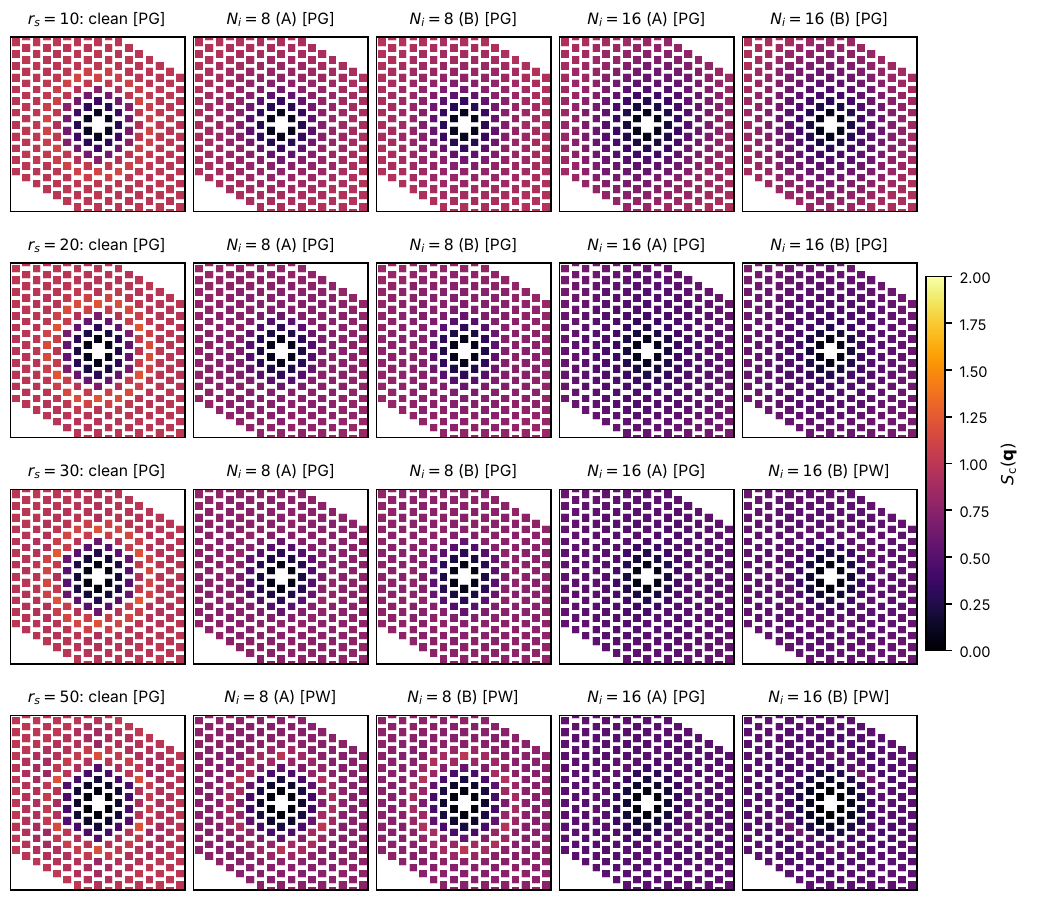}
\caption{Connected structure factor $\Sc(\bq)=S(\bq)-|\av{\rho_\bq}|^2/N$ for the same states (model (i), $Z=\pm1$, $d=\aB$; color scale 0--2). This is the part of $S(\bq)$ carried by many-body correlations. Disorder suppresses it at every $\rs$; at $\rs=30$ and 50 the Bragg star of the clean crystal is gone from $\Sc$ in every disordered panel.}
\label{vmc:fig:z1conn}
\end{figure*}

Figures~\ref{vmc:fig:z1dens}--\ref{vmc:fig:z1conn} and Table~\ref{vmc:tab:z1} contain the results for model (i): unit charges of both signs one Bohr radius from the plane, at $\rs=10$, 20, 30, and 50, for two impurity densities and two realizations. The physics is the same in all sixteen disordered Hamiltonians and is dictated by the energy scales of Sec.~\ref{vmc:sec:scales}.

{\small
\begin{longtable}{llccccccccc}
\caption{Model (i): $Z=\pm1$, $d=\aB$, at four densities. For every Hamiltonian and trial family: energy per electron (statistical error in the last digits; the lowest-energy family of each Hamiltonian is marked $^*$), mean number of electrons within $3\aB$ of an attractive impurity $N_{\rm b}$ (uniform-density value $36/\rs^2$ for $N_i=8$ and $72/\rs^2$ for $N_i=16$), total, static, and connected structure factor at the pristine lattice star $\{\bG_1\}$, local and global sixfold bond order of all electrons, and local bond order of the unbound electrons only ($\av{|\psi_6|}_{\rm m}$, computed with the bound electrons removed from the neighbor lists).}
\label{vmc:tab:z1}\\
\hline\hline
$\rs$ & case & family & $E$ ($\Ha$) & $N_{\rm b}$ & $S(\bG_1)$ & $\Sst(\bG_1)$ & $\Sc(\bG_1)$ & $\av{|\psi_6|}$ & $|\Psi_6|$ & $\av{|\psi_6|}_{\rm m}$\\
\hline
\endfirsthead
\multicolumn{11}{l}{\emph{Table \ref{vmc:tab:z1} (continued)}}\\
\hline\hline
$\rs$ & case & family & $E$ ($\Ha$) & $N_{\rm b}$ & $S(\bG_1)$ & $\Sst(\bG_1)$ & $\Sc(\bG_1)$ & $\av{|\psi_6|}$ & $|\Psi_6|$ & $\av{|\psi_6|}_{\rm m}$\\
\hline
\endhead
\hline
\multicolumn{11}{r}{\emph{continued on next page}}\\
\endfoot
\hline\hline
\endlastfoot
\input{table_i.tex}
\end{longtable}
}

\begin{figure}[!htbp]
\includegraphics[width=\columnwidth]{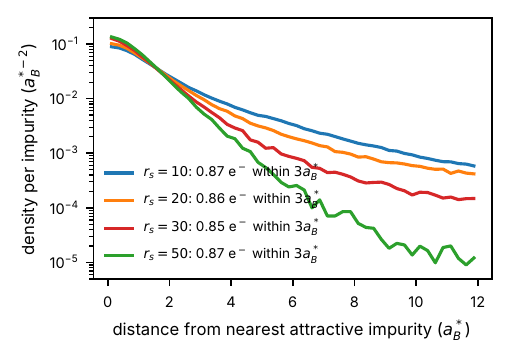}
\caption{Radial electron density around the nearest attractive impurity in model (i) ($Z=\pm1$, $d=\aB$), landscape 16A, at the four densities, in physical units. The bound orbit is identical at all four; the tails differ because the unbound background has a different mean density. The occupation within $3\aB$ is $0.86$--$0.88$ electrons per impurity in every case.}
\label{vmc:fig:core}
\end{figure}

\emph{Each attractive impurity captures one electron, independent of $\rs$.} The number of electrons within $3\aB$ of an attractive impurity is $3.4$--$3.5$ for four attractive impurities and $6.4$--$7.0$ for eight, i.e., $0.80$--$0.88$ per impurity, at all four densities alike (Table~\ref{vmc:tab:z1}); the probability of double occupancy is below 10\% at $\rs=10$ and below 3\% at $\rs\ge30$. Figure~\ref{vmc:fig:core} shows the radial density of the bound electron around its impurity on the $\aB$ scale: it is the same hydrogenic orbit, of rms radius $1.5$--$1.8\aB$, at $\rs=10$, 20, 30, and 50, because the binding energy ($\sim0.5\,\Ha$) is one to two orders of magnitude larger than every interaction scale. The bright spots in Fig.~\ref{vmc:fig:z1dens} are these orbits, and they are the same spots at $\rs=10$ and at $\rs=50$. Their size is $\aB$, ten to fifty times smaller than $a$; a density image of this state shows an array of resolved ``electrons'' whose number is $N_i/2$ and whose positions are $\{\mathbf{R}_I: Z_I>0\}$, with no reference to the electron density $n$ at all.

\emph{The remaining electrons are the clean system at a lower density in a landscape of holes.} The $N-N_i/2$ unbound electrons ($12.5$ or $9$) see the bare repulsive impurities, each of which carves a hole of radius $\sim a$ (the potential $1/\sqrt{r^2+d^2}$ equals $e^2/a$ at $r\simeq a$), and the neutral impurity--electron complexes, whose residual field is dipolar and weak. At $\rs=10$, 20, and 30 these electrons remain a liquid: the density between the holes is smooth, with no resolvable maxima (Fig.~\ref{vmc:fig:z1dens}), their local bond order $\av{|\psi_6|}_{\rm m}$ ($0.35$--$0.37$) is at or below the clean value of the lowest-energy clean state ($0.37$, $0.38$, $0.40$ at $\rs=10$, 20, 30), and the spread of their nearest-neighbor distances is wider than in the clean liquid. Nothing in the unbound population is more crystalline than the clean liquid; the only new static structure is the impurity-bound spots and the holes.

\emph{The connected crystalline correlation is suppressed at every $\rs$.} $\Sc(\bG_1)$ falls from $1.1$--$1.2$ (clean, lowest-energy family, $\rs=10$--30) to $0.72$--$0.83$ with 8 impurities and $0.45$--$0.59$ with 16; the static part grows from $0$ (PW) or $0.1$--$1.0$ (PG, trial-lattice imprint) to $0.3$--$1.3$. The full $S(\bq)$ maps of the disordered states (Fig.~\ref{vmc:fig:z1tot}) contain many sharp features, and all of them are static: they are the autocorrelation of the impurity positions. Subtracting the static part (Fig.~\ref{vmc:fig:z1conn}) leaves a structureless connected map with less weight at the crystal wave vector than the clean liquid has. At $\rs=30$, where the clean lattice-Gaussian state is already the lower-energy clean state and carries an incipient-crystal star ($\Sst=0.98$, $\Sc=1.21$), the same impurities reduce the connected star to $0.76$--$0.83$ (8 impurities) and $0.50$--$0.56$ (16 impurities), the same values as at $\rs=10$ and 20 within the realization-to-realization spread.

\emph{At $\rs=50$ the same impurities destroy the crystal.} The clean PG crystal has $S(\bG_1)=3.5$, of which $2.3$ is static (the pinned position of the finite crystal in the trial function) and $1.2$ connected; the clean PW liquid-like trial, which cannot pin, has $\Sc(\bG_1)=1.66$. With 16 impurities $\Sc(\bG_1)$ is $0.5$--$0.6$ in both families and both realizations, below the value of the clean $\rs=20$ liquid; with 8 impurities it is $0.8$--$1.0$. Realization 16B at $\rs=50$ retains a large \emph{static} star ($\Sst=1.7$--$2.1$) and a visibly deformed local lattice around the pinned spots, with the highest all-electron $\av{|\psi_6|}$ of the data set ($0.41$--$0.42$, the clean crystal value being $0.43$); this is what a pinned, deformed Wigner crystal looks like, and it is the one case in model (i) where the structure beyond the impurity sites is organized by the interaction. Its connected star is nevertheless $0.5$: the positions are pinned, the coherence is gone. Its unbound-electron bond order is $0.34$, the liquid value. Where the trial families disagree at $\rs=50$ (PW below PG in energy for the 8-impurity cases and for 16B, by $2$--$4\times10^{-4}\,\Ha$), they disagree about how much static lattice remains, not about the connected part, which is the same in both.

The conclusion of model (i) is foreseeable from the energy scales and the calculation confirms it in every detail at every density: strong point-like charged disorder converts a liquid into a density image with $N_i/2$ resolved electrons at the impurity positions, leaves the remaining electrons a liquid that is slightly less correlated than before, and converts an incipient or fully formed crystal into a pinned, deformed texture with no coherent Bragg star. At no density and in no family does it create crystalline correlations that were not there.

\subsection{From perturbative to trapping disorder: the impurity-strength scan at $\rs=10$, 20, 30, 50}
\label{vmc:sec:zscan}

\begin{figure*}[!htbp]
\includegraphics[width=\textwidth]{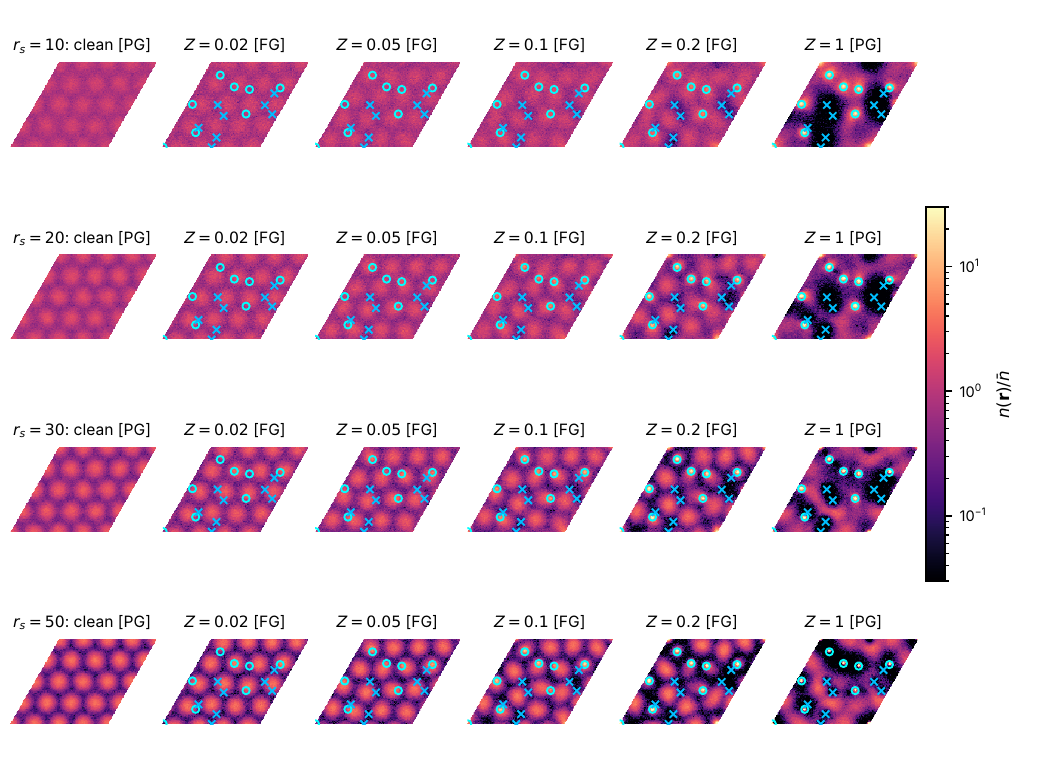}
\caption{Model (ii): density $n(\br)/\bar n$ (log scale $0.03$--$30$) for landscape 16A with impurity charge $\pm Z$ at $d=\aB$ scanned from $1/50$ to $1$ (columns, with the clean state first), at $\rs=10$, 20, 30, 50 (rows), lowest-energy trial family in each panel. Circles: attractive sites; crosses: repulsive sites. The lattice modulation in the clean and weak-disorder panels of the Gaussian families is the trial-orbital imprint at $\rs=10$ and 20 and the (incipient) crystal at $\rs=30$ and 50. Resolved spots at the attractive sites appear at $Z\simeq0.2$ at every density and are fully formed at $Z=1$.}
\label{vmc:fig:zdens}
\end{figure*}

\begin{figure*}[!htbp]
\includegraphics[width=\textwidth]{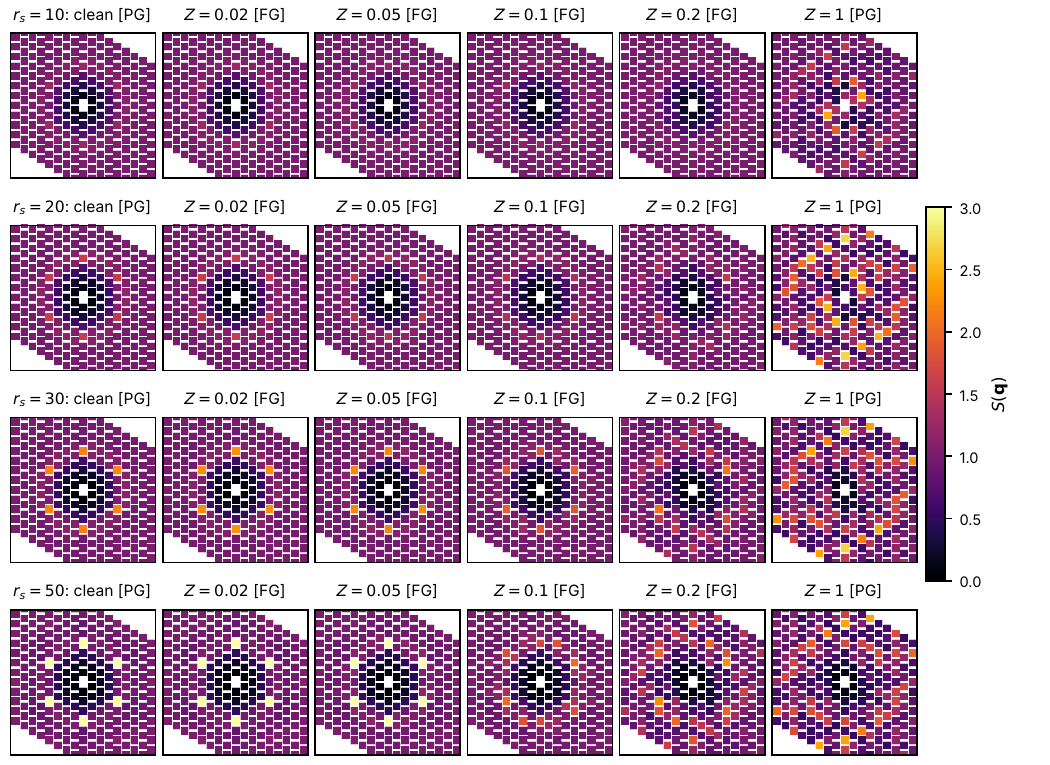}
\caption{Total structure factor $S(\bq)$ (0--3) for the states of Fig.~\ref{vmc:fig:zdens} (model (ii), $\pm Z$ at $d=\aB$). The six-spot star of the clean $\rs=30$ and $50$ states is progressively replaced by the irregular, static pattern of the impurity positions.}
\label{vmc:fig:ztot}
\end{figure*}

\begin{figure*}[!htbp]
\includegraphics[width=\textwidth]{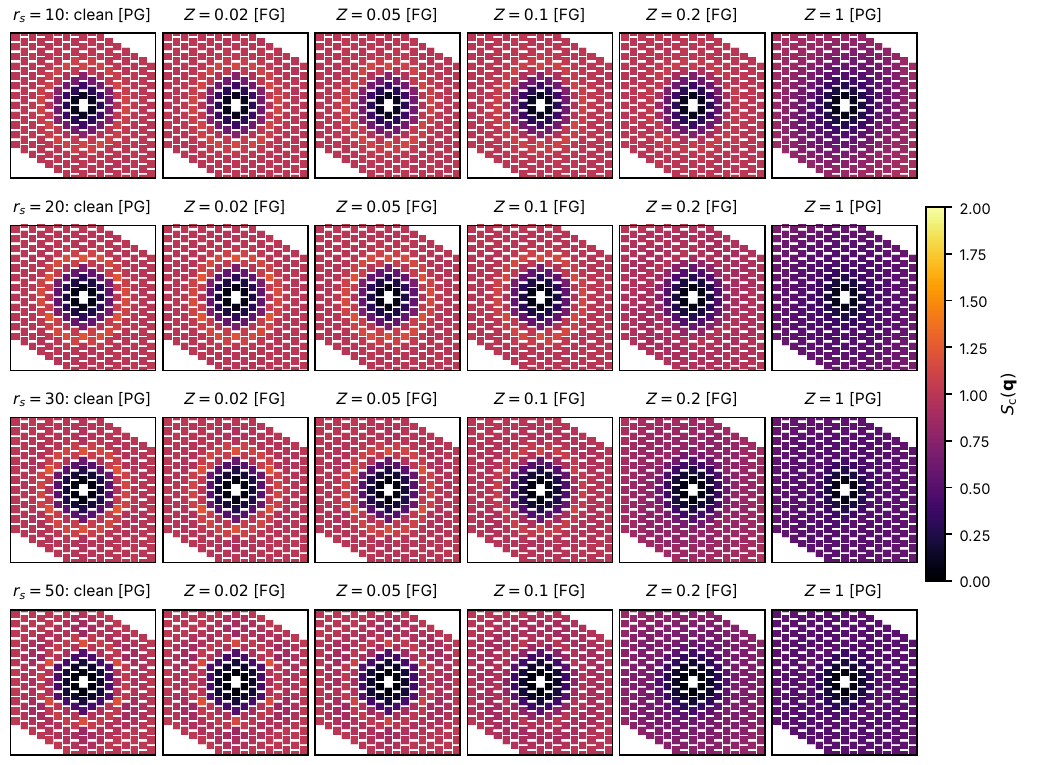}
\caption{Connected structure factor $\Sc(\bq)$ (0--2) for the states of Fig.~\ref{vmc:fig:zdens} (model (ii), $\pm Z$ at $d=\aB$). In every row the connected signal at the crystal wave vector decreases monotonically from left to right.}
\label{vmc:fig:zconn}
\end{figure*}

\begin{figure*}[!htbp]
\includegraphics[width=\textwidth]{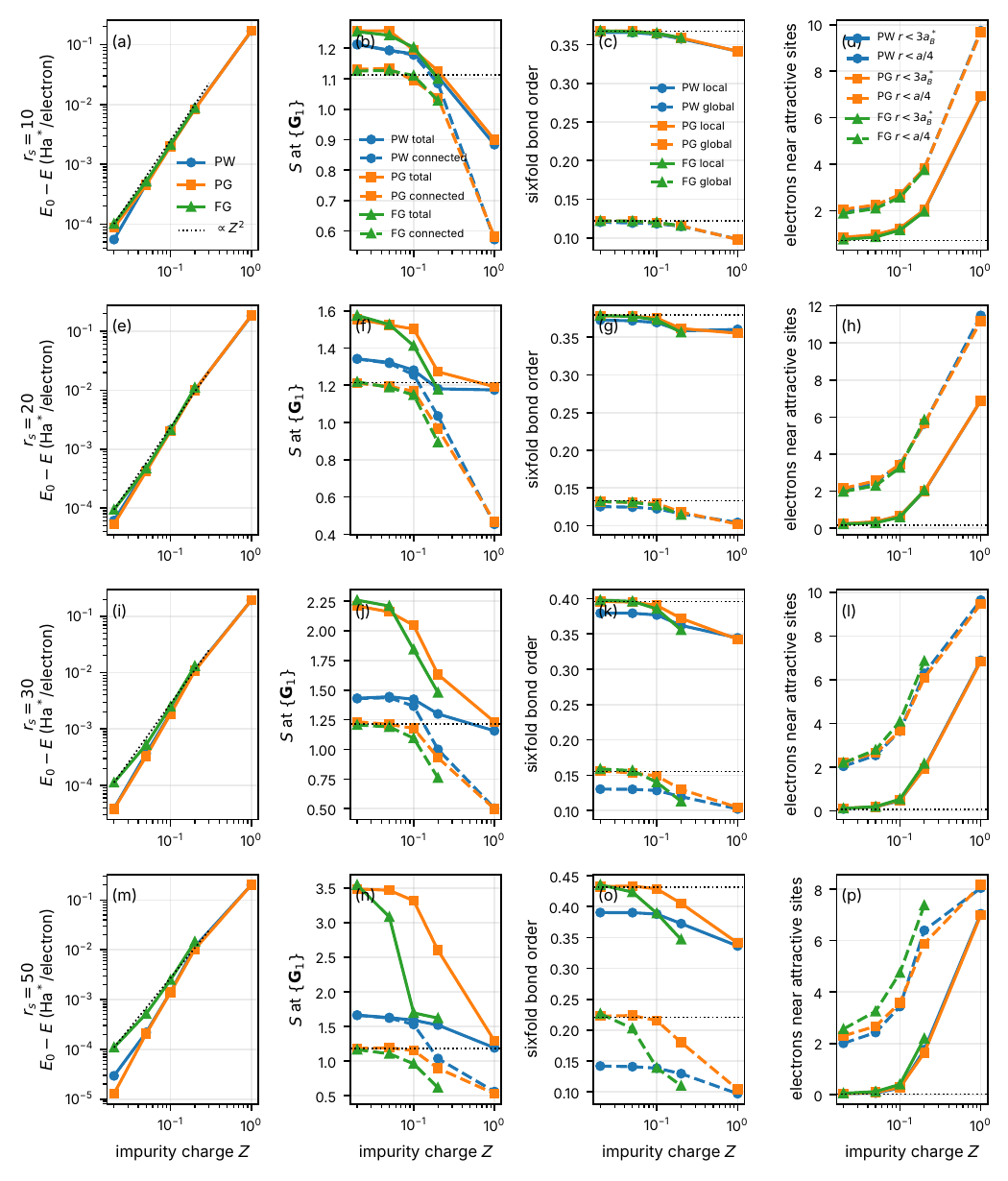}
\caption{Model (ii) diagnostics versus impurity charge $Z$ ($\pm Z$ at $d=\aB$, landscape 16A) at $\rs=10$, 20, 30, 50 (rows), all three trial families (PW circles, PG squares, FG triangles). First column: energy gain relative to the clean state of the same family (PW to clean PW; PG and FG to clean PG); the dotted line is $\propto Z^2$ through the $Z=0.02$ FG point. Second: total (solid) and connected (dashed) structure factor at the lattice star; dotted line, the connected value of the clean ground state. Third: local (solid) and global (dashed) sixfold bond order; dotted lines, clean values. Fourth: mean number of electrons within $3\aB$ (solid) and $a/4$ (dashed) of the eight attractive impurities; dotted line, the uniform-density value for $3\aB$ ($72/\rs^2$). The $Z=1$ points are the model (i) 16A states (PW and PG only).}
\label{vmc:fig:zdiag}
\end{figure*}

\begin{figure*}[!htbp]
\includegraphics[width=\textwidth]{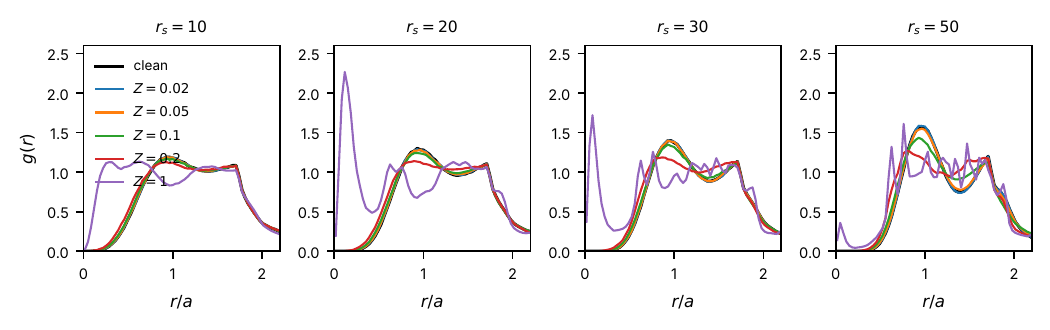}
\caption{Radial pair correlation in model (ii) ($\pm Z$ at $d=\aB$, landscape 16A) at the four densities, lowest-energy family at each $Z$. The first-neighbor peak is lowered and broadened, and the correlation hole fills in, as $Z$ increases; at $\rs=30$ and $50$ the shell structure of the clean state is lost by $Z=0.2$. The $Z=1$ peak at $r\simeq0.1a$ is the pair of electrons bound to two adjacent attractive impurities.}
\label{vmc:fig:zgr}
\end{figure*}

Model (i) is the strong-coupling limit, and one may object that it answers a question nobody asked: with wells fifty times deeper than $e^2/a$, of course the impurities win. Model (ii) answers the question that is asked. We fix landscape 16A and scan the impurity charge from $Z=1/50$, where the impurity binding energy $\sim2Z^2$ is a hundred times smaller than $e^2/a$ and the bound-orbit radius $\sim\aB/2Z$ is comparable to $a$, to $Z=1$, at each of the four densities. At $Z\simeq0.1$--$0.2$ the two energies cross; this is the regime in which a disorder-assisted crystallization, if it existed, would have to show up. The free-center family FG is included so that it can. Figures~\ref{vmc:fig:zdens}--\ref{vmc:fig:zgr} and Table~\ref{vmc:tab:zscan} give the results. The four densities answer four different versions of the question: $\rs=10$ is a liquid far from crystallization, $\rs=20$ is the density of the imaging experiments, $\rs=30$ is just below the crystallization point of the polarized system (the clean lattice-Gaussian state is already the lower-energy clean state there, by $1\times10^{-4}\,\Ha$, with a static star of $0.98$ from its own lattice), and $\rs=50$ is the clean crystal.

{\small
\begin{longtable}{llccccccc}
\caption{Model (ii): 16 impurities with $Z_I=\pm Z$ at $d=\aB$, landscape A, four densities (and landscape B at $\rs=20$, $Z=0.1$). Columns as in Table~\ref{vmc:tab:z1}; $N_{\rm b}$ counts electrons within $3\aB$ of an attractive impurity, with the uniform-density value $72/\rs^2$ given in the $Z=0$ rows. The $Z=1$ rows are the model (i) 16A states. The lowest-energy family of each Hamiltonian is marked $^*$.}
\label{vmc:tab:zscan}\\
\hline\hline
$Z$ & family & $E$ ($\Ha$) & $N_{\rm b}$ & $S(\bG_1)$ & $\Sst(\bG_1)$ & $\Sc(\bG_1)$ & $\av{|\psi_6|}$ & $|\Psi_6|$\\
\hline
\endfirsthead
\multicolumn{9}{l}{\emph{Table \ref{vmc:tab:zscan} (continued)}}\\
\hline\hline
$Z$ & family & $E$ ($\Ha$) & $N_{\rm b}$ & $S(\bG_1)$ & $\Sst(\bG_1)$ & $\Sc(\bG_1)$ & $\av{|\psi_6|}$ & $|\Psi_6|$\\
\hline
\endhead
\hline
\multicolumn{9}{r}{\emph{continued on next page}}\\
\endfoot
\hline\hline
\endlastfoot
\input{table_ii.tex}
\end{longtable}
}

\emph{The response is approximately quadratic in $Z$ over the measured interval $Z=0.02$--$0.2$.} The energy gain relative to the clean state of the same trial family grows by about two decades across this interval, consistent with a linear-response polarization law to within statistical and finite-system scatter (Fig.~\ref{vmc:fig:zdiag}, first column): at $\rs=20$ from $5\times10^{-5}$ to $1\times10^{-2}\,\Ha$, at $\rs=30$ from $1\times10^{-4}$ to $1.3\times10^{-2}\,\Ha$, at $\rs=50$ from $1\times10^{-4}$ to $1.5\times10^{-2}\,\Ha$. We quote this as approximate rather than exact: fitted exponents from a straight-line log--log fit over the five sampled $Z$ values lie within approximately $\pm0.2$ of 2 in all three families at all four densities within statistical uncertainty, and a departure from this response is not resolved within the measured range. The departure to the strong-coupling regime occurs only between $Z=0.2$ and $Z=1$, where the gain jumps to $0.17$--$0.20\,\Ha$ per electron, which is the hydrogenic binding of model (i).  Within the sampled interval and the trial-wave-function families used, no change of state consistent with disorder-selected crystallization is seen; approximate quadratic lowering in a finite variational system is not, of course, a general exclusion test.

\emph{Weak-disorder changes are small or unresolved; clear suppression of the connected crystalline correlation sets in at $Z\gtrsim0.1$.} In the lowest-energy family, $\Sc(\bG_1)$ runs, for $Z=0$, 0.02, 0.05, 0.1, 0.2, 1: at $\rs=10$, $1.11$, $1.13$, $1.13$, $1.11$, $1.03$, $0.58$; at $\rs=20$, $1.21$, $1.22$, $1.19$, $1.15$, $0.89$, $0.47$; at $\rs=30$, $1.21$, $1.21$, $1.19$, $1.09$, $0.76$, $0.50$; at $\rs=50$, $1.18$, $1.16$, $1.11$, $0.96$, $0.62$, $0.53$ (Fig.~\ref{vmc:fig:zdiag}, second column; Table~\ref{vmc:tab:zscan}). At $Z\le0.05$ the values track the clean reference within $\sim0.02$ at every density, which is within the statistical and finite-system scatter and is consistent with either weak enhancement or weak suppression; from $Z=0.1$ on every density loses connected weight, and the loss is largest where there was most to lose, at $\rs=50$, where the clean crystal's star is reduced by half at $Z=0.2$. The maximum of $\Sc(\bq)$ over all wave vectors, which would detect a crystal of any orientation, follows the same course and remains at the pristine star until it is no longer a maximum at all. The static part grows at every density. The local bond order (third column) is unchanged within $\pm0.003$ at $Z\le0.05$ and then decreases: at $Z=0.2$ it is $0.359$, $0.356$, $0.356$, $0.347$ at the four densities, against clean values of $0.368$, $0.380$, $0.396$, $0.432$. The pair correlation (Fig.~\ref{vmc:fig:zgr}) loses height at its first peak and the correlation hole fills in as $Z$ increases, at every density; at $\rs=30$ and 50 the shells of the clean (incipient) crystal are gone by $Z=0.2$. At $Z=1$ the pair correlation has a large peak at $r\simeq4\aB$ from two electrons bound to two adjacent attractive impurities; their mutual repulsion $e^2/4\aB=0.25\,\Ha$ is smaller than their binding, and for them the interaction has become irrelevant.

\emph{The disorder-adapted family is the lowest in energy wherever the disorder matters, and it is the least crystalline.} From $Z=0.05$ on, the FG family, whose orbital centers are obtained from a classical minimization in the disorder potential with a prescribed Gaussian smearing and then frozen during VMC optimization of the remaining Jastrow and impurity-factor parameters (Appendix~\ref{sec:SM_vmc}), has the lowest energy of the three at every density (Table~\ref{vmc:tab:zscan}), by margins that grow with $Z$ and with $\rs$: at $Z=0.2$ by $4\times10^{-4}$, $1.0\times10^{-3}$, $1.9\times10^{-3}$, $3.2\times10^{-3}\,\Ha$ per electron at $\rs=10$, 20, 30, 50 (statistical errors $4$--$6\times10^{-5}$). The electrons do prefer to adapt their positional structure to the disorder, and they prefer it most where the clean system is most crystalline. The adapted state has the lowest connected star and the lowest local bond order of the three families at every density where it wins: at $Z=0.2$, $\Sc(\bG_1)=1.03$, $0.89$, $0.76$, $0.62$ in FG against $1.04$--$1.08$, $0.97$--$1.03$, $0.93$--$1.00$, $0.90$--$1.04$ in PW and PG. Within the variational space the present family explores, the freedom that would have allowed a disorder-selected crystal to appear is used instead to follow the impurities, and the resulting state has lower connected lattice weight and lower local bond order than the clean ground state in the same family.  This bears on the comparison with the local liquid--solid competition under weak short-range disorder analyzed by Joy and Skinner~\cite{Joy2025}: within the specific trial-wave-function families and disorder models studied here (fixed-charge Coulomb impurities at setback $\aB$), the branch that gains most variational energy from the disorder is the branch whose density tracks the impurities rather than a disorder-selected lattice.  The models of Ref.~\cite{Joy2025} and of the present calculation are distinct in disorder spectrum and in the family of competing states considered; the agreement or disagreement between the two theoretical pictures must therefore be discussed in matched models, not read off from the present finite-system numerics.

\emph{Pinning sets in at $Z\simeq0.2$ at every density.} The number of electrons within $3\aB$ of the eight attractive impurities rises from the uniform value ($72/\rs^2$) by a factor of $1.1$--$2$ at $Z=0.02$, $1.2$--$4.5$ at $Z=0.05$, $1.6$--$14$ at $Z=0.1$, $3$--$76$ at $Z=0.2$ (the larger factors at the lower densities, where the uniform value is smallest), and reaches $6.9$--$7.0$, one electron per impurity, at $Z=1$ (Fig.~\ref{vmc:fig:zdiag}, fourth column). At $Z=0.2$ the density develops the first resolvable bright spots at the attractive sites and the first depletion around the repulsive ones (Fig.~\ref{vmc:fig:zdens}); by $Z=1$ there are eight spots and eight holes at every density. The onset of a ``solid-looking'' image coincides with the onset of trapping and with the steepest decline of $\Sc$, at all four densities. Realization 16B at $\rs=20$, $Z=0.1$ (Table~\ref{vmc:tab:zscan}) has a larger static star than 16A ($0.5$--$0.6$ versus $0.3$ in the Gaussian families) because its impurity positions happen to correlate better with the reference lattice; its connected star is the same, $1.14$--$1.15$, below the clean value. The static part depends on the landscape; the connected part does not.

Two numbers in Table~\ref{vmc:tab:zscan} could be misread and should be read correctly. At $\rs=30$ and $50$ the FG and PG bond orders at $Z\le0.05$ ($0.396$--$0.398$ and $0.424$--$0.435$) exceed the clean PW values ($0.380$ and $0.392$); the clean PW state is not the clean ground state at those densities, and compared with the clean state of lowest energy (PG, $0.396$ and $0.432$) the disordered values are equal within $0.003$ or lower. There is no $Z$, at any density, at which any family has a connected star or a bond order above the clean ground state by more than the statistical scatter.

Model (ii) therefore shows what model (i) could not: there is no intermediate disorder strength, at any density from the liquid through the crystallization point to the crystal, at which charged impurities enhance crystalline correlations. The system responds perturbatively, then it traps. The density image becomes more solid-looking at exactly the point where the many-body crystalline correlation starts its steepest decline, and that point is set by the ratio of the impurity binding energy to $e^2/a$, which is the same $Z\simeq0.2$ at every $\rs$ studied.

\subsection{Donor-layer disorder: liquid, puddles, and the amorphous solid}
\label{vmc:sec:donor}

\begin{figure*}[!htbp]
\includegraphics[width=\textwidth]{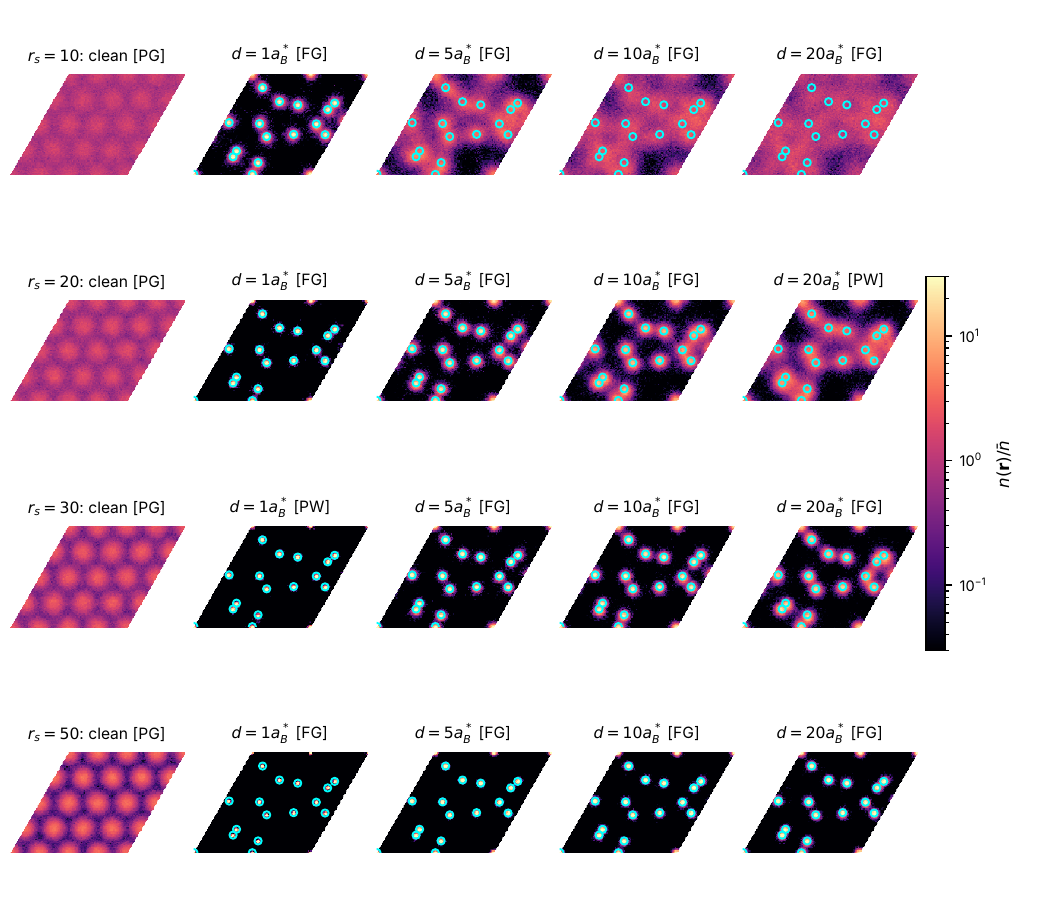}
\caption{Model (iii): density $n(\br)/\bar n$ (log scale $0.03$--$30$, saturating) for 16 electrons and 16 unit-charge donors (circles) at setback $d=1$, 5, 10, $20\aB$ (columns) and $\rs=10$, 20, 30, 50 (rows), lowest-energy family in each panel, with the clean state in the first column. The grid is organized by $d/a$: moving down (lower density) or left (smaller setback) takes the system from a connected inhomogeneous liquid to isolated puddles to one electron bound per donor. The solid-looking state appears along a diagonal set by the disorder, not along a row set by $\rs$.}
\label{vmc:fig:donordens}
\end{figure*}

\begin{figure*}[!htbp]
\includegraphics[width=\textwidth]{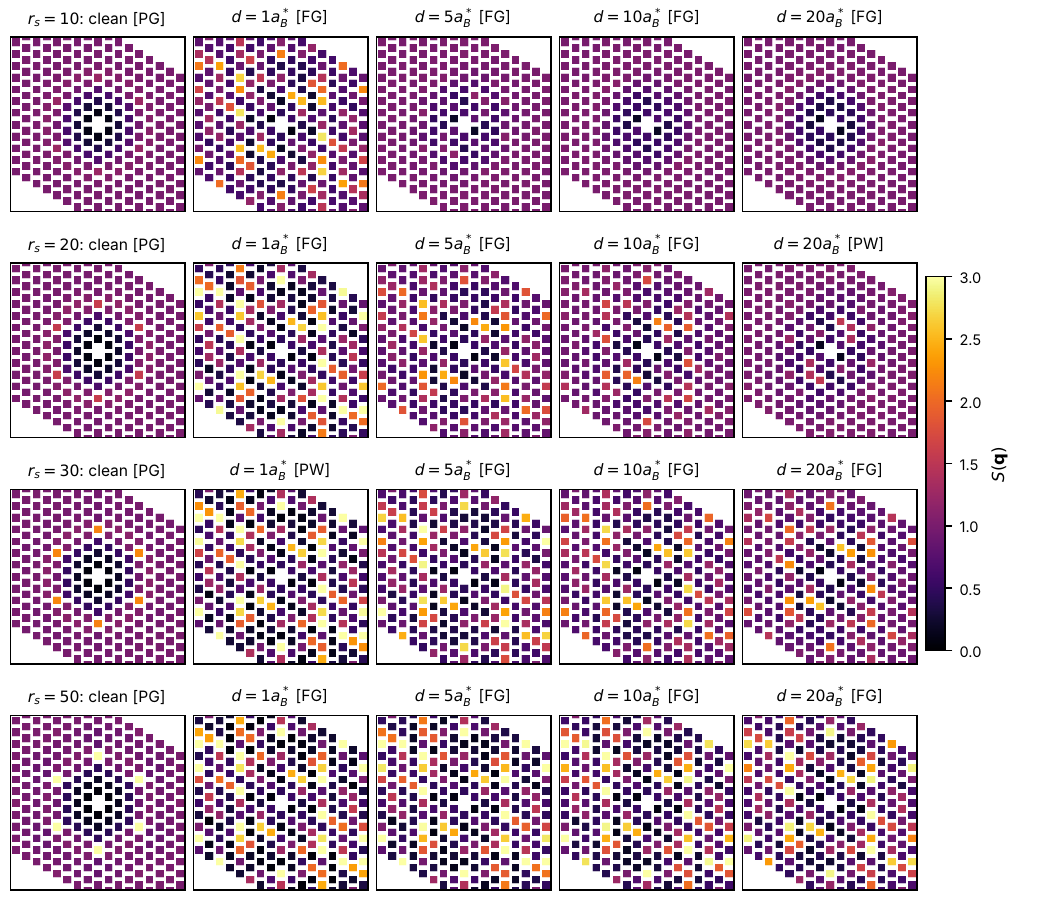}
\caption{Total structure factor $S(\bq)$ (0--3) for the states of Fig.~\ref{vmc:fig:donordens}. The strong spots in the trapped states are at the wave vectors of the donor arrangement.}
\label{vmc:fig:donortot}
\end{figure*}

\begin{figure*}[!htbp]
\includegraphics[width=\textwidth]{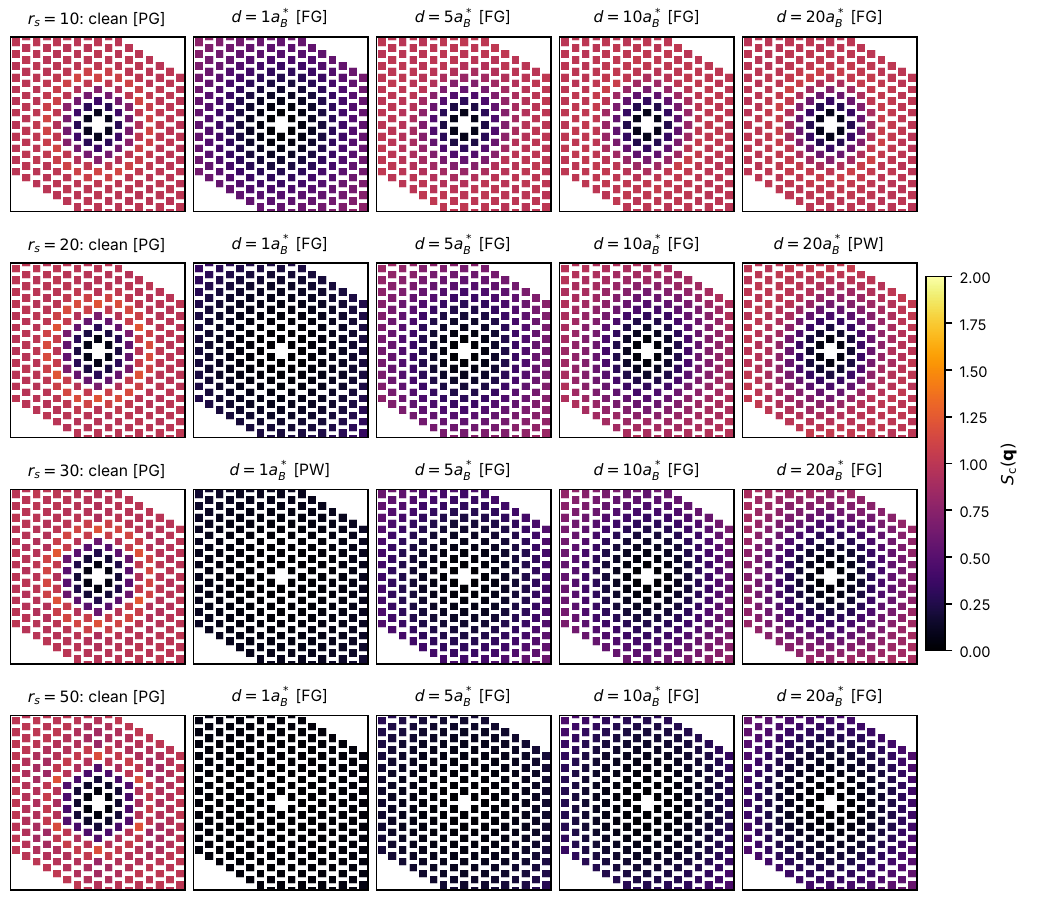}
\caption{Connected structure factor $\Sc(\bq)$ (0--2) for the states of Fig.~\ref{vmc:fig:donordens}. The crystalline star of the clean $\rs=30$ and $50$ states is absent at every setback; the liquid ring of the clean $\rs=10$ and $20$ states survives only in the upper right of the grid.}
\label{vmc:fig:donorconn}
\end{figure*}

\begin{figure*}[!htbp]
\includegraphics[width=\textwidth]{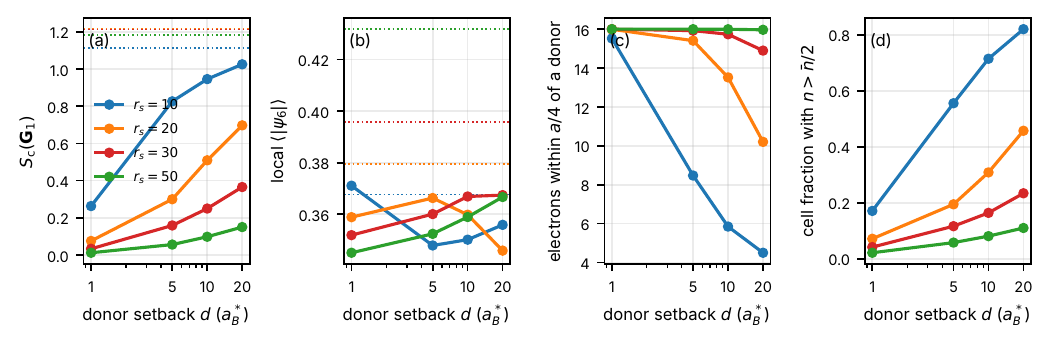}
\caption{Model (iii) diagnostics versus donor setback at the four densities, lowest-energy family: (a) connected structure factor at the lattice star (dotted lines: clean values); (b) local sixfold bond order (dotted: clean values); (c) number of electrons within $a/4$ of a donor; (d) fraction of the cell with $n>\bar n/2$. Every curve in (a) and (b) lies below its clean value; every curve in (c) and (d) moves toward trapping as $d$ decreases or $\rs$ increases.}
\label{vmc:fig:donordiag}
\end{figure*}

\begin{figure*}[!htbp]
\includegraphics[width=\textwidth]{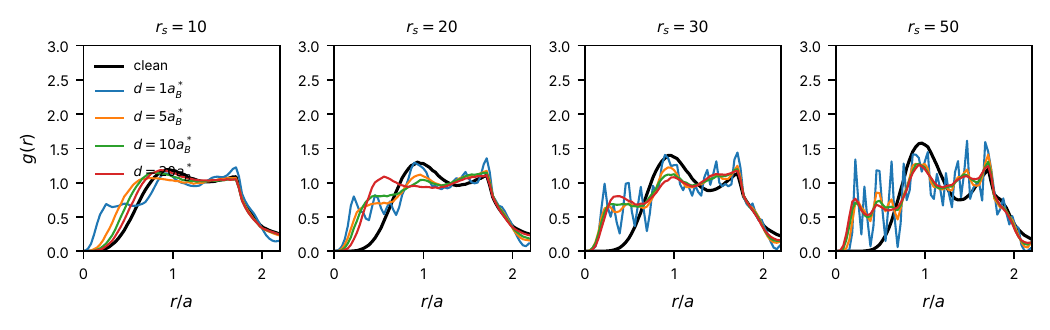}
\caption{Radial pair correlation in the donor-layer model at the four setbacks compared with the clean state at the same $\rs$. The first-neighbor peak of the clean liquid is lowered and broadened as $d$ decreases; the shell structure of the clean crystal at $\rs=30$ and $50$ is replaced, at every setback, by the pair distribution of the donor positions.}
\label{vmc:fig:donorgr}
\end{figure*}

Point charges one Bohr radius from the plane are the short-range limit of charged disorder. The long-range limit is a layer of remote ionized donors, which in a modulation-doped structure are also the source of the electrons and neutralize them. Model (iii) is this system: 16 electrons, 16 donors of charge $+1$ at the positions of landscape A, at setback $d$, computed at $d=1$, 5, 10, and $20\aB$ for each of the four densities. The resulting potential is smooth on the scale $d$, and the natural objects in it, when $d$ is comparable to the donor spacing, are the puddles of nonlinear screening theory \cite{Efros1988}, whose connectivity controls the semiclassical percolation transport of low-density 2D systems \cite{DasSarma2005}. Figures~\ref{vmc:fig:donordens}--\ref{vmc:fig:donorgr} and Table~\ref{vmc:tab:donor} give the results; the FG family has the lowest energy in every case but two ($\rs=20$, $d=20\aB$ and $\rs=30$, $d=\aB$, where PW is lower by $<1\times10^{-4}\,\Ha$, within the statistical error), by $1$--$24\times10^{-4}\,\Ha$ per electron.

{\footnotesize
\begin{longtable}{llcccccccccccc}
\caption{Model (iii): 16 electrons, 16 donors ($Z=+1$) at setback $d$, landscape A, four densities and four setbacks. $E$, $\av{V_{ee}}$, and $\av{V_{\rm imp}}$ per electron ($\av{V_{\rm imp}}$ omits the $\bq=0$ background constant and is comparable only within a given $\rs$ and $d$); $N_{a/4}$: mean number of electrons within $a/4$ of a donor; $\delta r$: rms distance of those electrons from their donor, in $\aB$; $A_{1/2}$: fraction of the cell with $n(\br)>\bar n/2$; span: whether that region spans the periodic cell; then the structure factor at the lattice star and the local bond order. The lowest-energy family of each Hamiltonian is marked $^*$.}
\label{vmc:tab:donor}\\
\hline\hline
$\rs$ & $d/\aB$ & family & $E$ ($\Ha$) & $\av{V_{ee}}$ & $\av{V_{\rm imp}}$ & $N_{a/4}$ & $\delta r$ & $A_{1/2}$ & span & $S(\bG_1)$ & $\Sst(\bG_1)$ & $\Sc(\bG_1)$ & $\av{|\psi_6|}$\\
\hline
\endfirsthead
\multicolumn{14}{l}{\emph{Table \ref{vmc:tab:donor} (continued)}}\\
\hline\hline
$\rs$ & $d/\aB$ & family & $E$ ($\Ha$) & $\av{V_{ee}}$ & $\av{V_{\rm imp}}$ & $N_{a/4}$ & $\delta r$ & $A_{1/2}$ & span & $S(\bG_1)$ & $\Sst(\bG_1)$ & $\Sc(\bG_1)$ & $\av{|\psi_6|}$\\
\hline
\endhead
\hline
\multicolumn{14}{r}{\emph{continued on next page}}\\
\endfoot
\hline\hline
\endlastfoot
\input{table_iii.tex}
\end{longtable}
}

The sixteen panels of Fig.~\ref{vmc:fig:donordens} are the central result, because they contain the whole crossover in one place and because the control parameter is visible by inspection: the state depends on $d/a$, which increases to the right and decreases downward, and not on $\rs$ alone.

\emph{$d=\aB$: the trapping limit at every density.} At $d=\aB$ a donor is a point charge and binds an electron hydrogenically, exactly as the attractive impurities of model (i) do. At all four densities $15.4$--$16.0$ of the 16 electrons sit within $a/4$ of a donor, at an rms distance of $1.9$--$2.1\aB$ (the same orbit as in Fig.~\ref{vmc:fig:core}), $83$--$98\%$ of the cell is below $\bar n/2$, and the connected star is $0.26$, $0.08$, $0.03$, $0.01$ at $\rs=10$, 20, 30, 50, against clean values of $1.1$--$1.2$. The donor energy is seven to forty times the electron--electron energy. The density image is sixteen spots at the sixteen donor positions, at every density; the structure factor is entirely static. This column is the $n/n_i=1$ version of model (i), and it is the state to which the entire grid flows as $d/a\to0$.

\emph{$\rs=10$: from trapped to liquid as the setback grows.} Along the top row, $d/a=0.05$, $0.26$, $0.53$, $1.05$. At $d=5\aB$ the donor wells are already too shallow and too broad to hold electrons individually: $8.5$ of 16 are within $a/4$ of a donor at $\delta r=3.1\aB$, the occupied region covers $57\%$ of the cell and spans it, and the connected star is $0.83$, i.e., a liquid that is strongly modulated but connected. At $d=10\aB$ the occupied region is $72\%$ of the cell, spanning, with maximum density $3.6\bar n$, no static star ($0.01$), and $\Sc(\bG_1)=0.95$; at $d=20\aB$ it is $82\%$, the static star is zero, and $\Sc(\bG_1)=1.03$, within $10\%$ of the clean liquid. The local bond order is $0.35$--$0.36$ throughout, at or just below the clean $0.37$. This row is the puddle regime of a high-density sample: the landscape is smooth, the liquid fills the valleys and spans the cell, and the electron--electron energy exceeds the donor energy from $d=5\aB$ on.

\emph{$\rs=20$: isolated puddles at $d=5$--$10\aB$, reconnection at $d=20\aB$.} In the second row, $d/a=0.03$, $0.13$, $0.26$, $0.53$. At $d=5\aB$ the system is trapped ($15.4$ within $a/4$, $19\%$ occupied, not spanning, $\Sc=0.30$). At $d=10\aB$ it is the isolated-puddle state: $13.5$ of 16 electrons sit within $a/4$ of a donor at $\delta r=5.3\aB$ (consistent with the single-donor harmonic length $5.6\aB$), closely spaced donors share puddles, the occupied region is $31\%$ of the cell and does not span it while the depleted region does, and $\Sc(\bG_1)=0.51$, less than half the clean value. At $d=20\aB$ the puddles have widened and merged: $10$ electrons within $a/4$, $46\%$ occupied, static star down to $0.2$--$0.3$, $\Sc=0.70$ — still not spanning, still well below the clean $1.21$, but on its way back to the liquid. The density image at $d=5$--$10\aB$ is an amorphous array of resolved maxima, one per donor or donor cluster, with large dark regions between them. This is the amorphous solid, at the density of the imaging experiments.

\emph{$\rs=30$: trapped at every setback up to $d=20\aB$.} In the third row, $d/a=0.02$--$0.35$. $15.7$--$16.0$ of the 16 electrons are within $a/4$ of a donor at $d\le10\aB$ and $14.9$ at $d=20\aB$; the occupied fraction is $4$--$23\%$ and never spans; the connected star is $0.03$, $0.15$, $0.25$, $0.37$ at $d=1$, 5, 10, $20\aB$, against $1.21$ for the clean state, which at this density is the incipient crystal. The clean $\rs=30$ state carries a six-spot star in $\Sc$ (Fig.~\ref{vmc:fig:z1conn}); at no setback does any trace of it survive in Fig.~\ref{vmc:fig:donorconn}. The pair correlation (Fig.~\ref{vmc:fig:donorgr}) has lost the shells of the incipient crystal at every setback and has acquired the pair distribution of the donor positions.

\emph{$\rs=50$: the crystal is gone at every setback.} In the bottom row, $d/a=0.01$--$0.21$. Every electron is on a donor at every setback; $\delta r$ grows from $2.0$ to $7.6\aB$ as the wells broaden, but the occupied fraction stays below $11\%$. The total star, $1.5$--$1.6$, looks ``crystalline'' and is almost entirely static ($1.4$--$1.6$); the connected part is $0.01$, $0.06$, $0.11$, $0.15$, against $1.18$ for the clean crystal (and $1.66$ for the clean PW trial). The electron--electron energy is between a third ($d=20\aB$) and a fortieth ($d=\aB$) of the donor energy. The clean Wigner crystal at $\rs=50$ does not exist as a correlated object in this model at any setback up to $20\aB$.

\emph{The crossover is set by $d/a$, i.e., by $n/n_i$ and the disorder length, not by $r_c$.} Reading Fig.~\ref{vmc:fig:donordens} along a column, lowering the density at fixed setback takes the system from a connected liquid through isolated puddles to one electron per donor; reading it along a row, increasing the setback at fixed density does the reverse. The states with the same $d/a$ look alike: $(\rs=10,d=5\aB)$, $(\rs=20,d=10\aB)$, and $(\rs=50,d=20\aB)$, with $d/a=0.26$, $0.26$, $0.21$, have $\Sc(\bG_1)=0.83$, $0.51$, $0.15$ — a sequence that reflects the increasing dominance of the donor energy over the interaction energy as $\rs$ grows, not any crystalline tendency, since the local bond order is $0.35$--$0.37$ in all three. The solid-looking state appears at $\rs\approx20$ for $d=10\aB$ because that is where $a$ exceeds $\sim3$--$4$ times $d$ and the smooth landscape resolves into individual wells; with $d=5\aB$ it appears at $\rs\approx10$--$20$ and with $d=20\aB$ at $\rs\approx30$. This is the $n/n_i$ argument of Sec.~\ref{sec:intro} realized in a calculation: the construction holds $N$ and $N_i$ fixed and uses the same fractional impurity coordinates as the cell is rescaled with $\rs$, so that $n_i/n$ is held fixed at unity throughout while the physical $n$ and $n_i$ both change together; lowering $\rs$ then takes the system monotonically from a liquid-side state through puddles to one electron per impurity, with the connected crystalline correlation decreasing at every step, at every setback.  This is a fixed-$n_i/n$, variable-$d/a$ path through parameter space; it is not an experimental fixed-sample gate sweep in which $n_i$ is held constant and $n$ is varied, which would require a different cell construction. Figure~\ref{vmc:fig:donordiag} shows the same thing as curves: at every density and every setback, $\Sc(\bG_1)$ and $\av{|\psi_6|}$ lie below their clean values, and the number of trapped electrons and the depleted fraction of the cell increase monotonically as $d$ decreases.

The sequence liquid $\to$ puddles $\to$ trapped electrons, driven by density at fixed disorder, is the sequence reported in the long-range-disorder regime of Ref.~\cite{Ge2026} (mixed solid--liquid phases, local re-entrant melting, Friedel-like oscillations around the puddles) and in Ref.~\cite{Xiang2024} (nanocrystalline domains at low density, a growing percolating liquid at higher density), where it is interpreted as a microemulsion of solid and liquid \cite{Spivak2004}. In the present model the coexisting regions are the puddles and the depleted regions of a random donor potential, and no interaction-driven phase separation is needed to produce them. The present calculation reproduces this sequence from a model in which the only ingredients are the electron--electron interaction and a layer of random donors, and in which the connected crystalline correlation decreases monotonically along it.

\subsection{Structural signatures of the two pictures}
\label{vmc:sec:stm}

\subsubsection{One-body density and the structural decomposition}
A time-averaged carrier-density image, however obtained, is a measurement of the one-body density $n(\br)$, equivalently of the static part $\Sst(\bq)$ of the structure factor of Eq.~\eqref{vmc:eq:decomp}.  The connected part $\Sc(\bq)$ encodes density fluctuations about the mean, and its interpretation as a crystallinity measure is subject to the caveats of Sec.~\ref{vmc:sec:Sc_caveat}.  A clean liquid of $N$ electrons in a periodic cell and a clean (unpinned) crystal of $N$ electrons both have translation-invariant mean density on average over the position zero-mode; in a finite system the trial wave function provides the symmetry breaking.  In the problem addressed here, what resolves individual electrons in the mean density of a disordered state is the external potential provided by the quenched impurities: the pinned density pattern carries structural information both about the disorder landscape and about the many-body response of the electrons to it.

The structural question we therefore pose, within the present approximations, is which features of the pinned density pattern are inherited from the impurity landscape and which from the electron--electron interaction, and whether the resulting pattern is better described as a disorder-broken Wigner crystal or as an impurity-organized amorphous solid.  Sections~\ref{vmc:sec:strong}--\ref{vmc:sec:donor} report, for each disorder model: the number of imaged maxima and whether it is set by the attractive-impurity count or by the electron count; the local sixfold bond amplitude of the maxima; the impurity-registered fraction; and the behavior of the connected lattice weight.  These observables are the inputs to the structural judgment.

\subsubsection{Local sixfold order}
A sixfold nearest-neighbor shell at the interparticle spacing is a property of any dense repulsive system in two dimensions: the clean liquid at $\rs=10$--20 has $\av{|\psi_6|}=0.37$--$0.38$ and a first-neighbor peak in $g(r)$ at $r\simeq a$, and the clean crystal at $\rs=50$ has $\av{|\psi_6|}=0.43$. A high local sixfold amplitude is therefore not a sharp discriminator between a pinned Wigner crystal and an amorphous solid; both have it on their first shell.  The sharper diagnostic in a finite cell is comparison between the local amplitude of the disordered state and the clean-state value at the same density within the same trial family: in every disordered state computed here on the liquid side of $r_c$, within the trial-wave-function families considered, the local amplitude is at or below the clean-state value and the connected lattice weight is at or below it.

\subsubsection{Impurity registration}
A density image of the $\rs=20$ donor-layer state at $d=10\aB$ (Fig.~\ref{vmc:fig:donordens}) shows an array of maxima located over the donor cluster centers, with locally sixfold-looking surroundings imposed by the mean donor spacing $\sim a$ and by the mutual repulsion.  A density image of the $\rs=20$, $Z=1$ point-impurity state (Fig.~\ref{vmc:fig:z1dens}) shows eight maxima at the attractive-impurity sites.  In both model calculations the maximum count tracks the impurity count rather than the electron count, and the nearest-neighbor-distance distribution reflects the impurity-spacing distribution convolved with the inter-electron repulsion.  The parametric distinction between the two structural pictures expressed by this behavior is:
\begin{itemize}
\item \emph{Amorphous solid.}  Spot count tracks the attractive-impurity count; the pattern becomes more resolved as the disorder is increased at fixed electron density; registration with the impurity landscape is high.
\item \emph{Pinned Wigner crystal.}  Spot count tracks the electron count; the pattern becomes less ordered as the disorder is increased; registration with the impurity landscape is low (the mean positions sit on crystalline lattice sites, up to Larkin displacements).
\end{itemize}
These are structural signatures that an experimental analysis has access to.  Which signature a given sample displays is a question for the experimental data; we have not attempted a per-sample fit to the published images.

\subsubsection{Finite-size scope and the clean benchmark}
Our conclusions apply within the specific disorder models, parameter ranges, trial-wave-function families, and finite system sizes used.  On $N=16$ the paper cannot resolve a density coherence length beyond the first shell, so we test only whether coherence beyond the first shell is \emph{generated} by the disorder; within the present calculations the connected lattice weight does not exceed the clean-state value at the same density in any disordered state on the liquid side of $r_c$.  The clean-system Wigner $\rc$ depends on spin sector, dielectric environment, layer thickness, band structure, and quantum Monte Carlo method, with reported values spanning the range quoted in Sec.~\ref{sec:intro}; the amorphous-solid analysis does not require a particular value of $\rc$, only the generic facts that a clean $\rc$ exists for interaction-driven crystallization and that the density at which an impurity-organized pattern becomes resolvable is set by the disorder landscape of the sample.

The comparison with the neural-quantum-state VMC calculations reported in Ref.~\cite{Ge2026}'s supplement, and with the bond-orientational analysis there, is best discussed in matched disorder models and matched trial-wave-function spaces, and we do not attempt that quantitative comparison here.  We note that Ref.~\cite{Ge2026}'s simulation uses a finite-range dual-gate-screened Coulomb potential for its short-range defects rather than a literal contact potential, and that the structural analyses in that supplement bear directly on the questions raised here.

%% file: table_i.tex
10 & clean & PW & $-0.08441(1)$ & 0.00 & 1.21 & 0.00 & 1.21 & 0.366 & 0.120 & 0.366 \\
10 & clean & PG$^*$ & $-0.08466(1)$ & 0.00 & 1.24 & 0.13 & 1.11 & 0.368 & 0.122 & 0.368 \\
10 & 8\,A & PW & $-0.17434(15)$ & 3.42 & 0.95 & 0.17 & 0.78 & 0.350 & 0.109 & 0.349 \\
10 & 8\,A & PG$^*$ & $-0.17510(14)$ & 3.48 & 0.98 & 0.19 & 0.79 & 0.351 & 0.109 & 0.349 \\
10 & 8\,B & PW & $-0.17058(16)$ & 3.45 & 0.90 & 0.13 & 0.77 & 0.356 & 0.113 & 0.352 \\
10 & 8\,B & PG$^*$ & $-0.17159(15)$ & 3.41 & 0.89 & 0.12 & 0.76 & 0.356 & 0.113 & 0.353 \\
10 & 16\,A & PW & $-0.25333(20)$ & 6.91 & 0.88 & 0.31 & 0.57 & 0.342 & 0.099 & 0.346 \\
10 & 16\,A & PG$^*$ & $-0.25555(20)$ & 6.94 & 0.90 & 0.32 & 0.58 & 0.342 & 0.098 & 0.345 \\
10 & 16\,B & PW & $-0.21963(20)$ & 6.37 & 0.89 & 0.31 & 0.59 & 0.362 & 0.119 & 0.361 \\
10 & 16\,B & PG$^*$ & $-0.22146(22)$ & 6.48 & 0.96 & 0.37 & 0.59 & 0.365 & 0.122 & 0.363 \\
\hline
20 & clean & PW & $-0.04611(1)$ & 0.00 & 1.32 & 0.00 & 1.32 & 0.374 & 0.126 & 0.374 \\
20 & clean & PG$^*$ & $-0.04625(1)$ & 0.00 & 1.58 & 0.37 & 1.21 & 0.380 & 0.133 & 0.380 \\
20 & 8\,A & PW & $-0.13974(14)$ & 3.45 & 1.18 & 0.43 & 0.76 & 0.369 & 0.121 & 0.361 \\
20 & 8\,A & PG$^*$ & $-0.14109(16)$ & 3.42 & 1.36 & 0.63 & 0.74 & 0.372 & 0.117 & 0.360 \\
20 & 8\,B & PW & $-0.13905(13)$ & 3.42 & 1.08 & 0.33 & 0.75 & 0.381 & 0.128 & 0.366 \\
20 & 8\,B & PG$^*$ & $-0.13984(14)$ & 3.45 & 1.18 & 0.45 & 0.72 & 0.379 & 0.128 & 0.366 \\
20 & 16\,A & PW & $-0.22918(22)$ & 6.89 & 1.17 & 0.72 & 0.45 & 0.360 & 0.104 & 0.366 \\
20 & 16\,A & PG$^*$ & $-0.23184(20)$ & 6.88 & 1.19 & 0.72 & 0.47 & 0.355 & 0.102 & 0.359 \\
20 & 16\,B & PW & $-0.21084(18)$ & 6.79 & 1.30 & 0.82 & 0.48 & 0.385 & 0.147 & 0.381 \\
20 & 16\,B & PG$^*$ & $-0.21212(20)$ & 6.79 & 1.32 & 0.84 & 0.48 & 0.386 & 0.147 & 0.381 \\
\hline
30 & clean & PW & $-0.03184(1)$ & 0.00 & 1.45 & 0.00 & 1.45 & 0.380 & 0.130 & 0.380 \\
30 & clean & PG$^*$ & $-0.03193(1)$ & 0.00 & 2.20 & 0.98 & 1.21 & 0.396 & 0.155 & 0.396 \\
30 & 8\,A & PW & $-0.13006(14)$ & 3.44 & 1.21 & 0.38 & 0.83 & 0.365 & 0.117 & 0.358 \\
30 & 8\,A & PG$^*$ & $-0.13023(14)$ & 3.44 & 1.69 & 0.91 & 0.78 & 0.375 & 0.121 & 0.360 \\
30 & 8\,B & PW & $-0.12943(15)$ & 3.44 & 1.10 & 0.28 & 0.83 & 0.379 & 0.128 & 0.364 \\
30 & 8\,B & PG$^*$ & $-0.13009(15)$ & 3.44 & 1.47 & 0.71 & 0.76 & 0.377 & 0.128 & 0.364 \\
30 & 16\,A & PW & $-0.22429(19)$ & 6.88 & 1.16 & 0.65 & 0.50 & 0.344 & 0.102 & 0.361 \\
30 & 16\,A & PG$^*$ & $-0.22495(22)$ & 6.84 & 1.23 & 0.73 & 0.50 & 0.342 & 0.105 & 0.354 \\
30 & 16\,B & PW$^*$ & $-0.21496(20)$ & 6.92 & 1.64 & 1.08 & 0.56 & 0.385 & 0.151 & 0.353 \\
30 & 16\,B & PG & $-0.21467(18)$ & 6.88 & 1.81 & 1.29 & 0.52 & 0.387 & 0.161 & 0.356 \\
\hline
50 & clean & PW & $-0.01974(1)$ & 0.00 & 1.66 & 0.00 & 1.66 & 0.392 & 0.142 & 0.392 \\
50 & clean & PG$^*$ & $-0.01983(1)$ & 0.00 & 3.46 & 2.28 & 1.18 & 0.432 & 0.221 & 0.432 \\
50 & 8\,A & PW$^*$ & $-0.12111(10)$ & 3.53 & 1.44 & 0.46 & 0.97 & 0.370 & 0.123 & 0.359 \\
50 & 8\,A & PG & $-0.12074(11)$ & 3.51 & 2.93 & 2.13 & 0.80 & 0.410 & 0.189 & 0.362 \\
50 & 8\,B & PW$^*$ & $-0.12139(12)$ & 3.52 & 1.32 & 0.35 & 0.97 & 0.385 & 0.133 & 0.357 \\
50 & 8\,B & PG & $-0.12118(11)$ & 3.52 & 2.64 & 1.87 & 0.77 & 0.403 & 0.170 & 0.367 \\
50 & 16\,A & PW & $-0.22077(16)$ & 7.04 & 1.19 & 0.63 & 0.56 & 0.336 & 0.097 & 0.341 \\
50 & 16\,A & PG$^*$ & $-0.22119(17)$ & 6.99 & 1.29 & 0.76 & 0.53 & 0.341 & 0.104 & 0.343 \\
50 & 16\,B & PW$^*$ & $-0.21514(16)$ & 7.01 & 2.29 & 1.71 & 0.58 & 0.414 & 0.198 & 0.341 \\
50 & 16\,B & PG & $-0.21487(13)$ & 7.04 & 2.60 & 2.08 & 0.52 & 0.420 & 0.216 & 0.343 \\

%% file: table_ii.tex
$\rs=10$ & & & & & & & & \\
0 & PW & $-0.08441$ & 0.72 & 1.21 & 0.00 & 1.21 & 0.366 & 0.120 \\
0 & PG$^*$ & $-0.08466$ & 0.72 & 1.24 & 0.13 & 1.11 & 0.368 & 0.122 \\
0.02 & PW & $-0.08447(1)$ & 0.80 & 1.21 & 0.00 & 1.21 & 0.366 & 0.120 \\
0.02 & PG & $-0.08475(1)$ & 0.85 & 1.26 & 0.13 & 1.13 & 0.368 & 0.122 \\
0.02 & FG$^*$ & $-0.08476(1)$ & 0.76 & 1.25 & 0.13 & 1.13 & 0.368 & 0.122 \\
0.05 & PW & $-0.08488(1)$ & 0.93 & 1.19 & 0.00 & 1.19 & 0.366 & 0.119 \\
0.05 & PG & $-0.08511(1)$ & 0.97 & 1.26 & 0.12 & 1.13 & 0.368 & 0.122 \\
0.05 & FG$^*$ & $-0.08517(1)$ & 0.86 & 1.24 & 0.12 & 1.13 & 0.367 & 0.122 \\
0.1 & PW & $-0.08646(2)$ & 1.24 & 1.18 & 0.00 & 1.18 & 0.363 & 0.118 \\
0.1 & PG & $-0.08666(2)$ & 1.24 & 1.19 & 0.10 & 1.09 & 0.365 & 0.119 \\
0.1 & FG$^*$ & $-0.08677(2)$ & 1.16 & 1.20 & 0.09 & 1.11 & 0.365 & 0.120 \\
0.2 & PW & $-0.09293(4)$ & 2.02 & 1.11 & 0.02 & 1.08 & 0.358 & 0.115 \\
0.2 & PG & $-0.09296(4)$ & 2.04 & 1.13 & 0.09 & 1.04 & 0.359 & 0.116 \\
0.2 & FG$^*$ & $-0.09340(4)$ & 1.95 & 1.10 & 0.07 & 1.03 & 0.359 & 0.116 \\
1 & PW & $-0.25333(20)$ & 6.91 & 0.88 & 0.31 & 0.57 & 0.342 & 0.099 \\
1 & PG$^*$ & $-0.25555(20)$ & 6.94 & 0.90 & 0.32 & 0.58 & 0.342 & 0.098 \\
\hline
$\rs=20$ & & & & & & & & \\
0 & PW & $-0.04611$ & 0.18 & 1.32 & 0.00 & 1.32 & 0.374 & 0.126 \\
0 & PG$^*$ & $-0.04625$ & 0.18 & 1.58 & 0.37 & 1.21 & 0.380 & 0.133 \\
0.02 & PW & $-0.04617(1)$ & 0.23 & 1.34 & 0.00 & 1.34 & 0.373 & 0.125 \\
0.02 & PG & $-0.04630(1)$ & 0.24 & 1.55 & 0.34 & 1.21 & 0.378 & 0.132 \\
0.02 & FG$^*$ & $-0.04635(1)$ & 0.22 & 1.57 & 0.36 & 1.22 & 0.379 & 0.132 \\
0.05 & PW & $-0.04656(1)$ & 0.34 & 1.32 & 0.00 & 1.32 & 0.372 & 0.125 \\
0.05 & PG & $-0.04667(1)$ & 0.35 & 1.52 & 0.33 & 1.19 & 0.378 & 0.131 \\
0.05 & FG$^*$ & $-0.04672(1)$ & 0.29 & 1.53 & 0.34 & 1.19 & 0.378 & 0.131 \\
0.1 & PW & $-0.04824(1)$ & 0.68 & 1.28 & 0.02 & 1.26 & 0.369 & 0.122 \\
0.1 & PG & $-0.04828(2)$ & 0.67 & 1.50 & 0.33 & 1.17 & 0.375 & 0.130 \\
0.1 & FG$^*$ & $-0.04838(2)$ & 0.59 & 1.41 & 0.26 & 1.15 & 0.373 & 0.127 \\
0.2 & PW & $-0.05616(4)$ & 1.98 & 1.18 & 0.15 & 1.03 & 0.358 & 0.116 \\
0.2 & PG & $-0.05625(5)$ & 2.01 & 1.27 & 0.31 & 0.97 & 0.361 & 0.118 \\
0.2 & FG$^*$ & $-0.05729(5)$ & 2.05 & 1.18 & 0.28 & 0.89 & 0.356 & 0.114 \\
1 & PW & $-0.22918(22)$ & 6.89 & 1.17 & 0.72 & 0.45 & 0.360 & 0.104 \\
1 & PG$^*$ & $-0.23184(20)$ & 6.88 & 1.19 & 0.72 & 0.47 & 0.355 & 0.102 \\
0.1 (B) & PW & $-0.04806(2)$ & 0.62 & 1.30 & 0.04 & 1.26 & 0.371 & 0.123 \\
0.1 (B) & PG$^*$ & $-0.04853(1)$ & 0.70 & 1.71 & 0.56 & 1.15 & 0.381 & 0.136 \\
0.1 (B) & FG & $-0.04852(1)$ & 0.66 & 1.65 & 0.51 & 1.14 & 0.379 & 0.134 \\
\hline
$\rs=30$ & & & & & & & & \\
0 & PW & $-0.03184$ & 0.08 & 1.45 & 0.00 & 1.45 & 0.380 & 0.130 \\
0 & PG$^*$ & $-0.03193$ & 0.08 & 2.20 & 0.98 & 1.21 & 0.396 & 0.155 \\
0.02 & PW & $-0.03188(1)$ & 0.09 & 1.43 & 0.00 & 1.43 & 0.379 & 0.130 \\
0.02 & PG & $-0.03197(1)$ & 0.12 & 2.21 & 0.98 & 1.23 & 0.396 & 0.156 \\
0.02 & FG$^*$ & $-0.03204(1)$ & 0.12 & 2.26 & 1.05 & 1.21 & 0.398 & 0.159 \\
0.05 & PW & $-0.03221(1)$ & 0.19 & 1.44 & 0.01 & 1.44 & 0.379 & 0.130 \\
0.05 & PG & $-0.03226(1)$ & 0.19 & 2.16 & 0.95 & 1.21 & 0.395 & 0.153 \\
0.05 & FG$^*$ & $-0.03245(1)$ & 0.20 & 2.21 & 1.02 & 1.19 & 0.396 & 0.157 \\
0.1 & PW & $-0.03374(2)$ & 0.47 & 1.42 & 0.06 & 1.36 & 0.377 & 0.128 \\
0.1 & PG & $-0.03377(2)$ & 0.50 & 2.05 & 0.87 & 1.18 & 0.390 & 0.149 \\
0.1 & FG$^*$ & $-0.03441(2)$ & 0.54 & 1.84 & 0.75 & 1.09 & 0.385 & 0.140 \\
0.2 & PW & $-0.04306(6)$ & 1.97 & 1.30 & 0.30 & 1.00 & 0.362 & 0.120 \\
0.2 & PG & $-0.04270(5)$ & 1.93 & 1.63 & 0.70 & 0.93 & 0.372 & 0.130 \\
0.2 & FG$^*$ & $-0.04498(5)$ & 2.16 & 1.48 & 0.72 & 0.76 & 0.356 & 0.113 \\
1 & PW & $-0.22429(19)$ & 6.88 & 1.16 & 0.65 & 0.50 & 0.344 & 0.102 \\
1 & PG$^*$ & $-0.22495(22)$ & 6.84 & 1.23 & 0.73 & 0.50 & 0.342 & 0.105 \\
\hline
$\rs=50$ & & & & & & & & \\
0 & PW & $-0.01974$ & 0.03 & 1.66 & 0.00 & 1.66 & 0.392 & 0.142 \\
0 & PG$^*$ & $-0.01983$ & 0.03 & 3.46 & 2.28 & 1.18 & 0.432 & 0.221 \\
0.02 & PW & $-0.01977(1)$ & 0.05 & 1.66 & 0.00 & 1.66 & 0.390 & 0.142 \\
0.02 & PG & $-0.01984(1)$ & 0.05 & 3.49 & 2.30 & 1.18 & 0.433 & 0.223 \\
0.02 & FG$^*$ & $-0.01993(1)$ & 0.06 & 3.55 & 2.38 & 1.16 & 0.435 & 0.227 \\
0.05 & PW & $-0.01996(1)$ & 0.07 & 1.63 & 0.01 & 1.62 & 0.390 & 0.141 \\
0.05 & PG & $-0.02003(1)$ & 0.10 & 3.47 & 2.27 & 1.19 & 0.433 & 0.224 \\
0.05 & FG$^*$ & $-0.02034(1)$ & 0.13 & 3.08 & 1.98 & 1.11 & 0.424 & 0.203 \\
0.1 & PW & $-0.02116(2)$ & 0.26 & 1.59 & 0.06 & 1.53 & 0.388 & 0.138 \\
0.1 & PG & $-0.02122(2)$ & 0.29 & 3.32 & 2.16 & 1.15 & 0.429 & 0.215 \\
0.1 & FG$^*$ & $-0.02231(2)$ & 0.41 & 1.70 & 0.73 & 0.96 & 0.389 & 0.139 \\
0.2 & PW & $-0.03126(6)$ & 1.75 & 1.52 & 0.48 & 1.04 & 0.373 & 0.130 \\
0.2 & PG & $-0.03012(5)$ & 1.63 & 2.60 & 1.70 & 0.90 & 0.405 & 0.180 \\
0.2 & FG$^*$ & $-0.03441(5)$ & 2.19 & 1.62 & 1.00 & 0.62 & 0.347 & 0.110 \\
1 & PW & $-0.22077(16)$ & 7.04 & 1.19 & 0.63 & 0.56 & 0.336 & 0.097 \\
1 & PG$^*$ & $-0.22119(17)$ & 6.99 & 1.29 & 0.76 & 0.53 & 0.341 & 0.104 \\

%% file: table_iii.tex
10 & 1 & PW & $-0.46988(40)$ & $-0.0724$ & $-0.5673$ & 15.5 & 1.9 & 0.17 & no & 1.41 & 1.15 & 0.26 & 0.367 \\
10 & 1 & PG & $-0.46867(50)$ & $-0.0726$ & $-0.5620$ & 15.4 & 1.9 & 0.17 & no & 1.51 & 1.24 & 0.27 & 0.361 \\
10 & 1 & FG$^*$ & $-0.47938(32)$ & $-0.0743$ & $-0.5660$ & 15.5 & 1.9 & 0.17 & no & 1.42 & 1.16 & 0.26 & 0.371 \\
10 & 5 & PW & $-0.16902(5)$ & $-0.0797$ & $-0.1175$ & 8.7 & 3.1 & 0.57 & yes & 0.97 & 0.13 & 0.83 & 0.348 \\
10 & 5 & PG & $-0.16888(6)$ & $-0.0790$ & $-0.1169$ & 8.5 & 3.1 & 0.57 & yes & 0.99 & 0.16 & 0.83 & 0.348 \\
10 & 5 & FG$^*$ & $-0.17062(6)$ & $-0.0793$ & $-0.1176$ & 8.5 & 3.1 & 0.56 & yes & 0.97 & 0.15 & 0.83 & 0.348 \\
10 & 10 & PW & $-0.10668(4)$ & $-0.0854$ & $-0.0408$ & 6.0 & 3.2 & 0.72 & yes & 0.99 & 0.01 & 0.97 & 0.351 \\
10 & 10 & PG & $-0.10690(5)$ & $-0.0855$ & $-0.0409$ & 5.9 & 3.2 & 0.72 & yes & 0.99 & 0.03 & 0.95 & 0.351 \\
10 & 10 & FG$^*$ & $-0.10717(5)$ & $-0.0855$ & $-0.0411$ & 5.8 & 3.3 & 0.71 & yes & 0.96 & 0.01 & 0.95 & 0.350 \\
10 & 20 & PW & $-0.07876(4)$ & $-0.0900$ & $-0.0060$ & 4.6 & 3.3 & 0.84 & yes & 1.04 & 0.00 & 1.04 & 0.355 \\
10 & 20 & PG & $-0.07944(4)$ & $-0.0905$ & $-0.0059$ & 4.7 & 3.3 & 0.82 & yes & 1.05 & 0.02 & 1.03 & 0.357 \\
10 & 20 & FG$^*$ & $-0.07964(6)$ & $-0.0907$ & $-0.0058$ & 4.5 & 3.3 & 0.82 & yes & 1.04 & 0.01 & 1.03 & 0.356 \\
20 & 1 & PW & $-0.45592(24)$ & $-0.0376$ & $-0.5739$ & 16.0 & 2.1 & 0.07 & no & 1.57 & 1.49 & 0.07 & 0.358 \\
20 & 1 & PG & $-0.45499(23)$ & $-0.0375$ & $-0.5723$ & 16.0 & 2.1 & 0.07 & no & 1.62 & 1.55 & 0.07 & 0.358 \\
20 & 1 & FG$^*$ & $-0.45761(25)$ & $-0.0378$ & $-0.5713$ & 16.0 & 2.1 & 0.07 & no & 1.60 & 1.53 & 0.08 & 0.359 \\
20 & 5 & PW & $-0.15596(7)$ & $-0.0389$ & $-0.1480$ & 15.4 & 4.2 & 0.19 & no & 1.31 & 1.01 & 0.30 & 0.364 \\
20 & 5 & PG & $-0.15526(7)$ & $-0.0390$ & $-0.1473$ & 15.3 & 4.2 & 0.19 & no & 1.46 & 1.15 & 0.30 & 0.358 \\
20 & 5 & FG$^*$ & $-0.15834(5)$ & $-0.0392$ & $-0.1476$ & 15.4 & 4.2 & 0.19 & no & 1.32 & 1.02 & 0.30 & 0.366 \\
20 & 10 & PW & $-0.09187(5)$ & $-0.0393$ & $-0.0702$ & 13.7 & 5.2 & 0.30 & no & 1.17 & 0.66 & 0.51 & 0.360 \\
20 & 10 & PG & $-0.09170(7)$ & $-0.0394$ & $-0.0698$ & 13.5 & 5.2 & 0.31 & no & 1.26 & 0.76 & 0.50 & 0.354 \\
20 & 10 & FG$^*$ & $-0.09273(6)$ & $-0.0393$ & $-0.0699$ & 13.5 & 5.3 & 0.31 & no & 1.20 & 0.69 & 0.51 & 0.360 \\
20 & 20 & PW$^*$ & $-0.05211(4)$ & $-0.0382$ & $-0.0245$ & 10.2 & 6.1 & 0.46 & no & 0.94 & 0.24 & 0.70 & 0.346 \\
20 & 20 & PG & $-0.05192(5)$ & $-0.0378$ & $-0.0247$ & 10.0 & 6.1 & 0.45 & no & 1.01 & 0.31 & 0.70 & 0.345 \\
20 & 20 & FG & $-0.05202(6)$ & $-0.0378$ & $-0.0247$ & 10.0 & 6.1 & 0.46 & no & 0.90 & 0.21 & 0.69 & 0.346 \\
30 & 1 & PW$^*$ & $-0.44815(22)$ & $-0.0256$ & $-0.5762$ & 16.0 & 2.1 & 0.04 & no & 1.60 & 1.57 & 0.03 & 0.352 \\
30 & 1 & PG & $-0.44758(22)$ & $-0.0256$ & $-0.5751$ & 16.0 & 2.1 & 0.04 & no & 1.61 & 1.58 & 0.03 & 0.352 \\
30 & 1 & FG & $-0.44808(22)$ & $-0.0256$ & $-0.5751$ & 16.0 & 2.1 & 0.04 & no & 1.61 & 1.58 & 0.03 & 0.352 \\
30 & 5 & PW & $-0.15036(6)$ & $-0.0258$ & $-0.1522$ & 15.9 & 4.6 & 0.12 & no & 1.46 & 1.30 & 0.15 & 0.361 \\
30 & 5 & PG & $-0.15003(6)$ & $-0.0258$ & $-0.1522$ & 15.9 & 4.6 & 0.11 & no & 1.56 & 1.40 & 0.15 & 0.357 \\
30 & 5 & FG$^*$ & $-0.15126(5)$ & $-0.0258$ & $-0.1520$ & 15.9 & 4.6 & 0.12 & no & 1.50 & 1.34 & 0.16 & 0.360 \\
30 & 10 & PW & $-0.08857(6)$ & $-0.0260$ & $-0.0784$ & 15.6 & 5.9 & 0.17 & no & 1.36 & 1.10 & 0.26 & 0.366 \\
30 & 10 & PG & $-0.08847(6)$ & $-0.0260$ & $-0.0786$ & 15.6 & 5.9 & 0.16 & no & 1.47 & 1.23 & 0.25 & 0.359 \\
30 & 10 & FG$^*$ & $-0.08989(4)$ & $-0.0255$ & $-0.0794$ & 15.7 & 5.9 & 0.16 & no & 1.46 & 1.21 & 0.25 & 0.367 \\
30 & 20 & PW & $-0.04872(6)$ & $-0.0259$ & $-0.0336$ & 14.9 & 7.2 & 0.23 & no & 1.26 & 0.89 & 0.37 & 0.368 \\
30 & 20 & PG & $-0.04870(4)$ & $-0.0259$ & $-0.0337$ & 14.9 & 7.1 & 0.23 & no & 1.36 & 1.00 & 0.36 & 0.360 \\
30 & 20 & FG$^*$ & $-0.04930(4)$ & $-0.0255$ & $-0.0339$ & 14.9 & 7.3 & 0.23 & no & 1.37 & 1.01 & 0.37 & 0.367 \\
50 & 1 & PW & $-0.44187(25)$ & $-0.0157$ & $-0.5836$ & 16.0 & 2.0 & 0.02 & no & 1.61 & 1.59 & 0.01 & 0.345 \\
50 & 1 & PG & $-0.44143(20)$ & $-0.0157$ & $-0.5813$ & 16.0 & 2.0 & 0.02 & no & 1.61 & 1.60 & 0.01 & 0.346 \\
50 & 1 & FG$^*$ & $-0.44273(19)$ & $-0.0157$ & $-0.5789$ & 16.0 & 2.0 & 0.02 & no & 1.61 & 1.60 & 0.01 & 0.345 \\
50 & 5 & PW & $-0.14385(6)$ & $-0.0156$ & $-0.1552$ & 16.0 & 4.7 & 0.06 & no & 1.54 & 1.48 & 0.06 & 0.353 \\
50 & 5 & PG & $-0.14372(6)$ & $-0.0155$ & $-0.1556$ & 16.0 & 4.6 & 0.06 & no & 1.59 & 1.53 & 0.06 & 0.353 \\
50 & 5 & FG$^*$ & $-0.14468(4)$ & $-0.0156$ & $-0.1563$ & 16.0 & 4.5 & 0.06 & no & 1.58 & 1.52 & 0.06 & 0.352 \\
50 & 10 & PW & $-0.08395(6)$ & $-0.0154$ & $-0.0835$ & 16.0 & 6.2 & 0.08 & no & 1.51 & 1.40 & 0.11 & 0.358 \\
50 & 10 & PG & $-0.08397(6)$ & $-0.0154$ & $-0.0832$ & 15.9 & 6.3 & 0.08 & no & 1.57 & 1.47 & 0.11 & 0.357 \\
50 & 10 & FG$^*$ & $-0.08465(4)$ & $-0.0153$ & $-0.0840$ & 16.0 & 6.1 & 0.08 & no & 1.56 & 1.47 & 0.10 & 0.359 \\
50 & 20 & PW & $-0.04658(5)$ & $-0.0153$ & $-0.0411$ & 15.9 & 7.7 & 0.11 & no & 1.50 & 1.35 & 0.15 & 0.365 \\
50 & 20 & PG & $-0.04650(5)$ & $-0.0153$ & $-0.0410$ & 15.9 & 7.7 & 0.11 & no & 1.58 & 1.43 & 0.15 & 0.361 \\
50 & 20 & FG$^*$ & $-0.04699(4)$ & $-0.0150$ & $-0.0415$ & 16.0 & 7.6 & 0.11 & no & 1.57 & 1.42 & 0.15 & 0.367 \\

%% file: meta_sec4_lll.tex

\section{{W}igner Crystallization and Disorder in the Lowest {L}andau Level: {H}artree--{F}ock Theory}
\label{sec:lll}

The zero-field studies of Secs.~\ref{sec:hf_letter} and
\ref{sec:vmc} establish that charged disorder suppresses crystalline
correlation at every $\rs$ studied.  In the lowest Landau level the
kinetic energy is quenched and the clean HF state is already a
Wigner crystal at every partial filling~\cite{FPA1979,%
YoshiokaFukuyama1979,YoshiokaLee1983,MakiZotos1983}.  The LLL is
therefore a controlled setting in which the disorder-driven loss of
crystallinity can be isolated from the kinetic-energy question: the
only competition is between Coulomb crystallization and disorder
pinning, with $\eta_{\mathrm{dis}} = \nimpurity/n$ (which, in the
LLL, equals the dimensionless impurity-to-electron ratio
$\nimpurity/n = N_i/N$ on the torus) as the sole dimensionless
disorder parameter at fixed form factor.

\emph{A word on terminology.} In the discussion that follows we
speak of hills, basins, lakes, puddles and droplets, and we
deliberately avoid the word \emph{percolation}.  In the quantum Hall
context that word already denotes the semiclassical drift of
guiding centers along equipotentials and the plateau transition
constructed from it, which is a different question from the one
asked here and invites a confusion that is best avoided.  What is at
issue in this section is whether the electron density breaks up into
disconnected pools of charge; \emph{puddle} is the accurate word for
that, and we make no claim about any percolation threshold.

\input{body_lll}

%% file: body_lll.tex
The impurity potential of a charged impurity at a setback produces
a smooth hill or basin many magnetic lengths across, so the LLL
problem is one of electrons spreading across a landscape of hills
and basins rather than scattering from point centers; whether the
density at low filling forms a crystal across this landscape or
collects in its basins as puddles is the central question
below.  We take as a motivating experiment the Yazdani-group
scanning tunneling microscopy of Bernal bilayer graphene in a
strong perpendicular field~\cite{Tsui2024}, which images the LLL
electronic density directly and resolves three regimes: a
triangular Wigner crystal with sharp Bragg peaks over an
intermediate window of filling ($0.13 \lesssim \nu \lesssim 0.38$),
a localized state without discernible spatial order at lower
filling, and an isotropic liquid at higher density that nevertheless
retains the crystalline Bragg wavevector.  The low-filling regime
is the one that concerns us.  A localized state without spatial
order is an amorphous solid in the sense of the preceding
sections, and the natural agent is the disorder of the sample
rather than the electron--electron interaction, as argued for the
zero-field case in Refs.~\cite{Babbar2026a,HuangDS2026}.  Hartree--Fock in
a single Landau level contains no fractional quantum Hall liquid,
so the ground state is a crystal at every filling we compute and
the only competition left is between Coulomb crystallization and
disorder pinning.  We fix one disorder realization throughout, in
analogy with a single scanning-probe experiment on one sample, and
ask how the crystal is destroyed along the two cuts $\nu \to 0$ at
fixed $n_i$ and $n_i \to \infty$ at fixed $\nu$.  We take as
background the pinned-Wigner-crystal transport literature in high-
mobility GaAs 2DEGs at small LL filling~\cite{ShayeganRev1997,FertigRev1997,Wilson1981,Willett1988,Goldman1990,Jiang1990,Pan2002,Ye2002,Chung2022,Madathil2024}; the present calculation adds that in the regime where the insulator is observed, the crystalline coherence length has collapsed to the lattice spacing, so there is no long-range translational order left to pin.

\subsection{Formalism}
\label{lll:sec:form}

\subsubsection{Projected Hamiltonian and the magnetic translation algebra}

With all electrons in the LLL the Hamiltonian is, up to a constant,
\begin{equation}
H=\frac{1}{2A}\sum_{\bq\neq0}\bar v(q)\left[\rhob_{-\bq}\rhob_{\bq}-N\right],
\quad
\bar v(q)=\frac{2\pi e^2}{\epsilon q}e^{-q^2\ell^2/2},
\label{lll:eq:H}
\end{equation}
where $\rhob_{\bq}=\sum_j e^{-i\bq\cdot\bR_j}$ is built from the guiding centers
$\bR_j$, with $[X,Y]=-i\ell^2$. We work on a torus $L_x\times L_y$ with
$N_\phi=L_xL_y/2\pi\ell^2$ flux quanta, so the LLL is $N_\phi$ dimensional. The
magnetic translations $\tau_{\bq}=e^{-i\bq\cdot\bR}$ at the allowed wavevectors
$\bq_{s,t}=(2\pi s/L_x,\,2\pi t/L_y)$ are represented faithfully by clock and
shift matrices,
\begin{equation}
\tau_{s,t}=\omega^{st/2}\,U^{-s}V^{t},\qquad \omega=e^{2\pi i/N_\phi},
\label{lll:eq:tau}
\end{equation}
with $U|m\rangle=\omega^m|m\rangle$, $V|m\rangle=|m+1\rangle$, obeying
$\tau_{s,t}^{\dagger}=\tau_{-s,-t}$ and
$\mathrm{Tr}\,(\tau^{\dagger}_{s,t}\tau_{s',t'})=N_\phi\delta_{ss'}\delta_{tt'}$.
This is not a convenience. The $\tau$'s span the entire $N_\phi\times N_\phi$
one-body operator algebra, and that fact is what makes the rest of the argument
exact.

\subsubsection{The Hartree--Fock energy is a functional of the order parameters alone}

Let $P$ be the HF density matrix and define the order parameters
\begin{equation}
\Delta(\bq)=\frac{1}{N_\phi}\mathrm{Tr}\,\!\left[\tau_{\bq}P\right].
\end{equation}
Wick decomposition of Eq.~(\ref{lll:eq:H}) gives exactly
\begin{equation}
E=\frac{1}{2A}\sum_{\bq\neq0}\bar v(q)\Big[\,|\mathrm{Tr}\,\tau_{\bq}P|^2
-\mathrm{Tr}\,\tau^{\dagger}_{\bq}P\tau_{\bq}P\,\Big],
\end{equation}
the $-N$ term of Eq.~(\ref{lll:eq:H}) cancelling the self-exchange. Because the
$\tau$'s are complete, $P$ may be expanded in them, the exchange trace collapses,
and the energy becomes a functional of $\Delta$ alone \cite{CoteMacDonald1990}:
\begin{equation}
\frac{E}{N_\phi}=\frac12\sum_{\bG}\Big[U_H(G)\big(1-\delta_{\bG,0}\big)-X(G)\Big]
\,|\Delta(\bG)|^2,
\label{lll:eq:EDelta}
\end{equation}
with
\begin{align}
U_H(q)&=\frac{\bar v(q)}{2\pi\ell^2}
=\frac{e^2}{\epsilon\ell^2 q}e^{-q^2\ell^2/2},\\
X(q)&=\int\!\frac{d^2k}{(2\pi)^2}\,\bar v(k)\,e^{i\ell^2(\bm{k}\times\bq)_z}
\nonumber\\
&=\sqrt{\tfrac{\pi}{2}}\,\frac{e^2}{\epsilon\ell}\,e^{-q^2\ell^2/4}
I_0\!\left(\tfrac{q^2\ell^2}{4}\right).
\label{lll:eq:X}
\end{align}
Equation~(\ref{lll:eq:EDelta}) with $\Delta(0)=\nu$ and $\Delta(\bG\neq0)=0$ returns
$E/N=-\nu\sqrt{\pi/8}\,(\ee)$, the uniform reference, and at $\nu=1$ it returns
the exact $-\sqrt{\pi/8}\,(\ee)$.

\subsubsection{Why no state is uniform, and why the one we find is a crystal}
\label{lll:sec:always}

The order parameters are not free. Idempotency $P^2=P$ with
$\mathrm{Tr}\,P=\nu N_\phi$ gives the exact sum rule
\begin{equation}
\sum_{\bG}|\Delta(\bG)|^2=\nu
\qquad\Longrightarrow\qquad
\sum_{\bG\neq0}|\Delta(\bG)|^2=\nu(1-\nu).
\label{lll:eq:sumrule}
\end{equation}
Every Slater determinant in a single Landau level is therefore obliged to carry a
density-wave weight $\nu(1-\nu)$. The uniform state $P=\nu\mathbb{I}$ is not a
determinant at all. This is exact and it is the whole of what the sum rule gives:
it forbids homogeneity, and it says nothing about periodicity, about which
Bravais lattice, or about one electron per site. The remaining freedom is where
in $\bq$ space the mandatory weight sits, and the kernel that decides this is
\begin{equation}
W(q)=U_H(q)-X(q).
\end{equation}
$W$ is positive at small $q$, where the Hartree cost of charge modulation
diverges as $1/q$, crosses zero at $q^*\ell=0.7026$, and is negative for all
larger $q$, with a minimum $W_{\min}=-0.5568\,(\ee)$ at $q\ell=1.5677$; at large
$q$ the Hartree term dies as $e^{-q^2\ell^2/2}$ while $X\to(\ee)/q\ell$, so $W$
stays negative. Weight that must be placed somewhere can always be placed where
$W<0$. Hence the LLL Hartree--Fock ground state is inhomogeneous for every
$0<\nu<1$. This is special to the lowest level: in higher Landau levels the
oscillating form factor makes $W$ change sign repeatedly, and the HF ground state
is a stripe or bubble phase over much of the partial filling range
\cite{FKS1996}. In the lowest level $W$ has a single sign change.

Two further exact statements follow, and then the argument becomes numerical.
First, since $W(G)\ge W_{\min}$ everywhere, Eq.~(\ref{lll:eq:EDelta}) and the sum
rule give a rigorous variational floor,
\begin{align}
\frac{E}{N}&\;\ge\;-\tfrac12 X(0)\,\nu+\tfrac12 W_{\min}(1-\nu)\nonumber\\
&\;=\;-0.6267\,\nu-0.2784\,(1-\nu)
\label{lll:eq:floor}
\end{align}
in $\ee$, which no determinant in this level can beat. Second, that floor is not
attained: it gives $-0.4525$ at $\nu=1/2$ and $-0.3945$ at $\nu=1/3$ against the
computed $-0.4436$ and $-0.3885$. The gap is the price of idempotency. The
$\Delta(\bG)$ are not free amplitudes on a sphere of radius $\nu(1-\nu)$; $P^2=P$
couples them, so the weight cannot simply be piled at the minimum of $W$. The
sum rule alone therefore cannot select a lattice, and we do not claim that it
does. What selects one is the full minimization, and the answer it returns is a
one-electron-per-site crystal at every filling we computed, with the energy
within $2\%$ of the floor near half filling. By particle--hole symmetry the
energy gain is symmetric about $\nu=1/2$: the crystal is an electron crystal
below half filling and a hole crystal above it. This is the content of
Fig.~\ref{lll:fig:clean}(a) and (d).

\subsubsection{Quenched Coulomb impurities}

$N_i$ impurities of charge $\pm e$ in equal numbers sit at fixed random in-plane
positions $\bR_j$ with a setback $d$ out of the plane. The potential seen by an
electron and its transform are
\begin{align}
V(\br)&=-\sum_j s_j\frac{e^2/\epsilon}{\sqrt{|\br-\bR_j|^2+d^2}},\quad s_j=\pm1,\\
V(\bq)&=-\frac{2\pi e^2}{\epsilon q}e^{-qd}\sum_j s_j e^{-i\bq\cdot\bR_j}.
\label{lll:eq:Vq}
\end{align}
Neutrality kills the $\bq=0$ component, so no background term is needed.
Projected into the LLL this is a one-body operator in the same algebra,
\begin{equation}
h_{\rm imp}=\frac{1}{A}\sum_{\bq\neq0}V(\bq)\,e^{-q^2\ell^2/4}\,\tau^{\dagger}_{\bq},
\label{lll:eq:himp}
\end{equation}
so interaction and disorder are treated on exactly the same footing, with the
same form factors and the same matrices. The total energy is
$E=E_{\rm HF}[P]+\mathrm{Tr}\,(h_{\rm imp}P)$ and the HF single-particle
Hamiltonian is $h=\delta E_{\rm HF}/\delta P+h_{\rm imp}$. Two checks follow
immediately and are satisfied numerically: $h_{\rm imp}$ is traceless by
neutrality, and it drops out identically at $\nu=1$, where $P=\mathbb{I}$ is
forced.

\subsection{Clean system}

\begin{figure*}[!t]
\centering
\includegraphics[width=\textwidth]{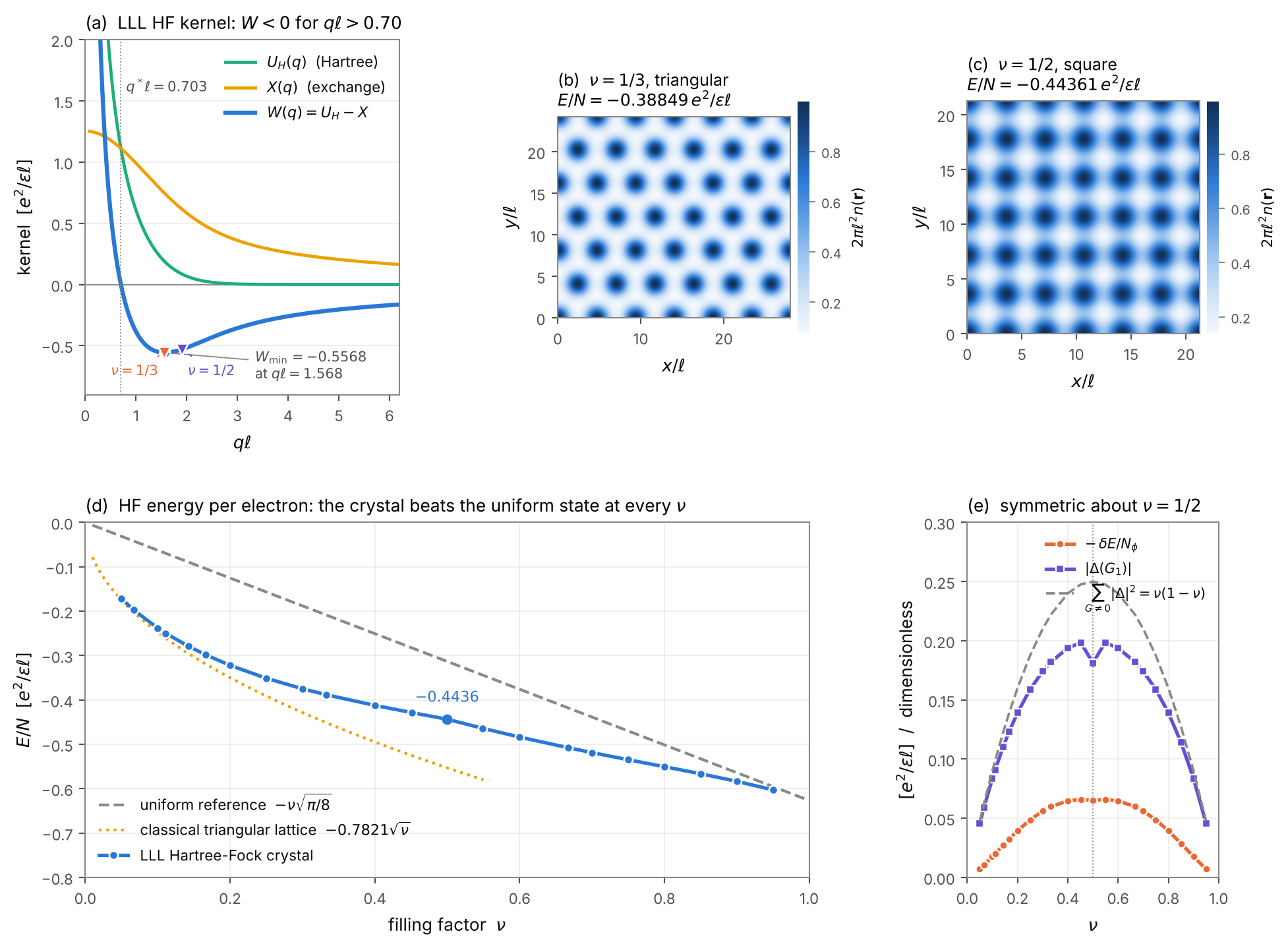}
\caption{Clean lowest-Landau-level Hartree--Fock.
(a) The kernel $W(q)=U_H-X$, negative for $q\ell>0.703$ with a minimum at
$q\ell=1.568$; triangles mark the first-shell $G_1$ of the triangular crystal at
$\nu=1/3$ and $\nu=1/2$.
(b),(c) Self-consistent densities.
(d) Energy per electron against filling, with the uniform reference
$-\nu\sqrt{\pi/8}$ and the classical triangular lattice $-0.7821\sqrt{\nu}$.
(e) The condensation energy per flux quantum and the first-shell order parameter,
both symmetric about $\nu=1/2$, together with the sum rule $\nu(1-\nu)$.}
\label{lll:fig:clean}
\end{figure*}

We solve the HF equations self-consistently on the torus in the Landau gauge
basis, seeding with coherent-state determinants on commensurate lattices
\cite{MakiZotos1983} and relaxing. Table~\ref{lll:tab:valid} lists the checks. The calculation reproduces the
two limits that are known exactly and the one published number that is directly
comparable, and it satisfies the particle--hole relation
$e(\nu)-e(1-\nu)=\sqrt{\pi/8}\,(1-2\nu)$, $e=E/N_\phi$, to machine precision.

\begin{table}[b]
\caption{Validation of the clean calculation. Energies in $\ee$.}
\label{lll:tab:valid}
\small

\begin{tabular}{@{}lll@{}}
\toprule
Quantity & This work & Exact / published \\
\midrule
$\nu=1$, $E/N$                 & $-0.626657068658$ & $-\sqrt{\pi/8}$ \\
Uniform reference, $E/N$       & $-\nu\sqrt{\pi/8}$ & exact \\
Sum rule $\sum_{\bG}|\Delta|^2$ & $\nu$ to $10^{-10}$ & $\nu$ \\
$h=\delta E/\delta P$          & $10^{-10}$ (fin. diff.) & --- \\
$\nu=1/3$ triangular, $E/N$    & $-0.38849$ & $-0.389$ \cite{YoshiokaLee1983} \\
$\nu\to0$, $E/N$ & $0.985\times$ cl., $\nu=0.04$ & $-0.7821\sqrt{\nu}$ \\
$e(\nu)-e(1-\nu)$              & to $10^{-17}$ & $\sqrt{\pi/8}(1-2\nu)$ \\
\bottomrule
\end{tabular}
\end{table}

At $\nu=1/2$ the crystal energy is $E/N=-0.4436\,(\ee)$ against $-0.3133$ for the
uniform reference, a condensation energy of $0.130$ per electron, which is
$29\%$ of the total. At $\nu=1/3$ we get $-0.38849$. The condensation energy per
electron peaks near $\nu\simeq0.21$ and vanishes at both ends, and per flux
quantum it is symmetric about $\nu=1/2$, as it must be.

One feature of the clean problem matters for the disordered one. We scanned the
Bravais lattice continuously using a centered-rectangular cell of aspect
$r=b/a$, which interpolates between the square lattice at $r=1$ and the
triangular lattice at $r=\sqrt3$. At $\nu=1/2$ the entire range $0.9\le r\le1.9$
spans only $1.5\times10^{-4}\,(\ee)$ in energy, and the square lattice lies
$1.2\times10^{-4}$ below the triangular one. The triangular lattice takes over
below $\nu\simeq0.45$, but by only $9\times10^{-4}$ there, reaching its largest
margin of $2.8\times10^{-3}$ near $\nu\simeq0.30$. The reason is
visible in Fig.~\ref{lll:fig:clean}(a): at $\nu=1/2$ the triangular $G_1\ell=1.905$
has overshot the minimum of $W$, while the square $G_1\ell=1.772$ sits closer to
it. The energy difference is far below anything physical. The significant
statement is not which lattice wins but that the crystal has almost no shear
rigidity near half filling.

This near-degeneracy has a consequence for the Larkin-length
analysis that follows and we flag it here. The Larkin coherence
length $R_c \sim \sqrt{c_{66}/\langle f^2\rangle}$ is set by the
shear modulus $c_{66}$ of the clean crystal. Over the filling
window $0.40 \lesssim \nu \lesssim 0.55$ the triangular LLL HF
crystal has $c_{66}$ that is small on the scale of the Coulomb
energy $e^2/\epsilon\ell$ by the ratio
$\simeq 10^{-3}$--$10^{-4}$ visible in the triangular--square
difference above, and the Larkin estimate correspondingly predicts
$R_c$ of order a lattice spacing at any accessible disorder.
In this filling window, therefore, the loss of structural
coherence seen in the disordered calculations below is
substantially set by the small clean shear modulus and only
partially by the disorder strength itself. Outside this window
($\nu \lesssim 0.35$) the clean $c_{66}$ is of order the full
Coulomb scale and the Larkin coherence length does respond to the
disorder as the collective-pinning estimate predicts.

\subsection{Numerical method with disorder}

Two points are worth recording because they change answers.

First, plain damped self-consistency is unreliable once the Coulomb disorder is
switched on. An imperfectly screened density modulation costs a Hartree energy
$\propto U_H(q)\sim1/q$, which is enormous at long wavelength, and the iteration
converges to stationary points far above the variational optimum; we obtained
positive $E/N$ this way. We therefore switch the impurity term on adiabatically,
$h=\delta E_{\rm HF}/\delta P+\lambda h_{\rm imp}$ with $\lambda:0\to1$ in ten
steps, re-converging at each step so that the state stays screened throughout,
and we keep a variational floor by evaluating every trial determinant, including
the unrelaxed starting ones.

Second, at low filling the HF landscape is glassy, which is unsurprising for an
amorphous system but has a practical consequence. Searching harder, with several
random starting determinants in addition to the lattice and impurity-eigenstate
seeds, finds solutions that are both lower in energy and markedly less ordered.
At $\nu=0.15$ with $n_i/n=1.87$, the harder search lowers $E/N$ by
$6\times10^{-4}\,(\ee)$ while raising the Bragg-ring $N_{\rm eff}$ defined below
from $23$ to $71$, that is, from a state that still looks partly crystalline to
one that does not. An under-converged Hartree--Fock calculation systematically
\emph{overstates} crystalline order at low filling. All results reported here use
the harder search.

\subsection{Disorder results}
\label{lll:sec:dis}

\begin{figure*}[!t]
\centering
\includegraphics[width=\textwidth]{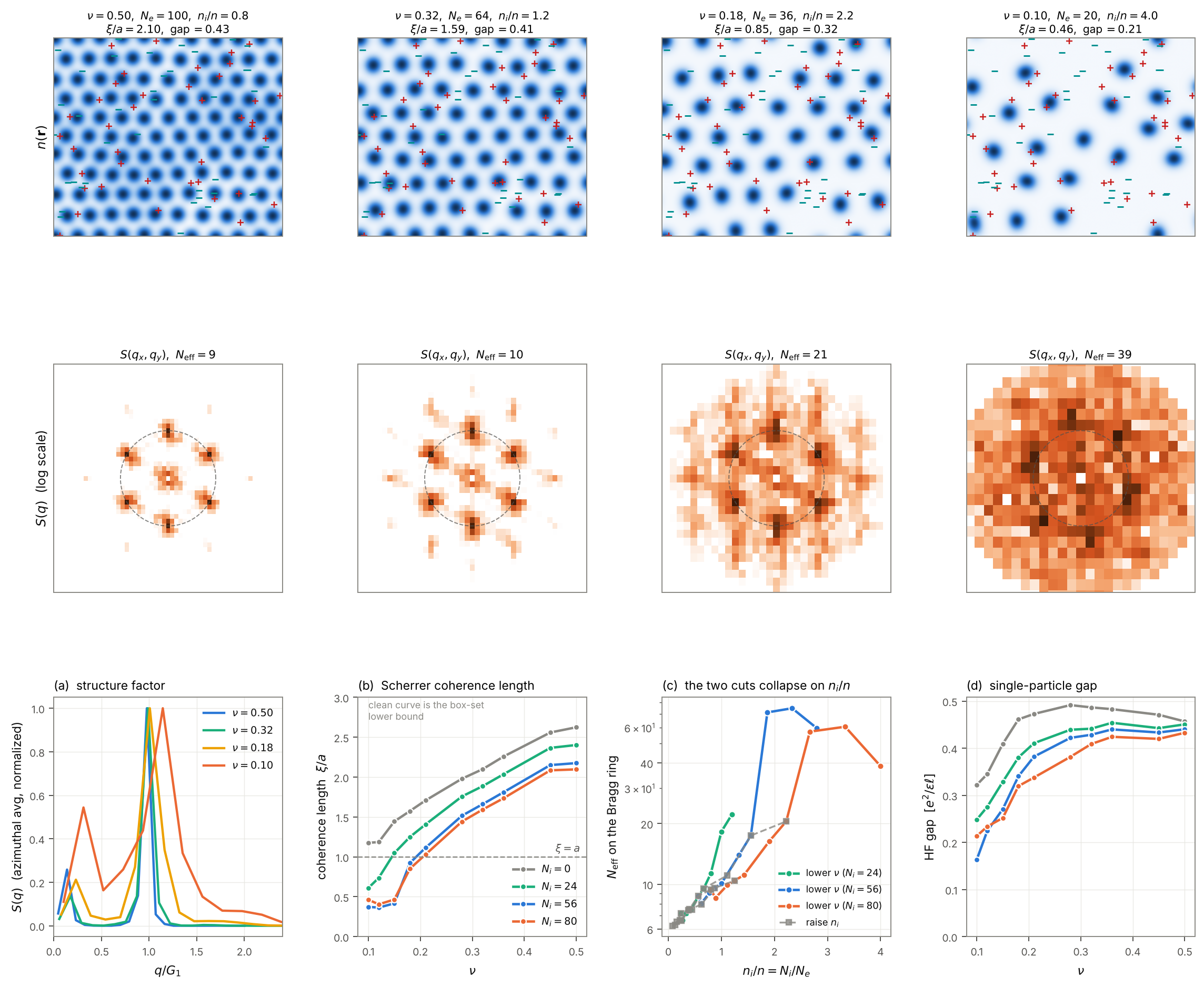}
\caption{Same sample, same field, same $N_i=80$ impurities ($\pm e$, $d=6\ell$,
$V_{\rm rms}=0.149\,\ee$); only the filling changes.
Top row, self-consistent density, with $+$ and $-$ marking the impurity charges.
Middle row, the coherent structure factor on a logarithmic scale; the dashed
circle is $|q|=G_1$. Six sharp Bragg spots at $\nu=0.50$ become a featureless
ring at $\nu=0.10$.
(a) Azimuthally averaged $S(q)$.
(b) The Scherrer coherence length of Eq.~(\ref{lll:eq:scherrer}) in units of the
lattice constant; the clean curve is the lower bound set by the box.
(c) The two cuts plotted against $n_i/n$.
(d) The single-particle gap.}
\label{lll:fig:dis}
\end{figure*}

All disorder calculations use one fixed torus, $N_\phi=200$, $38.09\times32.99\,
\ell$, area $1257\,\ell^2$, with $N_e$ between $20$ and $100$, and one fixed
realization of up to $80$ impurities whose positions and signs are never
redrawn. Smaller impurity numbers are nested subsets of the same list, so the
randomness is fixed as $n_i$ is varied. We characterize the states by:

\emph{(i)} the coherent structure factor
$S(\bq)=|\langle\rho_{\bq}\rangle|^2/N_e$ with
$\langle\rho_{\bq}\rangle=N_\phi\Delta(\bq)e^{-q^2\ell^2/4}$, which is what the
Fourier transform of a density image measures;

\emph{(ii)} the inverse participation ratio of that weight on the ring
$0.55\,G_1<|q|<1.55\,G_1$,
\begin{equation}
N_{\rm eff}=\frac{\big(\sum_{\rm ring}S\big)^2}{\sum_{\rm ring}S^2},
\label{lll:eq:neff}
\end{equation}
the effective number of reciprocal-space grid points over which the ring weight
is spread. \emph{This single number is the diagnostic we use throughout to say
whether a state is a crystal or an amorphous solid, and it is the one quantity
to watch.} A sharp six-spot Bragg pattern gives $N_{\rm eff}=6$ exactly, and we
obtain exactly $6.0$ for every clean state. As the weight spreads off those
spots and around the ring $N_{\rm eff}$ grows, reaching the full number of grid
points on the ring --- $152$ at $\nu=0.10$ and $754$ at $\nu=1/2$ on this torus
--- for a featureless ring. \emph{Larger $N_{\rm eff}$ means a more amorphous
system and smaller $N_{\rm eff}$ a more crystalline one.} Nothing in
Eq.~(\ref{lll:eq:neff}) assumes a lattice, a symmetry or a peak shape, and no fit
enters it, which is why we make it the primary measure and read everything else
against it;

\emph{(iii)} a length obtained from $N_{\rm eff}$ by the Scherrer relation. If
the ring weight sits in six broadened Bragg spots, it occupies $N_{\rm eff}$
reciprocal-space cells of area $(2\pi)^2/A$ each shared among the six, so each
spot has radius $\kappa=(2\pi/\sqrt{A})\sqrt{N_{\rm eff}/6\pi}$ and
\begin{equation}
\xi=\frac{1}{\kappa}=\frac{1}{2\pi}\sqrt{\frac{6\pi A}{N_{\rm eff}}},
\label{lll:eq:scherrer}
\end{equation}
quoted below as $\xi/a$ with $a$ the lattice constant. Two limits of this
estimate should be kept in mind. A spot confined to a single cell gives
$\xi=10.0\,\ell$ on this torus, so values at that ceiling are lower bounds set
by the box. At the opposite end the six-spot assumption fails: once the pattern
is an isotropic ring, angular spreading and radial peak width are different
quantities, $\xi$ becomes a monotone relabelling of $N_{\rm eff}$ rather than an
independent measurement, and it understates the true correlation length. We
therefore checked it against a direct real-space measure --- projecting the
density onto the dominant Bragg vector, keeping the slow envelope, and fitting
the decay of its autocorrelation --- which assumes no peak shape. That measure
is itself bounded by the box at $14.8\,\ell$ and is saturated wherever the
crystal survives; at the six most disordered states, $\nu=0.10$, $0.12$ and
$0.15$ with $N_i=56$ and $80$, it does decay inside the sample and gives
$\xi_{\rm dir}/a$ between $0.86$ and $1.44$ where
Eq.~(\ref{lll:eq:scherrer}) gives $0.36$ to $0.47$. The two agree that translational order is gone, and disagree on the
number by a factor of two to three. We therefore state the conclusion as a
correlation length of order one lattice constant, and treat $\xi$ as a
Scherrer estimate rather than a measured length;

\emph{(iv)} the Hartree--Fock single-particle gap at the Fermi level.

\subsubsection{Lowering the filling at fixed impurity density}

This is the cut in which the sample, the magnetic field and the impurity
configuration are all held fixed and electrons are removed, as a gate would do.
Table~\ref{lll:tab:cut1} and Fig.~\ref{lll:fig:dis} give the result. The crystal
degrades monotonically as $\nu$ falls. The structure factor passes from six sharp
Bragg spots, through spots that broaden and split while a background fills in,
to a ring with no angular structure. $N_{\rm eff}$ rises from $8.7$ to $59.4$,
the coherence length falls from above two lattice constants to of order one, and
the single-particle gap falls from $0.44$ to $0.16\,(\ee)$ without closing.

\begin{table}[b]
\caption{Lowering $\nu$ at fixed impurity density. One fixed sample,
$N_i=56$, $d=6\ell$, $V_{\rm rms}=0.137\,\ee$. Energies in $\ee$.}
\label{lll:tab:cut1}
\small
\begin{tabular}{@{}cccrrrr@{}}
\toprule
$\nu$ & $N_e$ & $n_i/n$ & $E/N$ & gap & $N_{\rm eff}$ & $\xi/a$ \\
\midrule
0.500 & 100 & 0.56 & $-0.44795$ & 0.440 & 8.7 & 2.18 \\
0.450 &  90 & 0.62 & $-0.43305$ & 0.433 & 8.0 & 2.15 \\
0.360 &  72 & 0.78 & $-0.40387$ & 0.440 & 9.1 & 1.81 \\
0.320 &  64 & 0.88 & $-0.38973$ & 0.428 & 9.6 & 1.66 \\
0.280 &  56 & 1.00 & $-0.37307$ & 0.422 & 10.1 & 1.52 \\
0.210 &  42 & 1.33 & $-0.33829$ & 0.382 & 14.0 & 1.11 \\
0.180 &  36 & 1.56 & $-0.32151$ & 0.340 & 17.4 & 0.92 \\
0.150 &  30 & 1.87 & $-0.30210$ & 0.271 & 71.2 & 0.42 \\
0.120 &  24 & 2.33 & $-0.28187$ & 0.224 & 74.6 & 0.36 \\
0.100 &  20 & 2.80 & $-0.26893$ & 0.164 & 59.4 & 0.37 \\
\bottomrule
\end{tabular}
\end{table}

There is a shallow non-monotonicity at the top of the range. In every disorder
setting we examined, at both setbacks, $N_{\rm eff}$ is slightly smaller at
$\nu=0.45$ than at $\nu=0.50$: $8.7\to8.0$ at $N_i=56$, $9.4\to8.6$ at $N_i=80$,
$7.2\to6.6$ at $N_i=24$, and $9.5\to8.5$ and $20.6\to15.8$ at the shorter
setback. Order improves marginally on moving off half filling before the loss of
order sets in. The mechanism is in the clean calculation. At $\nu=1/2$ the
Bravais lattice is nearly shape-degenerate, so the crystal has the least shear
rigidity it will ever have and is the easiest to deform; moving away from half
filling stiffens it, while $n_i/n$ grows as $1/\nu$ and eventually dominates.

\subsubsection{Raising the impurity density at fixed filling}

Here the field, the filling and the electron number are fixed and impurities are
added from the same nested list. At $\nu=1/2$, $N_{\rm eff}$ rises from $6.0$ to
$9.4$ and $\xi/a$ falls from $2.63$ to $2.10$ as $N_i$ goes from $0$ to $80$,
that is as $n_i/n$ goes from $0$ to $0.80$; at $\nu=0.18$ the same impurity
numbers take $N_{\rm eff}$ from $6.0$ to $20.5$ and $\xi/a$ from $1.58$ to
$0.85$, over $n_i/n$ from $0$ to $2.22$. The response is qualitatively identical
to the first cut.

\subsubsection{Which variable controls the loss of order}

The two cuts can be compared directly by asking which variable organizes all the
disordered states onto a single curve. We measure this by the root-mean-square
scatter about a running median, as a fraction of the full range of the response,
so that smaller is better. Within a fixed disorder potential the answer is
unambiguous: the number of impurities per electron,
\begin{equation}
\frac{n_i}{n}=\frac{N_i}{N_e},
\end{equation}
organizes the data about twice as well as the bare disorder amplitude
$V_{\rm rms}$, the root-mean-square of the one-body potential of
Eq.~(\ref{lll:eq:himp}). At $d=6\ell$, over $39$ states, the scatter on $n_i/n$ is
$0.17$, $0.16$ and $0.14$ for $N_{\rm eff}$, $\xi/a$ and the gap, against
$0.32$, $0.29$ and $0.30$ on $V_{\rm rms}$. At $d=3\ell$, over $28$ states, it is
$0.18$, $0.19$, $0.19$ against $0.33$, $0.30$, $0.20$. Matched pairs make the
same point directly. At $d=6\ell$, $(N_e,N_i)=(100,56)$ and $(42,24)$ sit at
$n_i/n=0.56$ and $0.57$ and give $N_{\rm eff}=8.7$ and $8.8$ although their bare
amplitudes differ by a factor of $1.5$; $(90,56)$ and $(64,40)$ sit at $0.62$ and
$0.63$ and give $8.0$ and $8.0$; $(72,56)$ and $(100,80)$ sit at $0.78$ and
$0.80$ and give $9.1$ and $9.4$. Above $n_i/n\simeq1$ the point-by-point matching
degrades, because the states there are glassy and $N_{\rm eff}$ fluctuates
strongly between neighboring points; the collapse quoted above is a statement
about the thirty-nine states taken together and not about any single pair.

Within a fixed disorder potential, therefore, lowering the filling at fixed
impurity density and raising the impurity density at fixed filling are to a good
approximation the same experiment, as conjectured. We put it no more strongly
than that: the scatter on $n_i/n$ is about half the scatter on $V_{\rm rms}$ over
the range studied, which makes $n_i/n$ a much better organizing variable rather
than an exact collapse, and the correspondence is tightest for $n_i/n\lesssim1$.
The physical content is that what matters is how many impurity wells are
available per electron, not how deep the potential is in absolute terms.

\subsubsection{Dependence on the impurity setback}
\label{lll:sub:setback}

\begin{figure*}[!t]
\centering
\includegraphics[width=0.96\textwidth]{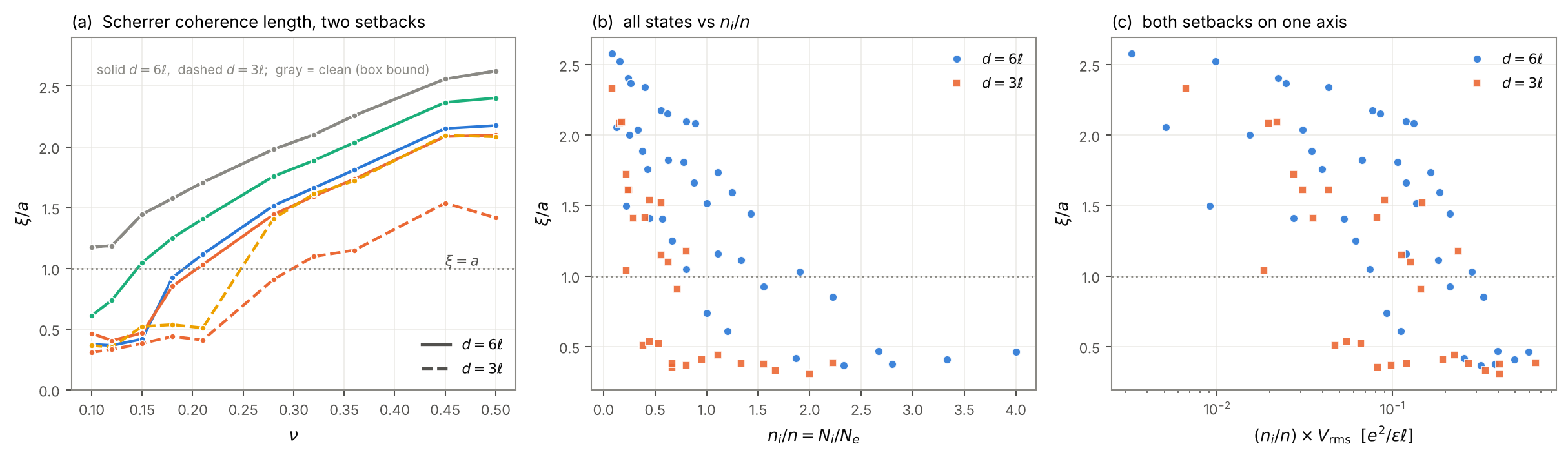}
\caption{Robustness check. Identical sample and identical impurity positions, two
setback distances. (a) The Scherrer coherence length of Eq.~(\ref{lll:eq:scherrer})
against filling; solid $d=6\ell$, dashed $d=3\ell$, gray the clean box-limited
bound. (b) All disordered states against
$n_i/n$; the two setbacks separate. (c) The same states against
$(n_i/n)\times V_{\rm rms}$.}
\label{lll:fig:setback}
\end{figure*}

We repeated the entire calculation with the setback changed from $d=6\ell$ to
$d=3\ell$, keeping the sample and the impurity positions identical, which changes
the shape of the disorder potential and not only its strength. Every qualitative
conclusion survives: monotonic loss of order on lowering $\nu$ at fixed $n_i$,
the same behavior on raising $n_i$ at fixed $\nu$, the superiority of $n_i/n$
over $V_{\rm rms}$ as the organizing variable within a fixed $d$, and the shallow
improvement on moving off half filling. Quantitatively the shorter-ranged
potential is more damaging at the same $n_i/n$, which it should be, since it has
more weight at the lattice wavevector.

The two setbacks do not, however, collapse onto one another on $n_i/n$ alone.
Over the combined set of $67$ disordered states the scatter on $n_i/n$ is $0.30$,
$0.23$ and $0.19$ against $0.28$, $0.28$ and $0.23$ on $V_{\rm rms}$, so that
neither is uniformly better, while the product
$(n_i/n)\times V_{\rm rms}$ gives $0.26$, $0.22$ and $0.18$ and is marginally the
best single variable, Fig.~\ref{lll:fig:setback}(c). The honest statement is that
$n_i/n$ controls the damage at fixed potential shape, and that both the count per
electron and the amplitude are needed when the shape is changed.

\subsubsection{The disorder landscape and the puddles}

Because the impurities are charged and set back from the plane, the potential
they produce is smooth: its correlation half-width is $7.2\,\ell$ at $d=6\ell$
and $6.1\,\ell$ at $d=3\ell$, so it is a landscape of hills and basins rather
than a set of point scatterers. Filling that landscape semiclassically to the
level $\mu_\nu$ that covers an area fraction $\nu$ gives the familiar picture of
lakes that shrink and separate into puddles as the filling drops. The electrons
of the full Hartree--Fock solution do move into the basins as $\nu$ falls --- the
mean landscape height at the electron positions goes from $-0.08\,V_{\rm rms}$ at
$\nu=1/2$ to $-0.51\,V_{\rm rms}$ at $\nu=0.10$ --- but they do so weakly, and at
no filling does the density look like a set of filled puddles.

A diagnostic numerical experiment is instructive here.  Removing
$U_H(G)$ from the Fock kernel while retaining exchange is a change of
Hamiltonian: it favors macroscopic charge accumulation by eliminating
the long-wavelength electrostatic cost.  Re-solving the modified theory
self-consistently gives a filled-droplet state: the electrons collapse
into the lake with $2\pi\ell^2n\simeq1$ inside and nearly zero outside,
and an energy per electron that is almost independent of filling.  This
illustrates the role of electrostatic stiffness in the full theory,
but it is not a proof that physical puddling requires removing the
Hartree term; a model with inhomogeneous occupied and depleted regions
can retain Hartree repulsion (as Sec.~\ref{sec:tf_irm} does with
nonlinear Thomas--Fermi), and the VMC donor-layer calculation of
Sec.~\ref{sec:vmc} likewise produces puddles while retaining full
electron--electron repulsion.  Within the LLL--HF search explored here,
however, the full solution does not approach the filled-droplet limit
at any filling studied; the amorphous solid it forms instead is the
compromise between following the landscape and paying the electrostatic
cost of long-wavelength density modulation.

\subsection{What the LLL calculation implies for the main picture}
\label{sec:lll_summary}

The LLL--HF crystal of the preceding subsections is not indifferent to
the disorder landscape because it fails to see it.  It sees it,
screens and in fact overscreens it, and at low filling it settles
measurably into its basins.  What it does not do is collapse into
them, and the reason is that collapsing costs Hartree energy.  The
Hartree kernel $U_H(q)\sim1/q$ is largest at exactly the long
wavelengths on which a Coulomb landscape varies, which is the same
statement as saying that a Coulomb landscape is slowly varying: the
slow variation that makes the puddle picture natural is also what
makes it expensive.  The amorphous solid found at low filling and at
strong disorder is therefore the compromise between following the
landscape and remaining uniform; the filled-droplet state is the
endpoint the system would reach if that compromise were not enforced.

This locates what would most change the conclusions: anything that
reduces the long-wavelength Coulomb cost \emph{relative to the
disorder}, such as a nearby metallic gate that cuts $U_H(q)$ off
below $q\sim1/d_{\mathrm{gate}}$ while leaving the impurity potential
of Eq.~(\ref{lll:eq:Vq}) comparatively intact, or correlations the
present mean-field treatment omits.  A uniform dielectric constant is
not such a knob: $\epsilon$ divides $U_H$, $X$ and $V$ alike, so in
the units used here it does not enter the problem at all and cannot
move the minimizing state at fixed geometry and filling.  Selective
screening, not an overall weakening of the Coulomb interaction, is
what moves the system toward the puddle limit.  Conversely, the
observation that the full HF solution stays so far from that limit
at every filling studied suggests that the amorphous solid, not a
set of isolated puddles, is the correct structural starting picture
for the insulating phase at low filling in a sample whose dominant
disorder is Coulombic.

Appendix~\ref{sec:SM_lll} gives the full computational details of
the landscape construction, the lake and droplet analyses, the
larger-sample check, and the setback scan.

%% file: meta_sec5_tf_irm.tex

\section{The 2{D} Metal--Insulator Transition: {T}homas--{F}ermi Crossover, Classical Percolation, and the {I}offe--{R}egel--{M}ott Density}
\label{sec:tf_irm}

The preceding sections establish, within three complementary
microscopic methods, that charged disorder converts the 2D Wigner
crystal into a short-range coherent amorphous solid.  The specific
energetic competition that controls each of them differs by regime:
in the zero-field HF of Sec.~\ref{sec:hf_letter} the dominant
competition is between the exchange-enhanced liquid response and
the Coulomb pinning potential; in the VMC of Sec.~\ref{sec:vmc} it is
between the impurity binding energy and the inter-electron Coulomb
energy, with the Fermi energy entering through $\rs$; in the LLL HF
of Sec.~\ref{sec:lll} the kinetic energy is quenched entirely and
the competition is between the Coulomb crystallization energy and
the impurity pinning potential at fixed form factor.  The present
section turns to a different observable --- the critical carrier
density $\nc$ at which 2D electron systems in Si MOSFETs and
GaAs/AlGaAs heterostructures undergo the metal--insulator transition
observed experimentally for more than three
decades~\cite{Kravchenko1994} --- in which the natural energetic
condition is instead $E_{\mathrm{dis}}\sim E_F$, the carrier density
below which the disorder potential becomes comparable to the Fermi
energy.  Within a controlled model of long-range Coulomb disorder
at setback we compare two diagnoses of that transition, the
half-depletion density of the semiclassical nonlinear-Thomas--Fermi
landscape and the quantum Ioffe--Regel--Mott (IRM) condition
$\kF \ell_{\mathrm{MFP}} \sim 1$, and we show that the two scales
track each other to within a slowly varying factor of order unity
over the physically relevant parameter range, with the agreement
originating in their shared Fermi-energy condition rather than in a
theorem.

\input{body_tf}

\bigskip
\noindent
\emph{What this section has established.}  Within the specific
nonlinear-Thomas--Fermi model of long-range Coulomb disorder studied
above, the half-depletion density of the semiclassical landscape and
the Ioffe--Regel--Mott density of Born--RPA transport are reached at
the same energetic condition $E_{\mathrm{dis}}\sim E_F$ and track
each other to within a factor of order unity, with a slow
$\bar{n}_i^{-1/10}d^{-1/5}$ spread.  Modest independent short-range
transport channels can shift the ratio by factors of order two.  The
near-coincidence is not a theorem; it reflects the common energetic
scale that also controls the structural crossover of
Secs.~\ref{sec:hf_letter}--\ref{sec:lll}.  Section~\ref{sec:pan_qhf}
now returns to the structural coherence question with a complementary
microscopic approach that this paper's own tools do not provide.

%% file: body_tf.tex
\subsection{Statement of the two central reference equations}

\begin{figure*}[!htbp]
\centering
\includegraphics[width=0.93\textwidth]{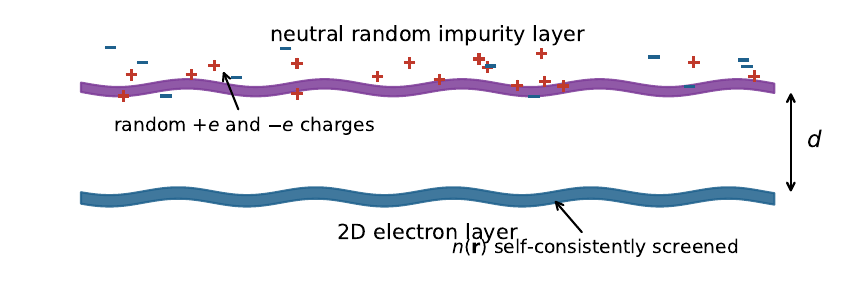}
\caption{Model geometry. A zero-thickness 2DEG lies at $z=0$. A
separate layer at $z=d$ contains random positive and negative unit
charges. Exact charge neutrality removes only the $q=0$ component;
the finite-wave-vector charge correlator is set by the total
density $\bar{n}_i$. The direct point-charge ensemble is used for the
illustrative puddle maps; the controlled threshold calculation uses
its continuum Gaussian limit.}
\label{tf:fig:schematic}
\end{figure*}

The classical reference point for the long-range-disorder regime is the
Efros--Pikus--Burnett (EPB)
construction~\cite{Efros1989,EPB1993}, from which
\begin{equation}
 n_c^{\rm EPB}=P\,\frac{\sqrt{\bar{n}_i}}{d},\qquad P=0.11,
\label{tf:eq:epb}
\end{equation}
where $\bar{n}_i$ is the total areal density of uncorrelated impurity
charges, $d$ is the spacer thickness, and the dimensionless
coefficient $P$ applies in the limit where the random charge in an
area $d^2$ is approximately Gaussian and the electron gas is a
nearly perfect metal on that scale.  Fogler developed a quantitative
nonlinear-screening treatment in the same
regime~\cite{Fogler2004}.  The local Thomas--Fermi (TF) crossover to
Eq.~(\ref{tf:eq:epb}) is derived below in the form
\begin{equation}
 n_{1/2}^{\rm LTF}=\frac{\sqrt{\bar{n}_i}}{d}F_{1/2}(q_{\rm TF}d),
\label{tf:eq:scalingresult}
\end{equation}
for continuum Gaussian remote disorder, where $n_{1/2}^{\rm LTF}$
is the density at which the depleted area is one half of the sample
and $F_{1/2}(x)$ is a controlled crossover function that recovers
the EPB coefficient in the $x\gg1$ limit.  This local-TF
construction is to be compared on equal footing with a fully
specified zero-temperature RPA--Boltzmann IRM calculation, from
which $n_{\rm IRM}$ is defined by $\rho(n_{\rm IRM})=h/e^2$,
equivalently $k_F\ell_{\rm MFP}=1$~\cite{IoffeRegel1960,Mott1972}.
The ratio $n_{\rm IRM}/n_{1/2}^{\rm LTF}$ is not a constant: it
crosses unity as $\bar{n}_i$ and $d$ are varied, and the comparison below
also introduces a controlled short-range transport channel as an
explicit separation test.  The model geometry is shown in
Fig.~\ref{tf:fig:schematic}.

\subsection{Model and density variables}
\label{tf:sec:model}

\subsubsection{Units and disorder correlator}

We use the effective Bohr radius and Coulomb energy
\begin{equation}
 a_{\rm B}=\frac{\kappa\hbar^2}{m e^2},\qquad
 E_{\rm B}=\frac{e^2}{\kappa a_{\rm B}}
 =\frac{\hbar^2}{m a_{\rm B}^2},
\label{tf:eq:units}
\end{equation}
as units of length and energy. The electron gas has spin degeneracy two and
no valley degeneracy, so
\begin{equation}
 q_{\rm TF}=\frac{2}{a_{\rm B}},\qquad
 n=\frac{k_F^2}{2\pi},\qquad
 r_s=\frac{1}{\sqrt{\pi n a_{\rm B}^2}}.
\label{tf:eq:variables}
\end{equation}

For a direct finite realization, the signed impurity density is
\begin{equation}
 \rho_i({\bf r})=\sum_{j=1}^{N_i/2}\delta({\bf r}-{\bf R}_j^+)
 -\sum_{j=1}^{N_i/2}\delta({\bf r}-{\bf R}_j^-).
\label{tf:eq:directrho}
\end{equation}
The positions are independent and uniform, and \(N_i/2\) charges have each
sign. Therefore \(\rho_i({\bf q}=0)=0\) identically, while for every nonzero
wave vector
\begin{equation}
 \frac{\overline{\rho_i({\bf q})\rho_i(-{\bf q})}}{\mathcal A}=\bar{n}_i,
\qquad q>0.
\label{tf:eq:correlator}
\end{equation}
The notation \(\bar{n}_i\) always denotes the total density of positive and
negative impurities, not the density of one sign.

The bare remote-impurity potential energy of an electron is
\begin{equation}
 V_{\rm imp}({\bf q})=-\frac{2\pi e^2}{\kappa q}
 e^{-q d}\rho_i({\bf q}),\qquad q>0.
\label{tf:eq:vimp}
\end{equation}
The sign convention has no statistical consequence because the distribution
of \(\rho_i\) is symmetric. Exact global neutrality likewise changes only
the removed uniform mode, not Eq.~\eqref{tf:eq:correlator} or the transport
correlator.

\subsubsection{Nonlinear Hartree Thomas-Fermi equation}

At fixed average density \(n\), the zero-temperature Hartree-TF functional is
\begin{align}
 {\cal E}[n({\bf r})]={}&\int d^2r\left[
 \frac{\pi\hbar^2}{2m}n^2({\bf r})+
 V_{\rm imp}({\bf r})n({\bf r})\right]\nonumber\\
 &+\frac{1}{2}\int d^2r\,d^2r'
 [n({\bf r})-n]v({\bf r}-{\bf r}')
 [n({\bf r}')-n],
\label{tf:eq:functional}
\end{align}
where \(v(q)=2\pi e^2/(\kappa q)\). The density is constrained to be
nonnegative. In the units of Eq.~\eqref{tf:eq:units}, stationarity yields
\begin{align}
 n({\bf r})&=\frac{1}{\pi}[\mu-U({\bf r})]_+,\label{tf:eq:tf}\\
 U({\bf r})&=V_{\rm imp}({\bf r})+V_H({\bf r}),\label{tf:eq:utotal}\\
 V_H({\bf q})&=\frac{2\pi}{q}n({\bf q}),\qquad q>0,
\label{tf:eq:hartree}
\end{align}
where \([x]_+=\max(x,0)\). Equation~\eqref{tf:eq:tf} is an obstacle problem:
the depletion boundary moves with the chemical potential and is not present
in a linear-screening calculation. Exchange, correlation, finite
quantum-well width, tunneling through depleted regions, and coherent
localization are not included.

\begin{figure*}[!htbp]
\centering
\includegraphics[width=0.98\textwidth]{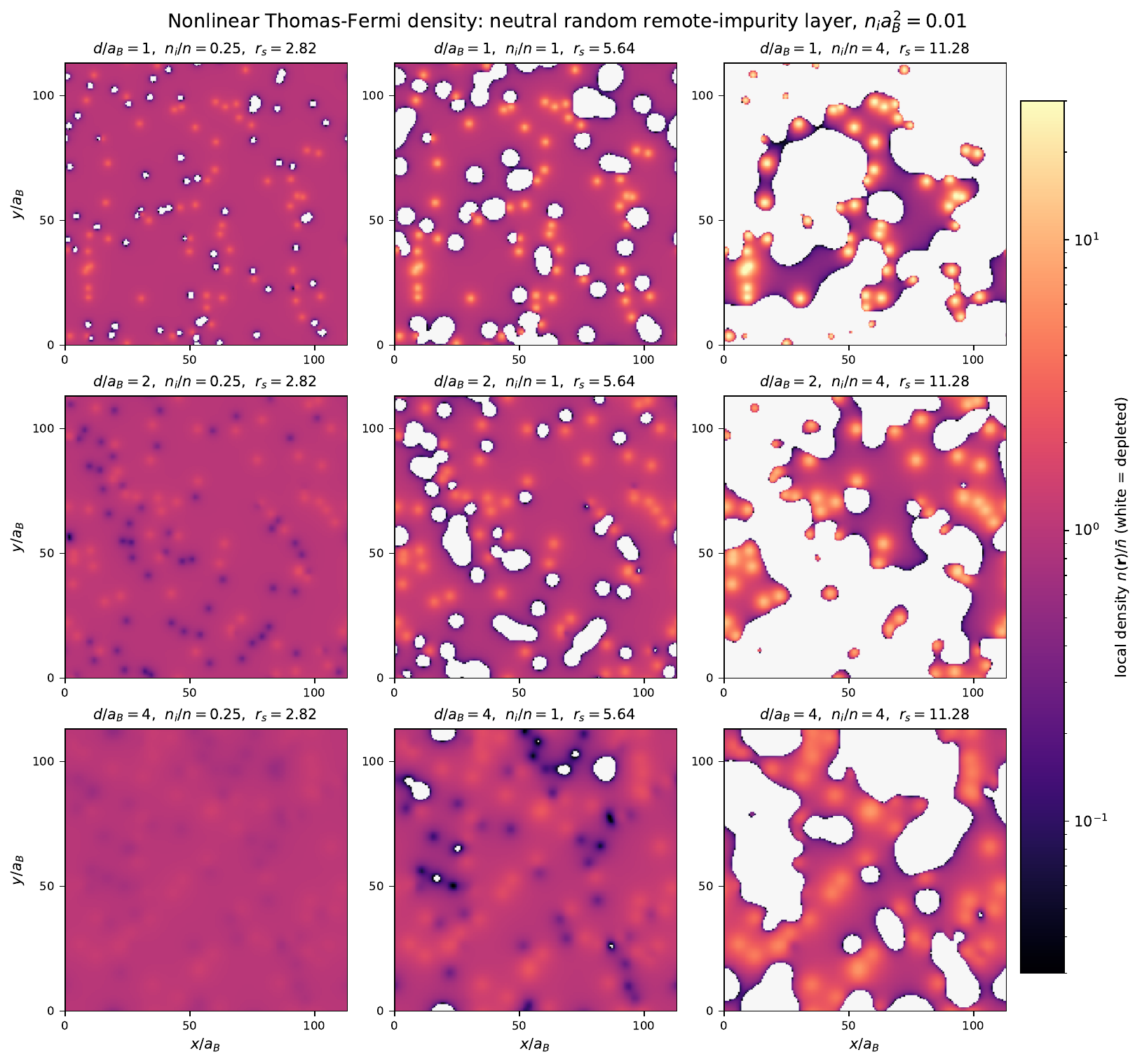}
\caption{Illustrative nonlinear TF density maps for a direct neutral
point-charge realization with \(\bar{n}_i a_{\rm B}^2=0.01\). Rows are
\(d/a_{\rm B}=1,2,4\); columns are \(\bar{n}_i/n=0.25,1,4\). White regions are
depleted. These maps demonstrate the puddle morphology of
Eqs.~\eqref{tf:eq:tf}-\eqref{tf:eq:hartree}; they are not used as a thermodynamic
percolation data set because their discrete-impurity parameter
\(\bar{n}_i d^2\) is not in the continuum-Gaussian regime.}
\label{tf:fig:puddles}
\end{figure*}
\clearpage

\subsection{Continuum-Gaussian limit and the classical benchmark}
\label{tf:sec:continuum}

\subsubsection{Why the Efros scaling has a restricted domain}

When
\begin{equation}
 \chi\equiv \bar{n}_i d^2\gg1,
\label{tf:eq:chi}
\end{equation}
many independent charges lie in an area of the disorder correlation length.
The central-limit theorem then turns the finite-\(q\) source in
Eq.~\eqref{tf:eq:correlator} into a Gaussian white-noise field. The remote
factor \(e^{-qd}\) smooths that field. Perfect screening gives
\begin{equation}
 \delta n({\bf q})=e^{-qd}\rho_i({\bf q}),\qquad q>0,
\label{tf:eq:perfectscreening}
\end{equation}
and hence
\begin{equation}
 \overline{\delta n^2}=
 \bar{n}_i\int\frac{d^2q}{(2\pi)^2}e^{-2qd}
 =\frac{\bar{n}_i}{8\pi d^2}.
\label{tf:eq:classicalvariance}
\end{equation}
The natural density scale is therefore \(\sqrt{\bar{n}_i}/d\). The symmetric
Gaussian landscape percolates at its median level, which is equivalent to a
depleted area of one half. EPB obtained the numerical coefficient
\(P=0.11\) in Eq.~\eqref{tf:eq:epb}; their simulation found \(0.11\), while the
earlier Efros estimate was \(0.10\) \cite{EPB1993}.

The finite-\(d/a_{\rm B}\) local-TF problem has one further dimensionless
parameter, \(x=q_{\rm TF}d=2d/a_{\rm B}\). Rescaling lengths by \(d\),
densities by \(\sqrt{\bar{n}_i}/d\), and potentials by \(\sqrt{\bar{n}_i}\) gives the
scaling form in Eq.~\eqref{tf:eq:scalingresult}. This collapse is exact for the
continuum Gaussian source within the local Hartree-TF model. A physical
point-charge layer has the more general dependence
\begin{equation}
 \frac{n_c d}{\sqrt{\bar{n}_i}}=
 {\cal F}\left(q_{\rm TF}d,\ \bar{n}_i d^2,\ \frac{L}{d},\
 \frac{\ell_{\rm uv}}{d},\ {\cal C}_{\rm perc}\right),
\label{tf:eq:generalscaling}
\end{equation}
where \(L\) is the simulation size, \(\ell_{\rm uv}\) is the ultraviolet
regularization, and \({\cal C}_{\rm perc}\) specifies the percolation
criterion. Equation~\eqref{tf:eq:scalingresult} is the
\(\chi,L/d\rightarrow\infty\) continuum limit of
Eq.~\eqref{tf:eq:generalscaling}, not a replacement for it.

\subsubsection{Finite-compressibility crossover and its small-\(d\) limitation}

Above depletion, the linearized version of Eq.~\eqref{tf:eq:tf} is
\begin{equation}
 \delta n({\bf q})=
 \frac{q_{\rm TF}}{q+q_{\rm TF}}e^{-qd}\rho_i({\bf q}).
\label{tf:eq:linearresponse}
\end{equation}
It gives a useful fluctuation diagnostic,
\begin{equation}
 \overline{\delta n^2}=
 \frac{\bar{n}_i}{2\pi d^2} I(x),\qquad
 I(x)=\int_0^\infty du\,u e^{-2u}
 \left(\frac{x}{u+x}\right)^2.
\label{tf:eq:variancecrossover}
\end{equation}
The integral tends to \(1/4\) for \(x\gg1\), reproducing
Eq.~\eqref{tf:eq:classicalvariance}. For \(x\ll1\),
\begin{equation}
 \overline{\delta n^2}\sim
 \frac{\bar{n}_i q_{\rm TF}^2}{2\pi}
 \ln\left(\frac{1}{q_{\rm TF}d}\right).
\label{tf:eq:smallxvariance}
\end{equation}
Equation~\eqref{tf:eq:smallxvariance} is not a percolation-threshold formula. It
demonstrates that the nominal zero-thickness, point-impurity \(d\to0\) limit
is ultraviolet sensitive even before the local-TF approximation itself is
questioned. Consequently, the finite-\(x\) curve calculated below is
presented as a local-TF crossover function, not as a universal microscopic
small-\(d\) critical density.

\subsubsection{Numerical protocol}

We discretize a Gaussian source with
\(\overline{\rho_i({\bf r})\rho_i({\bf 0})}=\bar{n}_i\delta({\bf r})\), remove its
zero mode, and use Eq.~\eqref{tf:eq:vimp}. The production calculations have
\(L/d=32\), \(\Delta x/d=1/4\), and therefore a \(128\times128\) mesh.
Twelve independent fields are used at each
\(x=0.5,1,2,4,8,16,32\). At every trial density the convex quadratic
functional, with the positivity constraint retained, is solved by a primal
active-set method. The largest direct local-TF fixed-point residual is
\(1.3\times10^{-7}\).

For each field we locate \(n_{1/2}\) by logarithmic bisection of the occupied
area through one half. We also record periodic wrapping in either direction
and in both directions at that same density. The latter diagnostic is
important: using first wrapping in either direction as the threshold would
systematically select a lower, geometry-dependent number. A separate
classical calculation sets the local kinetic coefficient \(\pi\) in
Eq.~\eqref{tf:eq:functional} to zero, while retaining the same long-range
Coulomb obstacle problem.

\begin{figure*}[!htbp]
\centering
\includegraphics[width=0.98\textwidth]{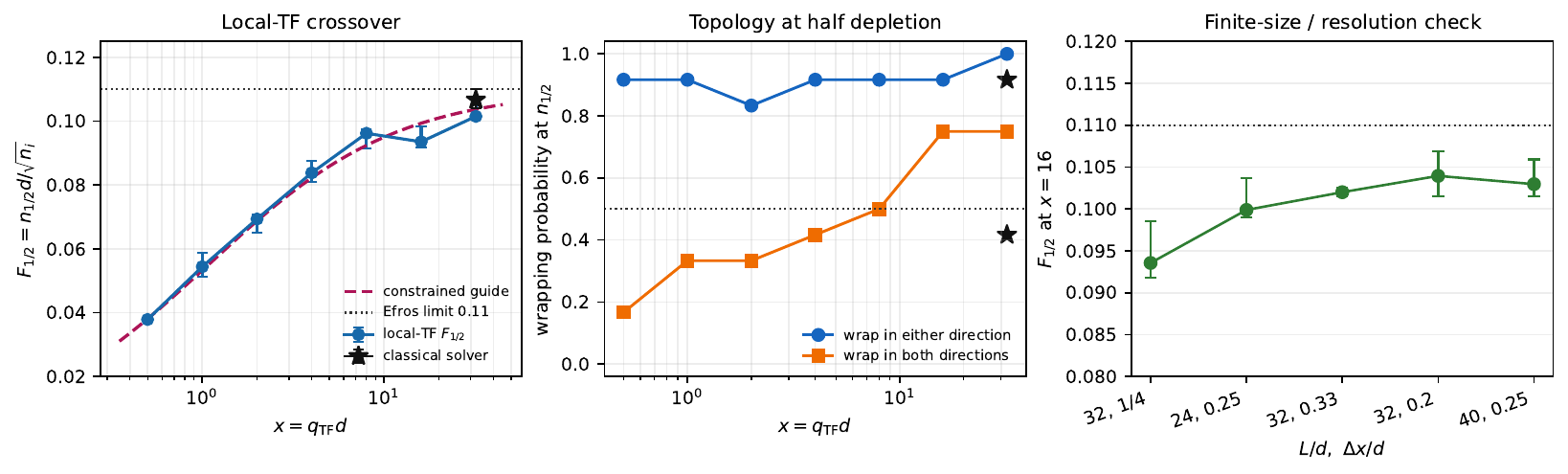}
\caption{Controlled local-TF crossover. Left: the numerical
\(F_{1/2}=n_{1/2}d/\sqrt{\bar{n}_i}\) versus \(x=q_{\rm TF}d\), with central
68\% bootstrap intervals for twelve disorder fields. The star is an
independent classical obstacle-problem calculation at \(x=32\):
\(F_{1/2}=0.1067[0.1040,0.1101]\). It agrees with the EPB limit
\(0.11\). The dashed line is a constrained two-parameter guide to the
local-TF data only. Middle: finite-cell winding diagnostics at the
half-depletion density. The two-direction probability for the classical
point is \(5/12\), statistically consistent with one half. Right:
independent changes of \(L/d\) and \(\Delta x/d\) at \(x=16\) give
\(F_{1/2}=0.094\)--\(0.104\), quantifying the remaining finite-ensemble and
discretization uncertainty.}
\label{tf:fig:crossover}
\end{figure*}
\clearpage

\subsection{Local-TF crossover results}
\label{tf:sec:tfresults}

Table~\ref{tf:tab:crossover} gives the numerical crossover function. It
increases from \(0.0379\) at \(x=0.5\) to a value near \(0.10\) by
\(x=32\). The independent classical result, rather than an extrapolation of
the local-TF points, is the benchmark for the limiting coefficient. The
finite-size and resolution variations displayed in Fig.~\ref{tf:fig:crossover}
show that the residual systematic uncertainty at \(x=16\) is of order ten
percent of the coefficient, comparable to the field-to-field spread.

\begin{table}[!htbp]
\caption{Continuum-Gaussian local-TF crossover function. Brackets are
central 68\% bootstrap intervals for the median over twelve fields. The
wrapping probabilities are diagnostic values at the half-depletion density,
not independent threshold definitions.}
\label{tf:tab:crossover}
\begin{ruledtabular}
\begin{tabular}{cccc}
\(x=q_{\rm TF}d\) & \(F_{1/2}\) & 68\% interval & \(P_{\rm both}\)\\ \hline
0.5 & 0.0379 & [0.0372, 0.0390] & 0.17\\
1 & 0.0544 & [0.0512, 0.0589] & 0.33\\
2 & 0.0694 & [0.0650, 0.0707] & 0.33\\
4 & 0.0838 & [0.0809, 0.0876] & 0.42\\
8 & 0.0962 & [0.0914, 0.0976] & 0.50\\
16 & 0.0936 & [0.0918, 0.0985] & 0.75\\
32 & 0.1015 & [0.1008, 0.1057] & 0.75
\end{tabular}
\end{ruledtabular}
\end{table}

For compact interpolation of the numerical results, the curve in
Fig.~\ref{tf:fig:crossover} uses
\begin{equation}
 F_{1/2}^{\rm fit}(x)=
 \frac{0.11}{1+(1.079/x)^{0.829}}.
\label{tf:eq:fit}
\end{equation}
The EPB coefficient \(0.11\) is fixed, not fitted. The two remaining
parameters are a least-squares representation of the seven numerical medians.
Equation~\eqref{tf:eq:fit} is useful for plotting the comparison with transport;
it is not an analytic prediction and should not be used to infer a
short-distance impurity cutoff.

The result resolves the Efros comparison directly. In the regime in which
Eq.~\eqref{tf:eq:epb} is supposed to apply, our calculation agrees with it. The
earlier apparent result
\(n_c\propto \bar{n}_i^{0.51}d^{-0.53}\) came from a finite direct-point-charge
\emph{first wrap in either direction} estimator. It is not a competing
asymptotic law. Its \(d^{-0.53}\) behavior and a coefficient smaller than
\(0.11\) are artifacts of mixing \(\bar{n}_i d^2\lesssim1\), variable \(L/d\), a
coarse cutoff at small \(d\), and a different topology criterion. The
self-consistent density fields from that calculation remain valid
illustrations of puddling, as in Fig.~\ref{tf:fig:puddles}; its reported
thresholds are removed from the present physical conclusions.

\subsection{RPA-Boltzmann IRM density}
\label{tf:sec:irm}

The transport calculation treats the same total impurity density \(\bar{n}_i\) and
setback \(d\), but assumes a homogeneous 2DEG. For elastic scattering on the
Fermi circle, \(q\le2k_F\). In this entire interval the zero-temperature
noninteracting 2D polarizability is constant \cite{Stern1967,Ando1982},
\begin{equation}
 \Pi_0(q,0)=\frac{m}{\pi\hbar^2},\qquad
 \epsilon(q)=1+\frac{q_{\rm TF}}{q}.
\label{tf:eq:screening}
\end{equation}
Thus the screened potential of a single remote unit charge is
\begin{equation}
 u(q)=\frac{2\pi e^2}{\kappa}
\frac{e^{-qd}}{q+q_{\rm TF}}.
\label{tf:eq:screenedpotential}
\end{equation}
The sign of the charge disappears after disorder averaging:
\begin{equation}
 \frac{\overline{|V_{\rm dis}({\bf q})|^2}}{\mathcal A}
 =\bar{n}_i|u(q)|^2,\qquad q>0.
\label{tf:eq:transportcorrelator}
\end{equation}

The Born transport time contains the transport factor rather than the
single-particle scattering rate,
\begin{equation}
 \frac{1}{\tau_t}=\frac{2\pi \bar{n}_i}{\hbar}
 \int\frac{d^2k'}{(2\pi)^2}|u(q)|^2
 (1-\cos\theta)\delta(\varepsilon_{k'}-\varepsilon_F).
\label{tf:eq:tau}
\end{equation}
Using \(q=2k_F\sin(\theta/2)\), setting \(t=\theta/2\), and measuring
\(k_F\) and \(d\) in \(a_{\rm B}^{-1}\) and \(a_{\rm B}\), respectively,
gives
\begin{equation}
 \frac{\rho}{h/e^2}=
 \frac{2\pi \bar{n}_i}{k_F^2}\int_0^\pi dt\,
 \frac{\sin^2t\,e^{-4k_Fd\sin t}}
 {[1+k_F\sin t]^2}.
\label{tf:eq:rho}
\end{equation}
The denominator is the exact static 2D RPA screening factor on the elastic
Fermi surface. Equation~\eqref{tf:eq:rho} is evaluated by Gauss-Legendre
quadrature and \(n_{\rm IRM}\) is the positive root of
\begin{equation}
 \rho(n_{\rm IRM};\bar{n}_i,d)=\frac{h}{e^2},
\qquad
 k_F(n_{\rm IRM})\ell(n_{\rm IRM})=1.
\label{tf:eq:irm}
\end{equation}
This is a crossover criterion, not a controlled Boltzmann result on the
localized side of \(k_F\ell=1\) \cite{Gold1988,DasSarma2014}.

\subsection{Comparison between the local-TF and IRM scales}
\label{tf:sec:comparison}

\begin{figure*}[!htbp]
\centering
\includegraphics[width=0.98\textwidth]{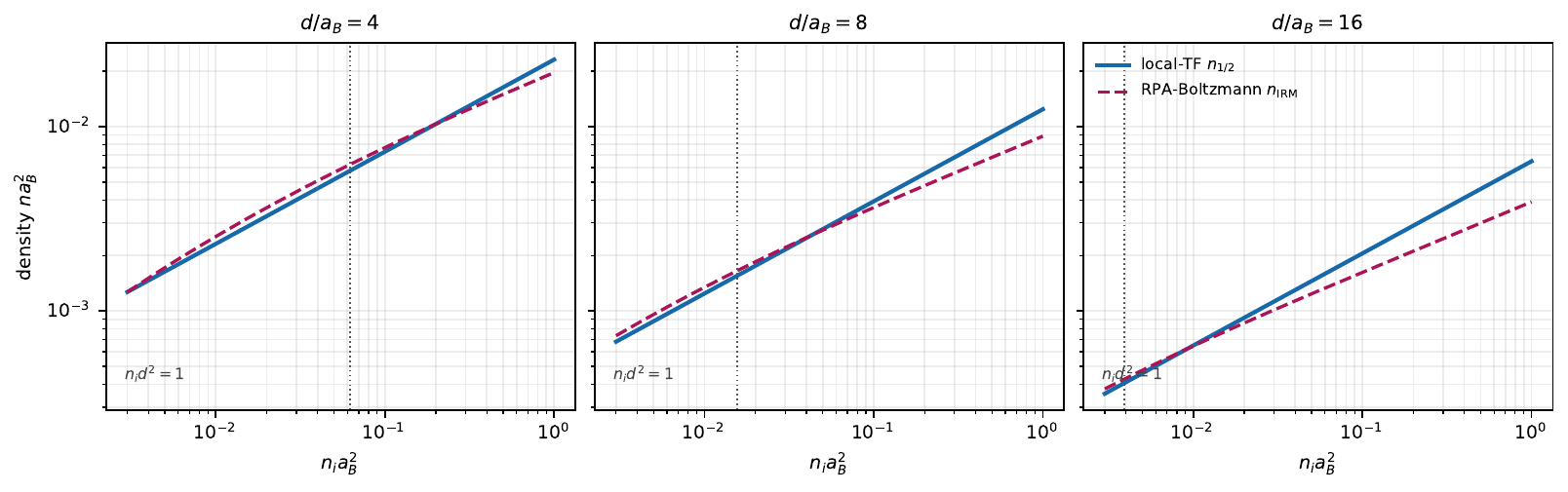}
\caption{Local-TF half-depletion scale and IRM density versus impurity
density at fixed setback. The vertical dotted lines mark \(\bar{n}_i d^2=1\).
To their right the continuum-Gaussian condition begins to hold; a stronger
\(\chi\gg1\) inequality is required for a controlled physical comparison.
The two curves cross, so neither construction is uniformly larger.}
\label{tf:fig:vsni}
\end{figure*}

\begin{figure*}[!htbp]
\centering
\includegraphics[width=0.98\textwidth]{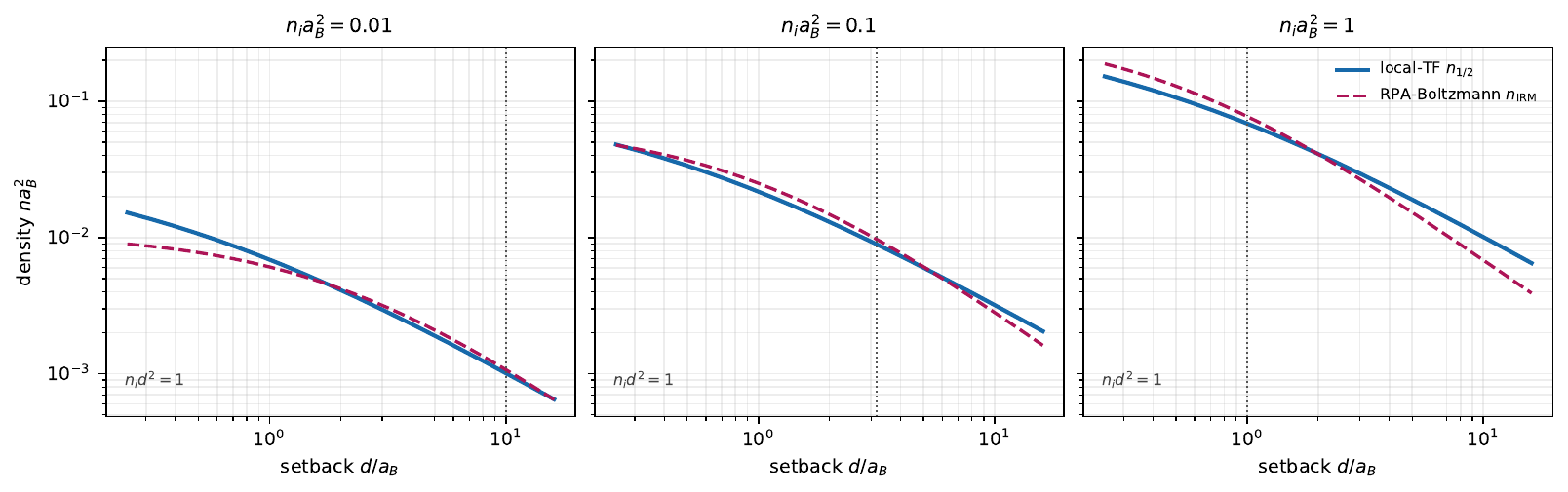}
\caption{Local-TF half-depletion scale and IRM density versus setback at
fixed impurity density. At large \(d\), the remote-scattering IRM scale falls
faster than the classical percolation scale. The dotted line again marks
\(\bar{n}_i d^2=1\), not a sharp physical boundary.}
\label{tf:fig:vsd}
\end{figure*}

\begin{figure*}[!htbp]
\centering
\includegraphics[width=0.98\textwidth]{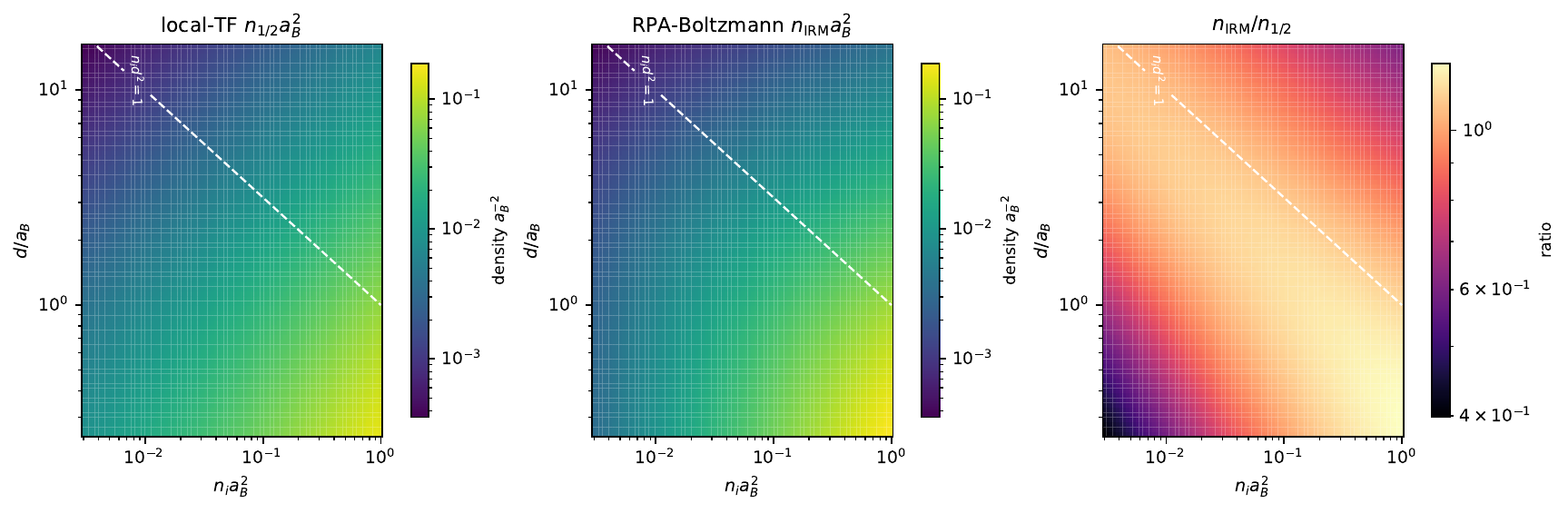}
\caption{False-color comparison in the \((\bar{n}_i,d)\) plane. Left:
\(n_{1/2}^{\rm LTF}\) from the numerical local-TF crossover interpolation;
middle: \(n_{\rm IRM}\) from the RPA-Boltzmann integral; right: their ratio.
The dashed contour is \(\bar{n}_i d^2=1\). The ratio crosses unity, demonstrating
that the two density scales are parametrically distinct.}
\label{tf:fig:falsecolor}
\end{figure*}

Figures~\ref{tf:fig:vsni}, \ref{tf:fig:vsd}, and \ref{tf:fig:falsecolor} compare
\(n_{1/2}^{\rm LTF}\), generated by Eq.~\eqref{tf:eq:fit}, with the direct
quadrature of Eq.~\eqref{tf:eq:rho}. No relative rescaling or fit connects the
two curves. The white contour in Fig.~\ref{tf:fig:falsecolor} is
\(\bar{n}_i d^2=1\); the continuum-Gaussian percolation proxy is quantitatively
controlled only well above that contour. The transport calculation itself is
defined for all plotted parameters within its Born-RPA assumptions.

\clearpage

Representative numerical values are listed in Table~\ref{tf:tab:comparison}.
For \(d/a_{\rm B}=4\) and \(\bar{n}_i a_{\rm B}^2=0.1\), the two scales nearly
coincide. At larger \(d\) or larger \(\bar{n}_i d^2\), the IRM root is lower than
the local-TF percolation proxy. This reversal is physical within the two
approximations, not a numerical inconsistency.

\begin{table*}[!htbp]
\caption{Representative comparison of the local-TF half-depletion and
RPA-Boltzmann IRM density scales. The local-TF values use
Eq.~\eqref{tf:eq:fit}; the IRM values are direct roots of
Eq.~\eqref{tf:eq:rho}. The continuum-Gaussian comparison is best controlled
for \(\bar{n}_i d^2\gg1\).}
\label{tf:tab:comparison}
\begin{ruledtabular}
\begin{tabular}{cccccc}
\(\bar{n}_i a_{\rm B}^2\) & \(d/a_{\rm B}\) & \(\bar{n}_i d^2\) &
\(n_{1/2}^{\rm LTF}a_{\rm B}^2\) &
\(n_{\rm IRM}a_{\rm B}^2\) & \(n_{\rm IRM}/n_{1/2}^{\rm LTF}\)\\ \hline
0.1 & 4 & 1.6 & \(7.31\times10^{-3}\) & \(7.70\times10^{-3}\) & 1.05\\
1.0 & 4 & 16 & \(2.31\times10^{-2}\) & \(1.96\times10^{-2}\) & 0.85\\
0.1 & 8 & 6.4 & \(3.93\times10^{-3}\) & \(3.64\times10^{-3}\) & 0.93\\
1.0 & 8 & 64 & \(1.24\times10^{-2}\) & \(8.87\times10^{-3}\) & 0.71\\
0.1 & 16 & 25.6 & \(2.05\times10^{-3}\) & \(1.62\times10^{-3}\) & 0.79\\
1.0 & 16 & 256 & \(6.48\times10^{-3}\) & \(3.90\times10^{-3}\) & 0.60
\end{tabular}
\end{ruledtabular}
\end{table*}

The large-setback trend is transparent analytically. When \(k_Fd\gg1\),
the integral in Eq.~\eqref{tf:eq:rho} is dominated by small angles. To leading
order,
\begin{equation}
 \frac{\rho}{h/e^2}\simeq
 \frac{\pi \bar{n}_i}{8 k_F^5d^3}.
\label{tf:eq:remoteasymptote}
\end{equation}
The IRM condition then gives
\begin{equation}
 n_{\rm IRM}\propto \bar{n}_i^{2/5}d^{-6/5},
\label{tf:eq:irmasymptote}
\end{equation}
whereas the classical percolation scale is
\(n_c\propto \bar{n}_i^{1/2}d^{-1}\). Hence their ratio carries a nontrivial
power-law dependence, \(n_{\rm IRM}/n_c\propto
\bar{n}_i^{-1/10}d^{-1/5}\), in this asymptotic regime. The crossing in the
figures is therefore expected; it is not sensible to impose a single
ordering of the two criteria over the entire parameter plane.

\subsection{Origin and limits of the near coincidence}
\label{tf:sec:whyclose}

The numerical order-unity proximity of \(n_{\rm IRM}\) and
\(n_{1/2}^{\rm LTF}\) is not a theorem and is not evidence that classical
percolation and quantum localization are the same transition. There is,
however, a clear physical reason why the two crossover densities can be
similar in a remote-Coulomb-disorder system. First, the present
\(n_{\rm IRM}\) calculation is \emph{not} a calculation of destructive
quantum interference. It is a homogeneous Born-RPA calculation stopped at
the phenomenological Ioffe-Regel condition. No maximally crossed diagrams,
localization scaling, or tunneling through depleted saddle points are
included. The classical and IRM constructions are therefore two distinct
semiclassical diagnoses of the breakdown of the same homogeneous metallic
description.

Both diagnostics are generated by the same screened disorder spectrum,
\begin{equation}
 {\cal S}_U(q)\equiv
 \frac{\overline{|V_{\rm dis}({\bf q})|^2}}{\mathcal A}
 =\bar{n}_i\left(\frac{2\pi e^2}{\kappa}
 \frac{e^{-qd}}{q+q_{\rm TF}}\right)^2.
\label{tf:eq:common_spectrum}
\end{equation}
The electrostatic criterion becomes nonlinear when the density fluctuation
in Eq.~\eqref{tf:eq:classicalvariance} is comparable to the mean density,
\begin{equation}
 \delta n_{\rm rms}\sim n,
 \qquad\Longleftrightarrow\qquad V_{\rm rms}\sim E_F.
\label{tf:eq:amplitudecriterion}
\end{equation}
The IRM condition expresses an analogous energetic statement. Since
\(\ell=v_F\tau_t\),
\begin{equation}
 k_F\ell=\frac{2E_F\tau_t}{\hbar}=1
 \qquad\Longleftrightarrow\qquad
 \frac{\hbar}{\tau_t}=2E_F.
\label{tf:eq:irmenergy}
\end{equation}
Thus the first construction tests whether the disorder \emph{amplitude}
breaks the fluid into depleted regions, while the second tests whether the
transport-weighted disorder rate has become of order \(E_F/\hbar\). For a
single smooth disorder scale \(d\), both are controlled by modes
\(q\sim d^{-1}\). Their common origin makes an order-unity relationship
natural, while their different spectral weights prevent an identity.

This point is particularly transparent in the asymptotic remote limit. Using
Eq.~\eqref{tf:eq:remoteasymptote}, \(n=k_F^2/(2\pi)\), and the EPB coefficient
\(P=0.11\), one obtains
\begin{align}
 n_{\rm IRM}a_{\rm B}^2&\simeq
 \frac{1}{2\pi}\left[\frac{\pi(\bar{n}_i a_{\rm B}^2)}
 {8(d/a_{\rm B})^3}\right]^{2/5},\label{tf:eq:irmcoefficient}\\
 \frac{n_{\rm IRM}}{n_c^{\rm EPB}}&\simeq
 \frac{(\pi/8)^{2/5}}{2\pi P}\,(\bar{n}_i d^2)^{-1/10}
 \simeq0.996\,(\bar{n}_i d^2)^{-1/10}.
\label{tf:eq:weakratio}
\end{align}
Equation~\eqref{tf:eq:weakratio} assumes \(q_{\rm TF}d\gg1\),
\(k_Fd\gg1\), and \(\bar{n}_i d^2\gg1\). It is not quantitatively imposed on
the finite-parameter data in Table~\ref{tf:tab:comparison}. Its lesson is
instead structural: the ratio carries only a tenth power of the number of
charges in an area \(d^2\). Even four decades of \(\bar{n}_i d^2\) change this
asymptotic factor by only about \(2.5\). Near equality over an experimentally
limited parameter range is therefore unsurprising, but it is not universal.

\subsubsection{Singular endpoint conventions}
\label{tf:sec:endpoint_conventions}

The finite-setback comparison should not be extrapolated as a regular
formula into either singular endpoint.  The endpoints are nevertheless a
useful separation test if the classical threshold is assigned its natural
extended-value convention.  In the ideal zero-thickness point-Coulomb model,
\(d=0\) removes the ultraviolet cutoff.  The logarithmic divergence in
Eq.~\eqref{tf:eq:smallxvariance} then implies that no finite,
cutoff-independent local-TF percolation scale exists.  The formal continuum
convention is \(n_{\rm perc}=\infty\): no finite mean density restores
classical connectivity.  A real finite width, impurity radius, or nonlocal
screening cuts off this singularity and gives a finite but nonuniversal
scale.

The opposite endpoint is equally instructive.  If the long-range Coulomb
component is absent and only an independent short-range \emph{transport}
correlator remains, there is no remote-scale TF puddle landscape.  The
uniform density field percolates at every \(n>0\), so the corresponding
convention is \(n_{\rm perc}=0\), not \(\infty\).  Strictly, this endpoint
has no classical percolation transition to calculate.  The IRM transport
root can nevertheless remain finite.  The two formal limits are summarized
in Table~\ref{tf:tab:endpoint_conventions}.

\begin{table}[!htbp]
\caption{Extended-value conventions for two singular disorder endpoints.
They are not finite universal critical densities: the \(d=0\) row is
ultraviolet singular, while the no-long-range-Coulomb row has no classical
percolation transition.  \(n_{\rm perc}\) refers only to connectivity of the
semiclassical density landscape.}
\label{tf:tab:endpoint_conventions}
\small
\begin{ruledtabular}
\begin{tabular}{lcc}
Endpoint & \(n_{\rm perc}\) & \(n_{\rm IRM}/n_{\rm perc}\)\\ \hline
\(d=0\), point Coulomb & \(\infty\) & \(0\)\\
Long-range Coulomb absent & \(0\) & \(\infty\)
\end{tabular}
\end{ruledtabular}
\end{table}

These entries do not diagnose the exact quantum spectrum.  Classical
percolation can be present while short-range disorder produces a finite IRM
scale and eventual quantum localization; conversely, quantum tunneling can
connect nominally depleted regions.  Their limited but decisive role is to
show that the order-unity agreement at finite remote Coulomb disorder is
contingent on the common smooth landscape, not a universal identity between
percolation and IRM transport.

\subsubsection{Short-range transport disorder: an explicit separation test}
\label{tf:sec:short_range}

We now make the distinction operational. Let an independent microscopic
short-range component be represented, after any microscopic screening and
band-structure renormalization, by an effective white-noise correlator
\begin{equation}
 \frac{\overline{|V_{\rm sr}({\bf q})|^2}}{\mathcal A}=W_{\rm sr}.
\label{tf:eq:shortcorrelator}
\end{equation}
For a spin-degenerate parabolic 2DEG, the Born result is
\begin{equation}
 \frac{\rho_{\rm sr}}{h/e^2}=\frac{n_{\rm sr}}{n},
 \qquad n_{\rm sr}\equiv\frac{m^2W_{\rm sr}}{2\pi\hbar^4}.
\label{tf:eq:shortrho}
\end{equation}
We use \(\gamma_{\rm sr}=n_{\rm sr}a_{\rm B}^2\) as the dimensionless
short-range strength. Matthiessen addition then gives the deliberately
minimal mixed-disorder test
\begin{equation}
 \frac{\rho_{\rm tot}}{h/e^2}=
 \frac{\rho_{\rm rem}(n;\bar{n}_i,d)}{h/e^2}
 +\frac{\gamma_{\rm sr}}{n a_{\rm B}^2}.
\label{tf:eq:mixedrho}
\end{equation}

The local-TF percolation calculation is held fixed in this test: it contains
the remote charged layer and therefore its long-wavelength puddle landscape,
but not the separate microscopic transport correlator in
Eq.~\eqref{tf:eq:shortcorrelator}. This is a controlled scale-separation
counterexample, not a claim that an arbitrary atomic defect has zero local
electrostatic effect. A finite correlation length, impurity chemistry, and
the associated density response would be needed to calculate that effect.
The purpose here is narrower and decisive: a channel which mainly changes
short-distance scattering changes the IRM root without changing the
long-distance real-space construction.

\begin{figure*}[!htbp]
\centering
\includegraphics[width=0.98\textwidth]{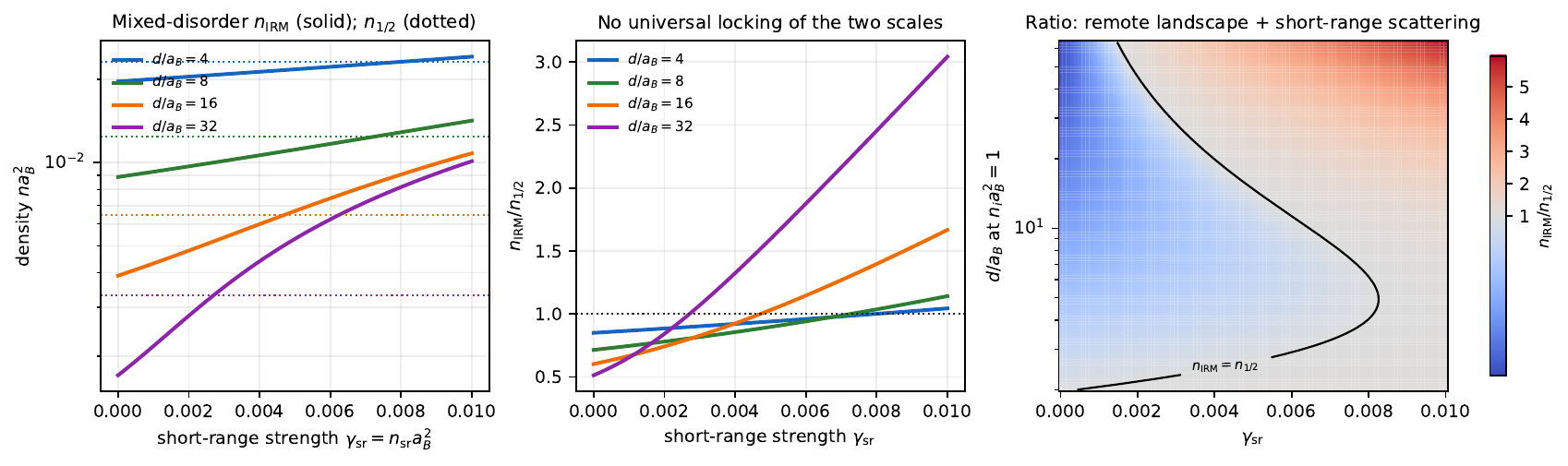}
\caption{Mixed-disorder separation test. The remote charged layer, with
\(\bar{n}_i a_{\rm B}^2=1\), is retained as the only smooth landscape entering the
local-TF half-depletion density \(n_{1/2}\) (dotted curves). An independent
effective white-noise short-range transport correlator of strength
\(\gamma_{\rm sr}=n_{\rm sr}a_{\rm B}^2\) is added only to the homogeneous
Born transport rate in Eq.~\eqref{tf:eq:mixedrho}. Left: the resulting
\(n_{\rm IRM}\) (solid) moves upward through the fixed \(n_{1/2}\).
Middle: the ratio is not locked to one. Right: false-color map of the ratio
versus setback and \(\gamma_{\rm sr}\); the black contour is unity. The
figure is a controlled counterexample to a universal relation, not a
microscopic model of a particular short-range defect.}
\label{tf:fig:short_range}
\end{figure*}

Figure~\ref{tf:fig:short_range} evaluates Eq.~\eqref{tf:eq:mixedrho} with the
same exact remote RPA quadrature used above, at \(\bar{n}_i a_{\rm B}^2=1\).
For example, at \(d/a_{\rm B}=8\), the remote-only values are
\(n_{1/2}a_{\rm B}^2=0.01242\) and
\(n_{\rm IRM}a_{\rm B}^2=0.00887\). Raising
\(\gamma_{\rm sr}\) from zero to \(0.008\) changes the latter to
\(0.01287\), crossing the unchanged smooth-landscape value. At
\(d/a_{\rm B}=32\), the same range moves the ratio from \(0.51\) to
\(2.45\). This is a large effect despite the fact that
Eq.~\eqref{tf:eq:shortcorrelator} supplies no additional remote-scale
electrostatic landscape.

\clearpage

%% file: meta_sec6_pan.tex

\section{Quantum {H}artree--{F}ock and Classical Molecular Dynamics:
Dynamical and Structural Coherence Lengths}
\label{sec:pan_qhf}

The microscopic calculations of
Secs.~\ref{sec:hf_letter}--\ref{sec:lll} report structural coherence
lengths and bond-order diagnostics on finite periodic cells and
tori, which the present toolset cannot by itself translate into
weak-pinning Larkin scalings, and they treat only quasi-static
structural correlations, which the present toolset cannot by itself
compare with finite-temperature dynamical response.  This section is
devoted to two further questions that bear on the physical picture
developed above but that neither the HF, VMC, nor LLL--HF
calculations of this paper address directly, and is placed as a
separate section rather than inside Secs.~\ref{sec:hf_letter}--%
\ref{sec:lll} because the microscopic tools used in those sections
could not, within the methodological scope of the present work,
provide the answers.  First, what is the parametric scaling of the
Larkin coherence length with disorder strength, in both the
amplitude and the concentration scans of a short-range disorder
model?  Second, how does the dynamical response of the pinned
amorphous solid differ from its static coherence --- that is, does
strong pinning suppress thermal rearrangement even when the
structural crystallinity has already been lost?  Both questions are
addressed by the recent independent study of Das Sarma and
Pan~\cite{DSPan2026}, whose main results we review here and connect
to the structural picture of the preceding sections.

Reference~\cite{DSPan2026} uses two complementary microscopic
descriptions, continuum Hartree--Fock in a periodic supercell and
classical molecular dynamics (MD) in an open square with a Langevin
thermostat.  The two descriptions use different Hamiltonians,
different disorder laws (an amplitude scan of a short-range Gaussian
random potential for HF; a concentration scan of compact attractive
wells for MD), different numerical checks, and different structural
diagnostics; both arrive at a mutually consistent picture that
complements and extends the one developed in the preceding sections
of the present paper.

\subsection{Quantum HF side}
\label{sec:pan_qhf_method}

The quantum HF side uses a periodic supercell of $N_c = 9$ electrons
spanning three primitive lattice spacings in each direction of a
triangular crystal, with the disorder entering as a short-range
Gaussian random potential of amplitude $W$ projected onto the
orbital basis.  The main scan covers $\rs = 2, 3, 4, 8, 16$ at
$W/E_F = 0, 0.5, 1, 2, 4$, with a direct-only Hartree calculation
kept as a control that isolates the role of exchange.  The structural
diagnostics are the density contrast
$C = \langle (n/\bar n - 1)^2 \rangle$ and the six-vector Bragg
fraction $f_6 = B_6/C$, together with the first-shell density-
coherence function $C_B(r)$ whose first $e^{-1}$ crossing defines
the operational translational coherence length $\xi_B$.  The HF
single-realization structural crossover diagram in the
$(\rs, W/E_F)$ plane spans the clean crystal ($f_6 \gtrsim 0.5$,
$\xi_B$ exceeding the supercell range) and the amorphous-pinned
solid ($f_6 \ll 0.5$, $\xi_B$ at the lattice spacing).

\subsection{Classical MD side}
\label{sec:pan_md}

The classical side uses $N = 288$ Coulomb-interacting electrons in
an open square with reflecting walls and a compensating positive
background, with short-range attractive wells of fixed depth
$D = 0.1 E_0$ and radius $R_p = 0.4 a$ placed by a nested Poisson
process at concentration $c = \bar{n}_i/n_0$ ($n_0 = 1/(\pi a_0^2)$).
The electron and impurity densities are tuned through
$q = n_e/n_0 \in \{0.5, 0.75, 1, 1.5, 2\}$ and
$c \in \{0, 1/64, 1/32, 1/16, 1/8, 1/4, 1/2, 1, 2, 4\}$.
Zero-temperature relaxation uses FIRE damped-inertial dynamics;
finite-temperature runs use Langevin dynamics at $\theta = k_B T/E_0$
from $0.001$ to $0.120$.

A common first-shell density-coherence construction is applied to
both calculations: the six pristine Bragg carriers are shifted to
zero before transforming the neighboring spectral weight, and the
first $e^{-1}$ crossing of the resulting $C_B(r)$ defines the
operational $\xi_B$.  The two independent calculations converge on
the same observable after this common construction.

\subsection{Main results}
\label{sec:pan_results}

\emph{Monotonic loss of structural coherence with disorder.}
For the HF calculation at $\rs = 4$, $\xi_B$ exceeds the accessible
range (above $1.5 a$) at $W/E_F = 0, 0.5, 1$, decreases to
approximately $1.1 a$ at $W/E_F = 2$, and reaches approximately
$0.9 a$ at $W/E_F = 4$.  At the largest disorder $W/E_F = 4$ the
corresponding values for $\rs = 3, 8, 16$ are approximately
$0.8 a, 1.0 a, 1.0 a$: nearest-neighbor coherence scales within the
stated spectral convention.  For the classical calculation at
$n_e = n_0$, the first-shell coherence extends beyond $5 a$ through
$c = 1/16$, decreases to approximately $2.7 a$ at $c = 1/8$,
$1.2 a$ at $c = 1/4$, $0.9 a$ at $c = 1/2$, and $0.8 a$ at $c = 4$.
Both descriptions reach microscopic coherence under sufficiently
strong pinning, in agreement with the structural picture of
Secs.~\ref{sec:hf_letter}--\ref{sec:lll}.

\emph{Weak-pinning Larkin estimates.}  Independent weak-pinning
estimates from the pristine elastic constants and the random-force
covariance give, with $\eta_{\mathrm{W}} \equiv W/E_F$ for the HF
scan (distinct from $\eta_{\mathrm{dis}} = \bar{n}_i/n$ of
Eq.~(\ref{eq:ratio})),
\begin{equation}
\frac{R_c^{\mathrm{HF}}}{a} \simeq \frac{1}{\eta_{\mathrm{W}}} \times
\begin{cases}
3.2, & \rs = 3,\\
2.5, & \rs = 4,\\
1.9, & \rs = 8,\\
2.3, & \rs = 16,
\end{cases}
\label{eq:pan_RcHF}
\end{equation}
and for the classical concentration scan at
$c = \bar{n}_i/n_0$,
\begin{equation}
\frac{R_c^{\mathrm{cl}}}{a} \simeq \frac{0.63}{\sqrt{c}},
\qquad
\frac{R_G^{(L)}}{a} \simeq \frac{0.87}{\sqrt{c}}.
\label{eq:pan_RcClassical}
\end{equation}
The weak-pinning scaling is therefore $R_c \propto W^{-1}$ for the
HF amplitude scan and $R_c \propto \bar{n}_i^{-1/2}$ for the
classical concentration scan, which reflect the different variances
of the random forces (quadratic in $W$, linear in $\bar{n}_i$) and
illustrate the Larkin mechanism from two microscopically distinct
starting points.

\emph{Decoupling of coherence from thermal mobility.}  A separate
finite-temperature classical result is that stronger pinning raises
the temperature for appreciable particle rearrangement even in
configurations whose crystalline order has already disappeared.  The
primary cage-escape bracket rises from
$k_B T/E_0 = 0.005$--$0.007$ in the clean system to
$0.020$--$0.030$ at $c = 4$, where the crystalline order is lost
at zero temperature.  Disorder therefore suppresses structural
coherence while stabilizing the resulting amorphous pinned assembly
against thermal rearrangement.  Crystalline coherence and resistance
to motion are distinct properties, and strong disorder decouples
them.

\subsection{Integration with the present work}
\label{sec:pan_integration}

The Das Sarma--Pan results integrate with the preceding sections in
four ways that are useful to state explicitly.

First, both the HF scan and the classical scan reach a coherence
length of order the lattice spacing at strong pinning, in
quantitative agreement with the structural
Secs.~\ref{sec:hf_letter}--\ref{sec:lll}: no finite-disorder
crystal-to-amorphous phase boundary is seen in any of the five
calculations (zero-field HF, VMC, LLL HF, Pan HF, Pan MD), as
required by Imry--Ma--Larkin.

Second, the two Larkin scalings
Eqs.~(\ref{eq:pan_RcHF}--\ref{eq:pan_RcClassical}) have a parametric
implication for the microwave pinning-mode resonance of the GaAs 2DEG
in strong perpendicular field~\cite{Ye2002,Chung2022,Madathil2024}.
In the magnetized collective-pinning picture of Chitra, Giamarchi
and Le Doussal~\cite{Chitra2002}, with cyclotron frequency
$\omega_c$ much larger than the zero-field pinning scale
$\omega_0 \sim c_t/R_c$, the low-frequency magnetophonon pole sits
at $\omega_p \simeq \omega_0^2/\omega_c \sim c_t^2/(\omega_c R_c^2)$.
At fixed field, density, elastic constant, and setback, this gives
$\omega_p \propto R_c^{-2}$ and therefore $\omega_p \propto W^{2}$ in
an HF disorder-amplitude scan and $\omega_p \propto \bar{n}_i$ in
a classical concentration scan.  These parametric statements are
subject to two caveats that the comparison with microwave data must
respect: the first-shell density-coherence length $\xi_B$ measured
in the structural calculations above is not automatically the
collective-pinning Larkin length that enters the magnetophonon
expression; and the weak-pinning elastic estimate cannot be carried
unchanged into the $R_c \sim a$ strong-pinning regime where several of
the amorphous-solid calculations of
Secs.~\ref{sec:hf_letter}--\ref{sec:lll} sit.  The predictions should
therefore be read as parametric expectations within the collective
weak-pinning picture, not as quantitative predictions for a specific
sample.

Third, the dynamical observation that stronger pinning raises the
cage-escape temperature even in the amorphous regime shows that
insulating-looking resistance and crystalline-looking local structure
are independent properties of the disordered 2D electron solid.  The
Ioffe--Regel--Mott transport scale $\nIRM$ of Sec.~\ref{sec:tf_irm}
and the loss of long-range crystalline order inferred from the
structural calculations above are set by different criteria within
the present framework and are not the same transition; a direct
microscopic calculation of a common HF-level localization-cum-
amorphization transition is not performed in this paper.

Fourth, the two disorder scans sample mutually orthogonal cuts
through the disorder parameter space: the HF amplitude scan at fixed
geometry probes the strength of the random potential at fixed
spatial statistics, while the classical concentration scan at fixed
well depth probes the spatial statistics at fixed per-well
strength.  Both scans yield the same qualitative behavior (clear suppression
of coherence as the disorder becomes appreciable, continuous
crossover, no finite-disorder phase boundary identified) and the
same weak-pinning limit (Larkin scaling with the appropriate
random-force variance).  Two independent microscopic models
yielding the same qualitative result are a non-trivial consistency
check on the Imry--Ma--Larkin picture of the disordered 2D Wigner
crystal within the scope of the two calculations.

%% file: meta_sec7_conclusion.tex

\section{Discussion}
\label{sec:discussion}

The structural, variational, Landau-level, and transport
calculations of Secs.~\ref{sec:hf_letter}--\ref{sec:pan_qhf}
converge on one physical picture of the two-dimensional electron
system in charged disorder: an impurity-organized amorphous solid
with short-range crystalline coherence and a metal--insulator
crossover whose two natural density scales coincide to within a
factor of order unity.  This section discusses two consequences of
that picture, placed within the specific scope of the present
microscopic framework: how it bears on the recent STM imaging
experiments on MoSe$_2$, and how it clarifies the long-standing
debate on the mechanism of the 2D MIT in Si and GaAs.

\subsection{Relation to the MoSe$_2$ STM imaging experiments}
\label{sec:disc_stm}

The STM imaging experiments on gate-tunable bilayer and monolayer
MoSe$_2$~\cite{Xiang2025,Ge2026,Xiang2024} report real-space density
patterns at carrier densities $\rs \approx 10$--$25$, below the clean
Wigner-crystal $\rc$ reported by quantum Monte Carlo
(Sec.~\ref{sec:intro}).  The experiments and their accompanying
supplementary analyses report density maxima at the interparticle
spacing, local sixfold coordination, bond-orientational analyses on
the imaged configurations, and model simulations of the disorder;
the published interpretations identify a disorder-affected
two-dimensional electron solid.  The hole experiment of
Ref.~\cite{Xiang2025} and the electron experiment of
Ref.~\cite{Ge2026} differ in carrier polarity and in sample details;
the complementary theoretical analysis of Joy and
Skinner~\cite{Joy2025} considers local liquid--solid competition
under weak short-range disorder.

Within the microscopic framework of the present paper---continuum
Hartree--Fock, variational Monte Carlo with the trial-wave-function
families of Sec.~\ref{sec:vmc}, and LLL Hartree--Fock on the
torus---the calculations find small or statistically unresolved
weak-disorder changes and clear suppression of the connected
crystalline correlation once the dimensionless disorder parameter
$\eta_{\mathrm{dis}} = \nimpurity/n$ becomes appreciable, at every
$\rs$ sampled and in every disorder model studied.  No window of
disorder-enhanced crystalline order is seen within these measures
and these methods.
Taken together with the Imry--Ma--Larkin argument (bearing on true
long-range translational order), these calculations suggest that
one plausible structural reading of the imaged states at
$\rs \lesssim 20$ is an impurity-organized amorphous solid: a
correlated electron state whose density pattern is organized by the
disorder of the sample rather than by the many-body crystalline
order parameter, with short-range coherence inherited from the
dense-liquid pair correlation and no long-range crystalline order.
The structural distinction we would propose to test is in the
width and persistence of Fourier peaks at the crystalline lattice
star versus at the impurity positions, in the first-shell coherence
length, and in the fraction of density maxima registered on the
impurity landscape.  The present paper offers these as structural
observables within its microscopic framework; the comparison with
the published image analyses, and with any future measurements of
Fourier peak widths and impurity registration, is a question for
the experimental literature to adjudicate.

The temperature dependence reported in Ref.~\cite{Ge2026}, that the
imaged state becomes more robust with increasing disorder density,
is compatible with the amorphous-solid structural picture: within
the classical molecular-dynamics calculation of
Ref.~\cite{DSPan2026}, stronger pinning raises the cage-escape
temperature of the impurity-templated texture even when its
structural coherence is at the lattice spacing (Sec.~\ref{sec:pan_qhf}).
We identify this as a parallel between the two pictures within the
limits of the two calculations, not as a decisive test.

The fact that disorder enhances the effective ``melting'' temperature
has already been pointed out in the literature~\cite{Vu2022}, and the
pristine melting temperature of a 2D Wigner crystal is very low
(10--100 mK for the 2D semiconductor layers under
consideration)~\cite{Hwang2001}.

\subsection{Interpretation of the 2D MIT in Si and GaAs}
\label{sec:disc_mit}

The emerging picture of the 2D MIT in Si MOSFETs and GaAs/AlGaAs
heterostructures~\cite{Kravchenko1994,DasSarma2005,DasSarma2013,%
DasSarma2014} from the analysis of Sec.~\ref{sec:tf_irm} is that
the half-depletion density $\nph$ of the semiclassical
nonlinear-Thomas--Fermi landscape (a geometric scale, not an
independently measured percolation threshold) and the
Ioffe--Regel--Mott density $\nIRM$ of Born--RPA transport track each
other to within a factor of order unity over the physically relevant
range of impurity density and setback distance, with a slow
$\nimpurity^{-1/10} d^{-1/5}$ spread.  The near-coincidence is
not a theorem.  Both densities correspond to the energetic
condition $E_{\mathrm{dis}} \sim E_F$ -- the carrier density
below which the disorder potential becomes comparable to the
Fermi energy -- and the ratio $\nIRM/\nph$ remains close to unity
precisely because both diagnostics are controlled by the same
screened spectrum at the same scale.  Over the sampled range of
Table~\ref{tf:tab:comparison} the ratio runs from $0.60$ to $1.05$,
and in the presence of an independent short-range transport channel
(Fig.~\ref{tf:fig:short_range}) the ratio can shift by factors of
order two.  The right statement is therefore that in a model with
the long-range Coulomb impurities providing both the screening
nonlinearity and the transport scattering, the two scales are
numerically close over the sampled remote-disorder range, with a
slowly varying ratio; this is not an identification of the two
constructions as the same transition, and the mixed-disorder
separation test of Sec.~\ref{tf:sec:short_range} supplies an
explicit counterexample to any such identification.

This observation is independent of the question of long-range order
in the low-density phase.  The 2D MIT, by the above analysis, is a
disorder-driven transport transition whose critical density is set
by the disorder scale; it is not a crystallization transition in
the Wigner sense.  The low-density insulating state is an amorphous
solid (in the structural sense of
Secs.~\ref{sec:hf_letter}--\ref{sec:lll}), and the metallic state
above $\nc$ is a strongly screened, strongly correlated liquid.
The absence of a Wigner-crystal interpretation of the 2D MIT does
not require a Wigner-crystal interpretation of the insulating state;
it merely identifies both as consequences of the same
Hamiltonian, with the organizing variable $\nimpurity/n$ playing
the role of the dimensionless disorder strength throughout.

\section{Conclusion}
\label{sec:conclusion}

The main conclusions of this paper separate into three logically
distinct statements that should not be collapsed into one.  First,
within the Imry--Ma--Larkin framework, quenched Coulomb disorder
destabilizes ideal long-range translational order in two dimensions
at any nonzero disorder strength; what survives in the
thermodynamic limit has finite translational coherence rather than
true Bragg peaks.  Second, at weak charged disorder the microscopic
calculations of Secs.~\ref{sec:hf_letter}--\ref{sec:pan_qhf} find
small or statistically unresolved structural response; weak
disorder does not by itself drive a correlated liquid into a solid,
and weakly disordered crystals can retain extended crystalline
domains that we do not resolve on the present finite cells.  Third,
once the disorder-to-electron density ratio becomes appreciable the
same calculations show clear suppression of the connected
crystalline correlation and a density pattern organized by the
impurity landscape rather than by the many-body crystalline order
parameter; this is the regime we have called the amorphous solid.
The dimensionless disorder-to-electron density ratio
$\nimpurity/n$ is the primary organizing variable for the structural
crossover, with impurity strength, setback, and disorder spectrum
entering as additional parameters that each section discusses in
its own context.  No window in the $(\rs, \nimpurity/n)$ plane
sampled by the microscopic calculations of
Secs.~\ref{sec:hf_letter}--\ref{sec:pan_qhf} exhibits
disorder-enhanced crystalline order within the measures used.

Within the Imry--Ma--Larkin framework, true long-range translational
order is unavailable in the thermodynamic limit at any nonzero
disorder; at weak disorder the coherence length can be large on the
scales accessible to our finite cells, and no finite-disorder
structural phase boundary is identified by the present calculations.
For the disorder strengths of experimental interest the coherence
length we measure is of order a few lattice spacings and often of
order the lattice spacing itself; the system is then in the
strong-pinning (individual-impurity) regime, where the elastic
description of a pinned Wigner crystal has broken down and the
amorphous solid is more naturally described as a glass of
impurity-trapped electrons.  The two descriptions agree on the
qualitative conclusion.  The finite-size calculations of this paper
cannot in principle distinguish $R_c = a$ from $R_c = 3a$ and do not
claim to.

The theoretical implications, within the stated approximations,
are:
\begin{enumerate}
\item The structural pattern of an impurity-organized amorphous solid
is a possibility consistent with the recent STM imaging experiments
on MoSe$_2$~\cite{Xiang2025,Ge2026,Xiang2024} at $\rs \approx 20$.
This is, in fact, the conclusion reached by a recent unpublished
Princeton STM study on 2D graphene layers~\cite{Yazdani2026}.
Specific structural observables that would discriminate between an
impurity-organized texture and a disorder-broken Wigner crystal
include Fourier peak widths at the crystalline lattice star versus
at the impurity positions, first-shell density coherence lengths,
and the fraction of imaged density maxima registered on the
impurity landscape.  We leave the comparison between these
observables and the experimental data to the experimental
literature.
\item On the mechanism of the 2D MIT in Si and
GaAs~\cite{Kravchenko1994}, the half-depletion density $\nph$ of the
semiclassical nonlinear-Thomas--Fermi landscape and the
Ioffe--Regel--Mott density $\nIRM$ of Born--RPA transport are reached
at the common energetic condition $E_{\mathrm{dis}} \sim E_F$
(Sec.~\ref{sec:tf_irm}); the two scales track each other to within a
factor of order unity in the long-range-Coulomb model studied, with
slow $\nimpurity^{-1/10} d^{-1/5}$ spread.  Modest independent
short-range transport channels can shift the ratio by factors of
order two.  These statements are made within the specific scope of
the two calculations and are not a theorem relating the geometric
depletion scale to a true percolation threshold.
\item In the magnetized collective-pinning picture of
Ref.~\cite{Chitra2002}, the microwave pinning-mode resonance of the
GaAs 2DEG in strong perpendicular field~\cite{Ye2002,Chung2022,%
Madathil2024} sits at $\omega_p \simeq c_t^2/(\omega_c R_c^2)$ when
$\omega_c \gg \omega_0$.  Combined with the two weak-pinning Larkin
scalings of Sec.~\ref{sec:pan_qhf}, this gives parametric expectations
$\omega_p \propto W^2$ in an HF disorder-amplitude scan and
$\omega_p \propto \bar{n}_i$ in a classical concentration scan, within
the collective weak-pinning regime.  Both the identification of the
first-shell density-coherence length with the Larkin length and the
extrapolation of weak-pinning elasticity into the strong-pinning
regime require independent verification before these parametric
statements can be used quantitatively.
\end{enumerate}

Open questions include the thermodynamic-limit extrapolation of the
coherence length from the present finite-cell calculations, the
role of spin and valley degeneracies in the amorphous solid, the
competition between the amorphous solid and the fractional quantum
Hall liquid in the LLL~\cite{DSPan2026,HuangDSLi2026} at specific
rational fillings, the finite-temperature phase diagram of the
pinned amorphous solid, and the extension of the analysis to
systems with screening by nearby gates and to electron-hole
bilayers.  The methodology of this paper---organized around the
dimensionless disorder-to-electron density ratio $\nimpurity/n$,
with the Imry--Ma--Larkin picture as the structural framework and
with HF, VMC, LLL HF, and nonlinear-TF as the complementary
microscopic tools---is directly applicable to each of these
extensions.

\subsection*{Acknowledgments}

This work is supported by the Laboratory for Physical Sciences
through the Condensed Matter Theory Center at the University of
Maryland.  The work has been assisted by AI through the use of
Anthropic Claude Fable 5 and OpenAI Astra 6 in producing
numerics/figures/checking/editing.  The author takes full
responsibility.

%% file: meta_appendix.tex

\section{Variational Monte Carlo computational details}
\label{sec:SM_vmc}

\subsection{Computational details}
\label{vmc:app:comp}

\emph{Cell and lattice.} The rhombic cell has primitive vectors $(L,0)$ and $(L/2,\sqrt3L/2)$ with $L=4a$ and $a=1.90463\,\rs\aB$, area $\pi N\rs^2(\aB)^2$; it contains a $4\times4$ triangular lattice whose first reciprocal star $\{\bG_1\}$ has $|\bG_1|=4\pi/\sqrt3a$. Structure factors are evaluated at all reciprocal vectors $\bq=m\mathbf{b}_1+n\mathbf{b}_2$ with $|m|,|n|\le8$. Electron--electron and electron--impurity interactions are 2D Ewald sums with the impurity plane offset by $d$; the $\bq=0$ background terms are constants and are omitted (so $\av{V_{\rm imp}}$ for the donor layer is defined up to a constant that depends on $\rs$ and $d$).

\emph{Trial functions and optimization.} The plane-wave determinant uses the 13 lowest closed-shell momenta plus a zero-total-momentum triplet at the fourth shell. The Gaussian orbitals are $\phi_\beta(\br)=\sum_{\mathbf{T}}\exp[-\beta|\br-\mathbf{T}|^2/2]$ over lattice translations $\mathbf{T}$, with one common optimized width. The Jastrow is $U=\sum_{i<j}a\,r_{ij}/(1+b\,r_{ij})$ with $a$ fixed by the cusp condition for the polarized system and $b$ optimized. The impurity factor has three parameters ($\kappa_+$, $\kappa_-$, $\ell$); for the donor layer $\kappa_-$ is inert. The impurity Coulomb singularity needs no regularization: the electron--impurity cusp is carried by $W$, the local energy is finite at coincidence, and the 2D Ewald kernel is integrable. Parameters are optimized by stochastic reconfiguration with 40--60 iterations of 64 blocks each, 16 chains, with the impurity amplitudes initialized proportional to $Z$; the final parameters are the average over the last 8--10 iterations. For the FG family the orbital centers are obtained by L-BFGS minimization of the classical energy of 16 point charges in the Ewald electron--electron potential plus the impurity potential convolved with the Gaussian orbital density $|\phi_\beta|^2$ (evaluated in reciprocal space, with $\beta$ taken from the clean PG optimization at the same $\rs$), from 13 starting arrangements (the lattice, six lattice-plus-noise, six random), keeping the lowest; the centers are then frozen and $\beta$, $b$, $\kappa_\pm$, $\ell$ are optimized by VMC as for the other families. For $Z\le0.05$ the relaxed centers are within $0.02a$ of the lattice sites; for $Z=1$ and for the donor layer they sit on or between the attractive impurities.

\emph{Sampling.} Production uses 24 Markov chains, 1200 blocks of 32 single-electron Metropolis sweeps with mixed local and global (impurity-centered) proposals and full Hastings ratios, after 120 equilibration blocks; every state yields 28\,800 stored configurations from which all diagnostics are computed. Quoted errors are the larger of the chain-to-chain standard error and the plateau of a time-reblocking analysis. Sampler correctness was verified against known targets and the local energy against finite-difference Laplacians of the full wave function at configurations including an electron inside an impurity core. The one-body densities are $96\times96$ histograms on the fractional cell; the sixfold bond order uses the six nearest minimum-image neighbors; the spanning test for the donor layer is a periodic connected-component analysis of the thresholded density on a $3\times3$ tiling of the cell.

\emph{Data.} The model (i) results at $\rs=10$, 20, and 50 and the pristine references at those densities were generated in an earlier stage of the project with the same code, and their stored production configurations were re-analyzed for this paper; every other state (the $\rs=30$ references and all $\rs=30$ disordered states, the complete model (ii) scans at four densities, the complete model (iii) grid of four setbacks at four densities, and the FG family throughout), the static/connected decomposition, the bound/unbound split, and the bond-order and morphology analyses were generated here from the same Hamiltonian and sampler. All results in the tables are from the stored production configurations. In all, 139 VMC ground states were computed and analyzed.

\section{Lowest Landau level HF --- disorder landscape and puddle picture}
\label{sec:SM_lll}

\label{lll:app:puddle}

This appendix collects the computational details of the LLL
landscape and puddle analyses summarized in Sec.~\ref{sec:lll}: the
form of the impurity landscape, its filling by lakes and puddles,
the behavior on a larger sample, where the Hartree--Fock electrons
sit on it, how well they screen it, how it changes with the
impurity setback, and the Hartree-deletion numerical experiment.
The Hartree-deletion calculation is a controlled diagnostic that
isolates the role of electrostatic stiffness in the full theory: it
is one route by which density inhomogeneity becomes favorable in a
model Hamiltonian, not a claim that physical puddles can arise only
when Hartree repulsion is removed.  Both the Thomas--Fermi
calculations of Sec.~\ref{sec:tf_irm} and the VMC donor-layer
calculations of Sec.~\ref{sec:vmc} produce density puddles with
full Hartree repulsion retained.  Within the LLL--HF search explored
here, however, the full solution does not approach a filled-droplet
state at any filling studied, and the structural interpretation
developed in Sec.~\ref{sec:lll_summary} follows.

\subsubsection{The potential landscape seen by a test electron}

\begin{figure*}[!t]
\centering
\includegraphics[width=\textwidth]{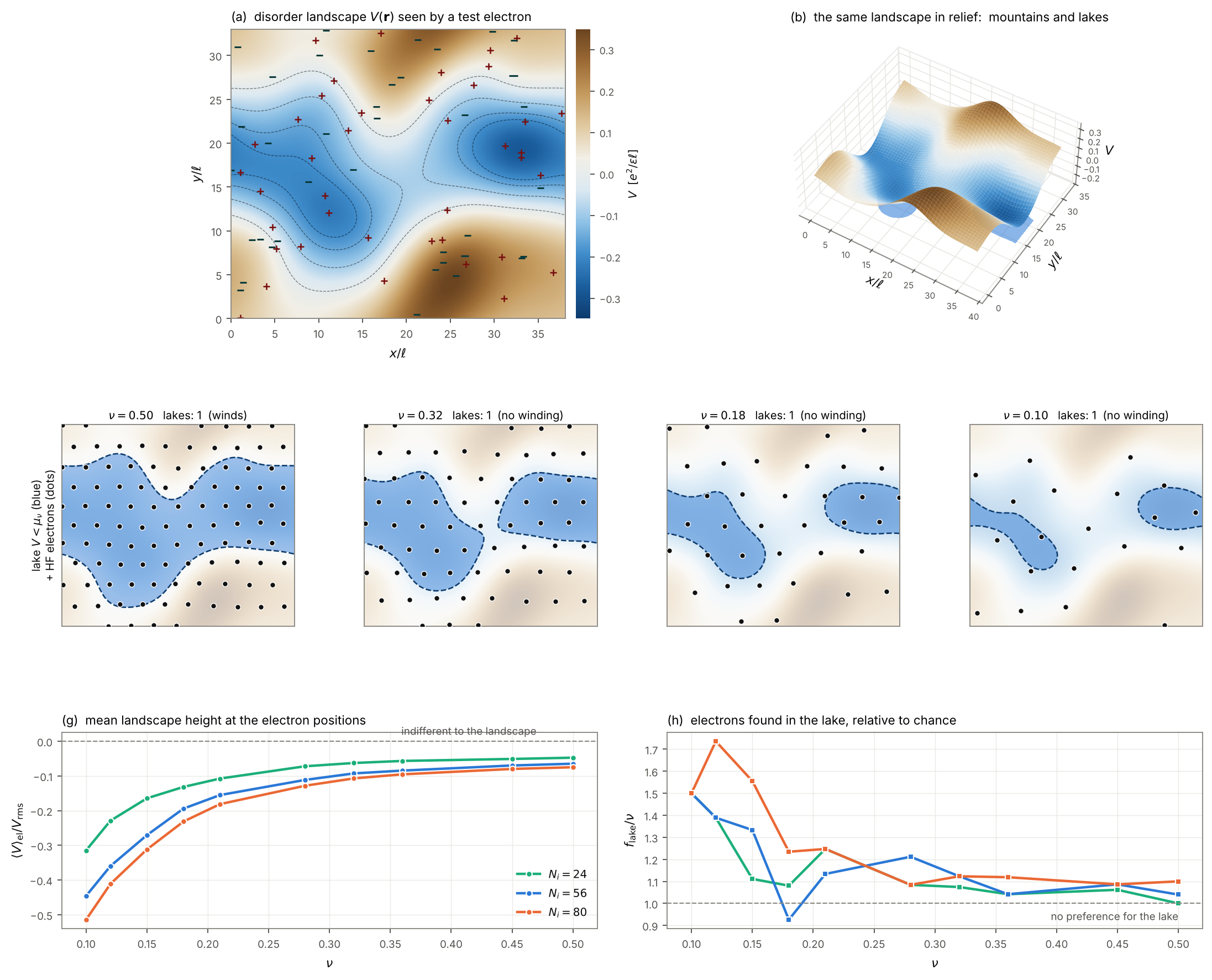}
\caption{The disorder landscape of the sample used throughout the paper
($N_i=80$, $\pm e$, $d=6\ell$, $V_{\rm rms}=0.149\,\ee$) and the Hartree--Fock
electrons on it. (a) The landscape seen by a test electron, Eq.~(\ref{lll:eq:Vreal}),
with the impurity charges marked and contours at the waterlines $\mu_\nu$ for
$\nu=0.10$, $0.18$, $0.32$, $0.50$. (b) The same function in relief. (c)--(f) The
lake $V<\mu_\nu$ (blue) at four fillings, with the density maxima of the
converged Hartree--Fock solution shown as dots. (g) Mean landscape height at the
electron positions, in units of the landscape rms; zero means no preference.
(h) Fraction of electrons inside the lake, relative to the chance value $\nu$.}
\label{lll:fig:land}
\end{figure*}

The one-body impurity term is Eq.~(\ref{lll:eq:himp}). Written in real space it is
the potential energy
\begin{equation}
V(\br)=\frac{1}{A}\sum_{\bq\neq0}V(\bq)\,e^{-q^2\ell^2/4}\,e^{i\bq\cdot\br}
\label{lll:eq:Vreal}
\end{equation}
of a test electron, with $V(\bq)$ from Eq.~(\ref{lll:eq:Vq}). It carries the same
Landau level form factor $e^{-q^2\ell^2/4}$ that enters the matrix elements, so
that structure finer than a magnetic length is absent, as it must be for an
electron confined to the lowest level. Figure~\ref{lll:fig:land}(a) shows this
landscape for the sample used throughout the paper, $N_i=80$ charges of both
signs at setback $d=6\ell$, and Fig.~\ref{lll:fig:land}(b) shows the same function in
relief, which is the form in which the hills and basins are easiest to read.

The amplitude is $V_{\rm rms}=0.1485\,(\ee)$ and the extreme values on this
sample are $-0.283$ and $+0.349\,(\ee)$. For comparison, the clean condensation
energy per electron is $0.130\,(\ee)$ at $\nu=1/2$ and at most $0.197\,(\ee)$
near $\nu=0.21$. The landscape is therefore not a weak perturbation: its typical
excursion is of the same order as the energy that holds the crystal together, and
its extremes are larger.

The landscape is also smooth, and this is the property that matters most for what
follows. Its autocorrelation function falls to one half at $7.2\,\ell$ and first
crosses zero at $12.2\,\ell$ for $d=6\ell$; at $d=3\ell$ the corresponding
numbers are $6.1\,\ell$ and $11.8\,\ell$. This is the precise sense in which
Coulomb disorder is slowly varying, and it is intrinsic rather than incidental: a
charged impurity at a setback $d$ produces a basin whose width is set by $d$ and
by the Coulomb tail, and no choice of impurity density changes that width. Only
the amplitude responds to the impurity density, through the usual
$\sqrt{N_i}$ of a random superposition.

What does change with filling is the length the landscape is being compared with.
The lattice constant of the crystal, $a=\ell\sqrt{4\pi/(\sqrt3\,\nu)}$, grows from
$3.81\,\ell$ at $\nu=1/2$ through $4.76\,\ell$ at $\nu=0.32$ and $6.35\,\ell$ at
$\nu=0.18$ to $8.52\,\ell$ at $\nu=0.10$. Measured in crystal sites, a region of
the correlation area $\pi\xi_V^2$ with $\xi_V=7.2\,\ell$ contains about $13$
sites at $\nu=1/2$, $8$ at $\nu=0.32$, $5$ at $\nu=0.18$ and $2.6$ at
$\nu=0.10$. A basin that at half filling accommodates a coherent patch of a dozen
crystal sites accommodates two or three at the lowest filling. In
the first case the disorder acts on the crystal collectively, displacing a patch
that remains internally ordered; in the second it acts on individual electrons.
This is an independent reason, additional to the growth of $n_i/n$, for the
collapse of the crystalline coherence length documented in
Sec.~\ref{lll:sec:dis}, and the two act in the same direction.

\subsubsection{Filling the landscape: the lake construction}
\label{lll:sub:lake}

The semiclassical construction is the obvious one and we use it only as a
reference, not as an approximation we believe. Let $\mu_\nu$ be the level whose
sub-level set $V(\br)<\mu_\nu$ covers an area fraction $\nu$, and call that
sub-level set the lake. Electrons that filled the landscape to a waterline while
ignoring one another entirely would occupy the lake exactly, with local filling
one inside it and zero outside. Figures~\ref{lll:fig:land}(c)--(f) show the lake at
$\nu=0.50$, $0.32$, $0.18$ and $0.10$.

Table~\ref{lll:tab:lakeown} gives the construction in full on the paper's own sample.
The waterline $\mu_\nu$ falls monotonically from $+0.085\,(\ee)$ at $\nu=0.70$ to
$-0.191\,(\ee)$ at $\nu=0.10$, as it must. The lake remains a single connected
region at every filling down to $\nu=0.10$; the entire lake area is always in one
component. What does change is that below $\nu\simeq0.40$ the lake no longer
winds around the torus, so that a path through the lake can no longer traverse
the sample.

This must not be over-read. The sample is $38\times33\,\ell$ and contains $80$
impurities, which on a landscape with a correlation length of $7\,\ell$ amounts
to a handful of hills and a handful of basins. A single-component lake down to
$\nu=0.10$ is a statement about this particular small sample and not about the
statistics of such landscapes. We therefore repeated the construction on larger
samples, which is the subject of the next subsection.

\subsubsection{Larger samples, and why we quote no threshold}

We generated landscapes at the same impurity density, $n_i=0.0637\,\ell^{-2}$,
on a torus five times larger in area, $A=6283\,\ell^2$ with $N_i=400$ impurities,
at both setbacks. Table~\ref{lll:tab:lakebig} gives the connectivity of the lake
there, and the bottom row of Fig.~\ref{lll:fig:terrain} shows it for $d=3\ell$.

Now the lake does break up. At $d=3\ell$ it goes from one component at
$\nu\ge0.40$ --- but for a single detached speck at $\nu=0.50$, which leaves
$0.994$ of the lake area in the main body --- to $3$, $4$, $7$ and $8$
disconnected puddles at $\nu=0.32$, $0.18$, $0.12$ and $0.10$, and the fraction
of the lake area carried by its largest component falls to $0.245$. At $d=6\ell$ the same trend is present but
weaker, reaching $3$ puddles and a largest-component fraction of $0.659$ at
$\nu=0.10$, which is what one expects of a smoother landscape with fewer and
broader basins. These larger samples are illustrative: no Hartree--Fock solution
was computed on them, and they appear here only to show what the same landscape
statistics do when given room.

It is tempting to look for a threshold in these numbers. We deliberately do not
quote one, and the reason is worth recording because we tested it. For a
statistically symmetric random field in two dimensions, a self-duality argument
places the connectivity threshold of the level set at area fraction exactly one
half, which would make $\nu=1/2$ a special filling. Equal numbers of positive and
negative impurities ought to produce such a symmetric field. They do not, on these
samples: the measured skewness of $V$ is $+0.289$ on the paper's own sample at
$d=6\ell$ and $+0.335$ at $d=3\ell$, and $+0.336$ and $+0.182$ on the five-times
larger samples. Nor is the behavior consistent with a one-half threshold in any
case: on the large $d=3\ell$ sample the lake still spans at $\nu=0.60$ and is
already fragmented at $\nu=0.32$, with spanning lost between $\nu=0.60$ and
$\nu=0.55$, while on the paper's own sample it survives to $\nu=0.40$. The duality argument requires short-range correlations, and the
Coulomb potential retains a long tail even at a setback, so its level sets are
not generic in the required sense. We therefore report the connectivity as
measured and attach no threshold to it.

\subsubsection{Where the Hartree--Fock electrons actually sit}

The dots in Figs.~\ref{lll:fig:land}(c)--(f) are the density maxima of the converged
Hartree--Fock solution at the same filling, on the same sample, with the same
impurities, taken from the production runs of Sec.~\ref{lll:sec:dis}. At $\nu=0.50$
they cover the sample in a crystal that takes almost no notice of the shoreline.
At $\nu=0.10$ they have moved into the basins. Table~\ref{lll:tab:sit} makes this
quantitative with three independent measures, each given for all three impurity
numbers.

The first is the correlation coefficient between the density $n(\br)$ and the
landscape $V(\br)$ over the whole sample. It is negative at every filling, as it
should be if electrons prefer basins, and it strengthens from $-0.180$ at
$\nu=0.50$ to $-0.249$ at $\nu=0.10$ for $N_i=80$. Its filling dependence is not
monotonic: it weakens to $-0.137$ near $\nu=0.28$ before growing again. The
reason is that this coefficient is diluted by the crystal's own density
modulation, which is large and has nothing to do with the disorder, and that
modulation is strongest in the middle of the filling range where the condensation
energy peaks. For this reason we regard it as the least clean of the three
measures and quote it mainly for completeness.

The second is the mean landscape height at the electron positions,
$\langle V\rangle_{\rm el}$, in units of $V_{\rm rms}$. It vanishes for electrons
placed without regard to the landscape and is negative for electrons in basins.
It is monotonic in filling at every disorder strength, which the correlation
coefficient is not, and it is the measure we use in the body of the paper. For
$N_i=80$ it runs $-0.073$, $-0.100$, $-0.233$, $-0.512$ at $\nu=0.50$, $0.36$,
$0.18$, $0.10$; for $N_i=56$ and $N_i=24$ the corresponding ranges are $-0.066$
to $-0.445$ and $-0.045$ to $-0.318$. More disorder traps the electrons more, and
lower filling traps them more, with no crossings between the three curves.

The third is the fraction $f_{\rm lake}$ of electrons found inside the lake, which
would equal $\nu$ if the electrons took no notice of it. The ratio
$f_{\rm lake}/\nu$ rises from $1.08$ at $\nu=0.50$ to $1.50$ at $\nu=0.10$ for
$N_i=80$. This measure is the noisiest of the three because it counts discrete
electrons --- at $\nu=0.10$ there are only twenty of them --- and it occasionally
dips below unity, as at $\nu=0.18$ for $N_i=56$, where $f_{\rm lake}/\nu=0.93$.
The trend is nevertheless unambiguous across all three impurity numbers.

Taken together the three measures say the same thing, and it is a quantitative
statement rather than a qualitative one. The trapping is real, it grows
monotonically as the filling falls and as the disorder grows, and it is modest.
At half filling the crystal is very nearly indifferent to a landscape whose rms
is comparable to the condensation energy per electron. Even at $\nu=0.10$ the
electrons sit only half a standard deviation below the mean landscape height, and
only fifty percent more of them lie in the lake than chance would put there.
Nothing in these numbers resembles a set of filled puddles.

\subsubsection{Screening of the landscape}

The electrons are not passive on this landscape; they rearrange and screen it.
Writing the self-consistent Hartree potential of the electron density as
$U_H(\br)$, the landscape a test electron actually experiences is $V+U_H$.
Table~\ref{lll:tab:screen} gives three measures of what the screening does.

The first is the bare ratio $V^{\rm tot}_{\rm rms}/V_{\rm rms}$. At $\nu=1/2$ it
is $0.33$ for $N_i=80$: the electrons flatten the landscape to a third of its
bare amplitude. As the filling falls the ratio rises, passes unity near
$\nu\simeq0.18$, and reaches $1.16$ at $\nu=0.10$. The total landscape is then
rougher than the bare one. This sounds like a failure of screening and is not: at
low filling the crystal's own density modulation is nearly complete, the density
contrast reaching unity, and the Hartree potential of that modulation is large
and has a period set by the lattice and not by the disorder. The ratio is
contaminated by the crystal's own field and we quote it only to explain why a
second measure is needed.

The second is the projection of the total landscape onto the bare one,
\begin{equation}
\alpha=\frac{\langle (V+U_H)\,V\rangle}{\langle V^2\rangle},
\label{lll:eq:alpha}
\end{equation}
which removes the part of $U_H$ that is uncorrelated with the disorder. It is
unity without screening and zero with perfect screening. We find $\alpha$
negative at every filling and every disorder strength: $-0.128$ at $\nu=1/2$ and
$-0.450$ at $\nu=0.10$ for $N_i=80$, with the same pattern for $N_i=24$ and
$N_i=56$. The electrons do not merely screen the disorder; they overscreen it,
and the residual landscape is anticorrelated with the bare one. The magnitude of
the overscreening is not monotonic in filling, passing through a minimum
between $\nu=0.32$ and $\nu=0.36$ before growing strongly at low filling.

The third is the same projection restricted to the slowly varying components
$q<G_1/2$, where the disorder has its weight and the crystal Bragg peaks do not.
This was computed as a check that the crystal's own field was not contaminating
$\alpha$. It is not: the restricted and unrestricted values agree to within a few
percent over most of the range and to $9\%$ at worst, the largest discrepancy
being $-0.458$ against $-0.416$ at $\nu=0.10$ for $N_i=24$. The overscreening is therefore a property of the
long-wavelength response and not an artifact of the Bragg peaks.

Within Hartree--Fock, overscreening is what one should expect. Linear screening
with a positive compressibility gives $0<\alpha<1$; a negative $\alpha$ requires
a negative dielectric function, which in this theory follows from the
exchange-driven negative compressibility of the lowest Landau level. That it
strengthens as the filling falls is consistent with the exchange energy becoming
relatively more important as the density drops.

\subsubsection{Terrain roughness at the two setbacks}

\begin{figure*}[!t]
\centering
\includegraphics[width=\textwidth]{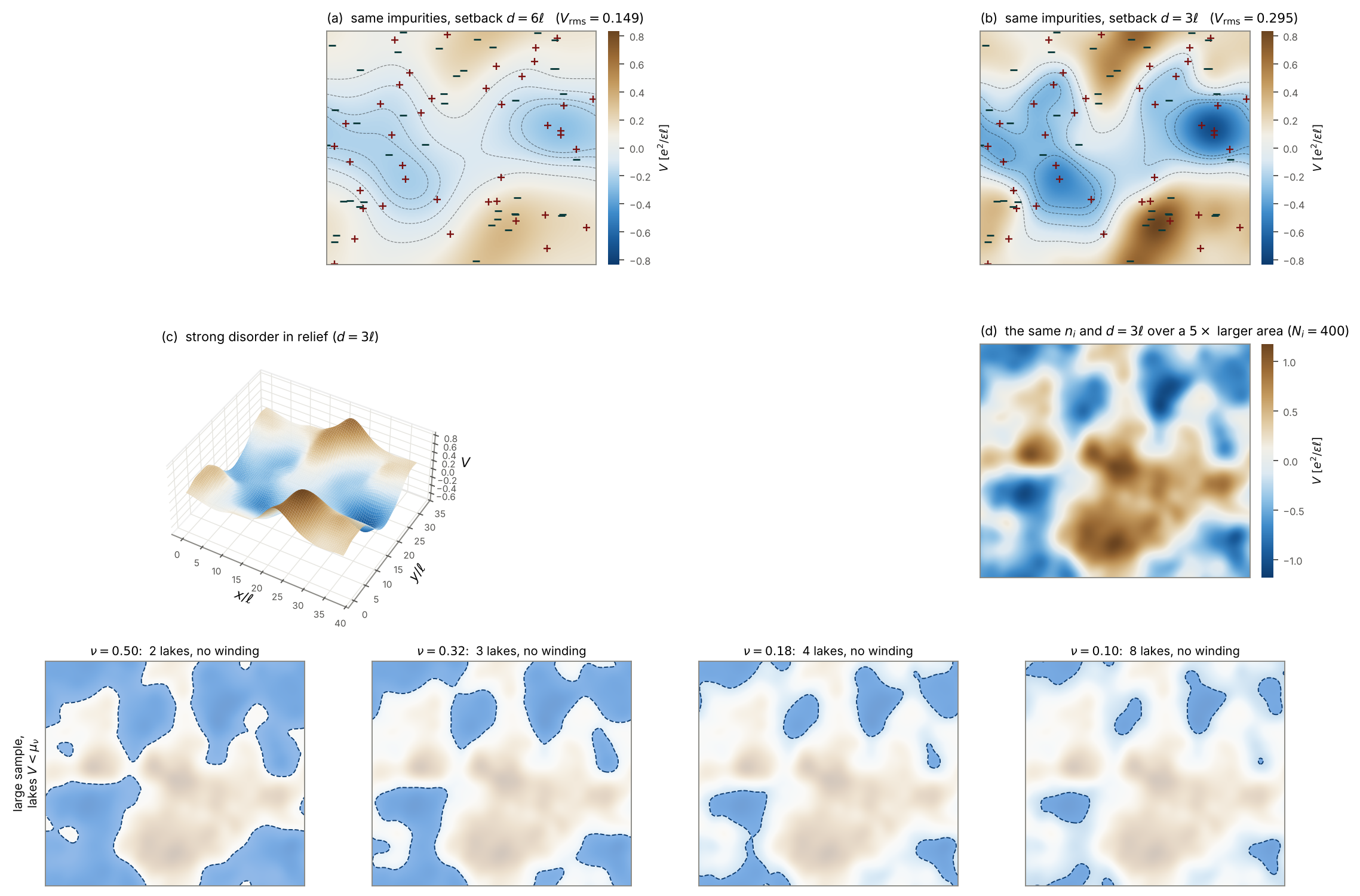}
\caption{Terrain roughness and the break-up into puddles. (a),(b) The same
impurity positions at $d=6\ell$ and $d=3\ell$ on a common color scale.
(c) The stronger landscape in relief. (d) A five times larger sample at the same
impurity density, $d=3\ell$. Bottom row: the lake in that larger sample at four
fillings, breaking into $2$, $3$, $4$ and $8$ disconnected puddles; the counts
for both setbacks are in Table~\ref{lll:tab:lakebig}. The larger sample is
illustrative, and no Hartree--Fock solution was computed on it.}
\label{lll:fig:terrain}
\end{figure*}

Figures~\ref{lll:fig:terrain}(a) and (b) show the same impurity positions at the two
setbacks on a common color scale, and Fig.~\ref{lll:fig:terrain}(c) shows the
stronger landscape in relief. Halving the setback doubles the amplitude,
$V_{\rm rms}=0.1485\to0.2952\,(\ee)$, and widens the range from
$[-0.283,+0.349]$ to $[-0.738,+0.833]\,(\ee)$. It sharpens the terrain much less:
the correlation half-width falls only from $7.2\,\ell$ to $6.1\,\ell$, and the
first zero from $12.2\,\ell$ to $11.8\,\ell$. Even at the shorter setback the
landscape is smooth on the scale of the lattice constant at every filling we
study. The setback is therefore mainly an amplitude knob and only weakly a
length knob, which is why the two setbacks give the same qualitative physics with
shifted thresholds, as reported in Sec.~\ref{lll:sub:setback}.

\subsubsection{Checks on the Hartree potential}

Because the sign of $\alpha$ is the one surprising result in this appendix, we
record the checks made on $U_H$. It is positive where the density is high, as a
repulsive potential must be: the correlation between $U_H(\br)$ and $n(\br)$ is
$+0.43$ at $\nu=0.50$ and $+0.73$ at $\nu=0.10$ for the full solution, and
$+0.90$ for the Hartree-free droplet discussed below. It vanishes identically for
the uniform state $P=\nu\mathbb{I}$, for which all $\Delta(\bG\neq0)$ are zero,
with a computed maximum of $0$ to machine precision. Its form factors are the
same as those of the impurity potential: the physical density already carries one
factor $e^{-q^2\ell^2/4}$, and the test electron that feels the resulting
potential carries a second, exactly as in Eq.~(\ref{lll:eq:Vreal}).

\subsubsection{Removing the Hartree term}

We come to the point of the appendix. The energy functional,
Eq.~(\ref{lll:eq:EDelta}), contains the Hartree kernel $U_H(G)$ and the exchange
kernel $X(G)$, and only the first penalizes charge inhomogeneity. At the first
shell of the $\nu=1/2$ crystal, $G_1\ell=1.905$, the two are $U_H=+0.0856$ and
$X=+0.6156\,(\ee)$, giving $W=U_H-X=-0.5300$. Deleting $U_H$ leaves the kernel
$-X(G)$, which is negative at every $q$ and, since $X$ decreases monotonically,
most negative as $q\to0$. The sum rule, Eq.~(\ref{lll:eq:sumrule}), continues to
require a density-wave weight $\nu(1-\nu)$, but there is now no reason not to
place all of it at the longest wavelength the sample allows. The expected
consequence is phase separation.

\begin{figure*}[!t]
\centering
\includegraphics[width=\textwidth]{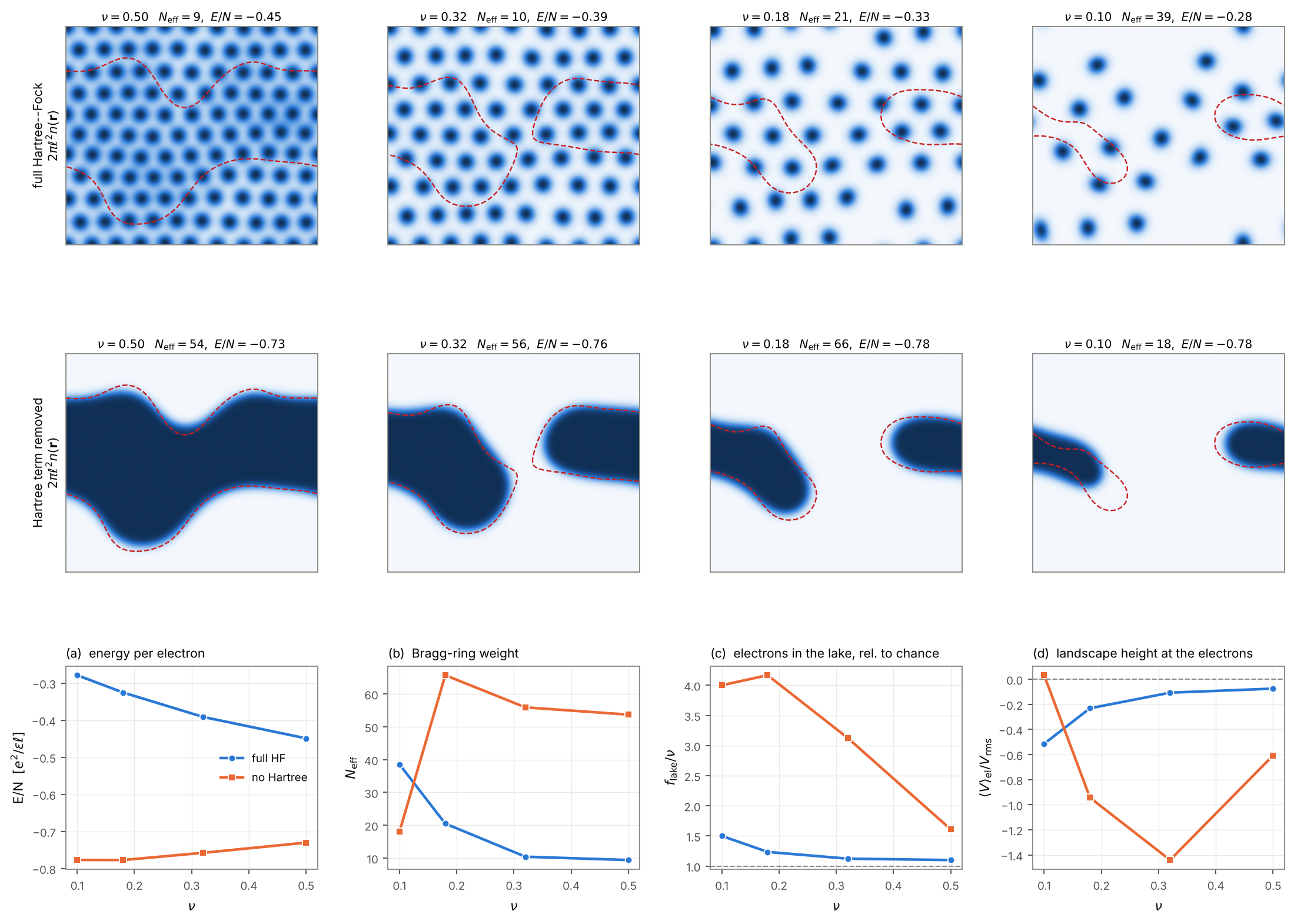}
\caption{What the Hartree term is doing. Same sample, same impurities
($N_i=80$, $d=6\ell$), same solver and same starting determinants; the lower row
deletes $U_H(G)$ from the kernel and nothing else. The red contour is the
semiclassical shoreline $V=\mu_\nu$, computed from the bare landscape with no
reference to the solution. The color scale is identical in all eight maps.
(a)--(d) Energy per electron, Bragg-ring weight, electrons in the lake relative
to chance, and mean landscape height at the electron positions.}
\label{lll:fig:noH}
\end{figure*}

That is what happens. Figure~\ref{lll:fig:noH} compares the two theories on the same
sample, with the same impurities, the same solver, the same adiabatic
continuation and the same set of starting determinants; the lower row deletes
$U_H(G)$ from the kernel and changes nothing else. The lower row is the puddle
state in its pure form. The electrons collapse into a single droplet that is
locally full, $2\pi\ell^2n=1$, with an edge one magnetic length wide, and the
boundary of that droplet coincides with the semiclassical shoreline $V=\mu_\nu$
drawn in red. That contour is computed from the bare landscape alone and has no
knowledge of the solution. The electrons fill the lake to the waterline, which is
precisely the lake construction of Sec.~\ref{lll:sub:lake} above.

Table~\ref{lll:tab:noH} gives the numbers alongside the full calculation. Four
features are worth drawing out.

First, the energy per electron without the Hartree term is $-0.730$, $-0.758$,
$-0.777$ and $-0.777\,(\ee)$ at $\nu=0.50$, $0.32$, $0.18$ and $0.10$. It is
almost independent of filling, and that independence is the signature of phase
separation: the energy per electron is set by the local density inside the
droplet, which is unity, and not by how many electrons the sample contains. The
full calculation, by contrast, gives an energy that tracks the filling closely,
from $-0.449$ to $-0.279\,(\ee)$ over the same range.

Second, the density contrast is exactly unity at every filling without the
Hartree term, whereas the full calculation reaches unity only below
$\nu\simeq0.18$ and is $0.749$ at half filling.

Third, the Bragg-ring weight $N_{\rm eff}$ rises from $9.4$ to $53.7$ at
$\nu=1/2$, from $10.4$ to $55.9$ at $\nu=0.32$ and from $20.5$ to $65.7$ at
$\nu=0.18$ when the Hartree term is removed. Here $N_{\rm eff}$ must be read
with care rather than as a crystallinity scale: a droplet has no crystalline
order at all, and the ring weight is simply measuring how far the density has
moved from six Bragg spots. At $\nu=0.10$ the ordering inverts, $38.5$ for the
full solution against $18.1$ without the Hartree term, because there the full
solution is the most amorphous state in the entire study while the droplet is a
single compact object whose weight has largely left the ring altogether.

Fourth, the trapping measures saturate. The fraction of electrons in the lake
rises to $1.61$, $3.12$, $4.17$ and $4.00$ times chance, against $1.10$ to $1.50$
for the full calculation, and $\langle V\rangle_{\rm el}/V_{\rm rms}$ reaches
$-1.441$ at $\nu=0.32$, against $-0.107$ for the full calculation at the same
filling.

Two cautions. These energies are not variational for the true Hamiltonian; they
are lower than the full ones only because a positive term has been deleted, and
nothing should be inferred from their magnitude. And at $\nu=0.10$ the droplet
contains only twenty electrons; exchange then dominates the landscape on the
scale of the droplet, and the droplet no longer settles in the deepest basin, so
that $\langle V\rangle_{\rm el}$ turns slightly positive, $+0.032$. This is why
the lower curve in Fig.~\ref{lll:fig:noH}(d) is not monotonic, and it is why the
comparison is most meaningful at the higher fillings where the droplet is large.

\begin{table}[b]
\caption{The lake construction on the sample used throughout the paper,
$N_i=80$, $d=6\ell$. $\mu_\nu$ is the waterline, in $\ee$; ``lakes'' is the number
of disconnected components; ``largest'' is the fraction of the lake area in the
largest component; ``winds'' records whether a path through the lake traverses the
torus.}
\label{lll:tab:lakeown}
\small
\begin{tabular}{@{}ccccc@{}}
\toprule
$\nu$ & $\mu_\nu$ & lakes & largest & winds \\
\midrule
0.70 & $+0.0847$ & 1 & 1.000 & yes \\
0.60 & $+0.0399$ & 1 & 1.000 & yes \\
0.55 & $+0.0166$ & 1 & 1.000 & yes \\
0.50 & $-0.0097$ & 1 & 1.000 & yes \\
0.45 & $-0.0364$ & 1 & 1.000 & yes \\
0.40 & $-0.0614$ & 1 & 1.000 & yes \\
0.32 & $-0.0931$ & 1 & 1.000 & no \\
0.25 & $-0.1227$ & 1 & 1.000 & no \\
0.18 & $-0.1552$ & 1 & 1.000 & no \\
0.12 & $-0.1826$ & 1 & 1.000 & no \\
0.10 & $-0.1907$ & 1 & 1.000 & no \\
\bottomrule
\end{tabular}
\end{table}

\begin{table}[b]
\caption{The same construction on a five times larger sample at the same
impurity density, $A=6283\,\ell^2$ with $N_i=400$, at both setbacks. No
Hartree--Fock solution was computed on these samples.}
\label{lll:tab:lakebig}
\small

\begin{tabular}{@{}ccccccc@{}}
\toprule
 & \multicolumn{3}{c}{$d=6\ell$} & \multicolumn{3}{c}{$d=3\ell$} \\
\cmidrule(lr){2-4}\cmidrule(lr){5-7}
$\nu$ & lakes & largest & winds & lakes & largest & winds \\
\midrule
0.70 & 1 & 1.000 & yes & 1 & 1.000 & yes \\
0.60 & 1 & 1.000 & yes & 1 & 1.000 & yes \\
0.55 & 1 & 1.000 & yes & 1 & 1.000 & no \\
0.50 & 1 & 1.000 & no & 2 & 0.994 & no \\
0.45 & 1 & 1.000 & no & 1 & 1.000 & no \\
0.40 & 1 & 1.000 & no & 1 & 1.000 & no \\
0.32 & 3 & 0.910 & no & 3 & 0.855 & no \\
0.25 & 2 & 0.932 & no & 3 & 0.868 & no \\
0.18 & 3 & 0.824 & no & 4 & 0.689 & no \\
0.12 & 4 & 0.653 & no & 7 & 0.363 & no \\
0.10 & 3 & 0.659 & no & 8 & 0.245 & no \\
\bottomrule
\end{tabular}
\end{table}

\begin{table*}[t]
\caption{Where the Hartree--Fock electrons sit on the landscape, for the three
impurity numbers. ${\rm corr}(n,V)$ is the correlation coefficient between the
density and the landscape; $\langle V\rangle_{\rm el}/V_{\rm rms}$ is the mean
landscape height at the electron positions; $f_{\rm lake}/\nu$ is the fraction of
electrons inside the lake relative to the chance value.}
\label{lll:tab:sit}
\small
\begin{tabular}{@{}cccccccccc@{}}
\toprule
 & \multicolumn{3}{c}{${\rm corr}(n,V)$}
 & \multicolumn{3}{c}{$\langle V\rangle_{\rm el}/V_{\rm rms}$}
 & \multicolumn{3}{c}{$f_{\rm lake}/\nu$} \\
\cmidrule(lr){2-4}\cmidrule(lr){5-7}\cmidrule(lr){8-10}
$\nu$ & $N_i=24$ & $56$ & $80$ & $N_i=24$ & $56$ & $80$ & $N_i=24$ & $56$ & $80$ \\
\midrule
0.500 & $-0.112$ & $-0.160$ & $-0.180$ & $-0.045$ & $-0.066$ & $-0.073$ & 1.00 & 1.02 & 1.08 \\
0.450 & $-0.101$ & $-0.144$ & $-0.163$ & $-0.050$ & $-0.072$ & $-0.081$ & 1.06 & 1.09 & 1.09 \\
0.360 & $-0.087$ & $-0.125$ & $-0.142$ & $-0.060$ & $-0.087$ & $-0.100$ & 1.04 & 1.04 & 1.16 \\
0.320 & $-0.084$ & $-0.122$ & $-0.138$ & $-0.068$ & $-0.098$ & $-0.112$ & 1.07 & 1.12 & 1.12 \\
0.280 & $-0.084$ & $-0.122$ & $-0.137$ & $-0.078$ & $-0.110$ & $-0.124$ & 1.08 & 1.21 & 1.08 \\
0.210 & $-0.091$ & $-0.131$ & $-0.150$ & $-0.109$ & $-0.158$ & $-0.181$ & 1.25 & 1.13 & 1.25 \\
0.180 & $-0.097$ & $-0.144$ & $-0.170$ & $-0.136$ & $-0.198$ & $-0.233$ & 1.08 & 0.93 & 1.23 \\
0.150 & $-0.108$ & $-0.171$ & $-0.198$ & $-0.171$ & $-0.270$ & $-0.312$ & 1.11 & 1.33 & 1.56 \\
0.120 & $-0.128$ & $-0.196$ & $-0.223$ & $-0.233$ & $-0.356$ & $-0.413$ & 1.39 & 1.39 & 1.74 \\
0.100 & $-0.153$ & $-0.215$ & $-0.249$ & $-0.318$ & $-0.445$ & $-0.512$ & 1.50 & 1.50 & 1.50 \\
\bottomrule
\end{tabular}
\end{table*}

\begin{table*}[t]
\caption{Screening of the landscape. $V^{\rm tot}_{\rm rms}/V_{\rm rms}$ is the
ratio of the rms of $V+U_H$ to that of $V$; $\alpha$ is the projection of
$V+U_H$ onto $V$, Eq.~(\ref{lll:eq:alpha}); $\alpha_{\rm LW}$ is the same projection
restricted to $q<G_1/2$. Values of $\alpha$ below zero indicate overscreening.}
\label{lll:tab:screen}
\small
\begin{tabular}{@{}cccccccccc@{}}
\toprule
 & \multicolumn{3}{c}{$V^{\rm tot}_{\rm rms}/V_{\rm rms}$}
 & \multicolumn{3}{c}{$\alpha$}
 & \multicolumn{3}{c}{$\alpha_{\rm LW}$} \\
\cmidrule(lr){2-4}\cmidrule(lr){5-7}\cmidrule(lr){8-10}
$\nu$ & $N_i=24$ & $56$ & $80$ & $N_i=24$ & $56$ & $80$ & $N_i=24$ & $56$ & $80$ \\
\midrule
0.500 & 0.482 & 0.345 & 0.329 & $-0.129$ & $-0.123$ & $-0.128$ & $-0.129$ & $-0.123$ & $-0.128$ \\
0.450 & 0.605 & 0.423 & 0.397 & $-0.115$ & $-0.109$ & $-0.112$ & $-0.115$ & $-0.109$ & $-0.112$ \\
0.360 & 0.862 & 0.593 & 0.551 & $-0.057$ & $-0.055$ & $-0.060$ & $-0.057$ & $-0.055$ & $-0.060$ \\
0.320 & 0.986 & 0.677 & 0.629 & $-0.054$ & $-0.057$ & $-0.063$ & $-0.054$ & $-0.057$ & $-0.062$ \\
0.280 & 1.157 & 0.793 & 0.735 & $-0.065$ & $-0.076$ & $-0.071$ & $-0.064$ & $-0.075$ & $-0.071$ \\
0.210 & 1.466 & 1.003 & 0.934 & $-0.124$ & $-0.122$ & $-0.137$ & $-0.123$ & $-0.121$ & $-0.136$ \\
0.180 & 1.587 & 1.090 & 1.013 & $-0.160$ & $-0.182$ & $-0.218$ & $-0.157$ & $-0.180$ & $-0.213$ \\
0.150 & 1.726 & 1.199 & 1.114 & $-0.232$ & $-0.329$ & $-0.360$ & $-0.228$ & $-0.323$ & $-0.350$ \\
0.120 & 1.850 & 1.263 & 1.157 & $-0.367$ & $-0.367$ & $-0.388$ & $-0.354$ & $-0.345$ & $-0.368$ \\
0.100 & 1.879 & 1.281 & 1.160 & $-0.458$ & $-0.413$ & $-0.450$ & $-0.416$ & $-0.384$ & $-0.426$ \\
\bottomrule
\end{tabular}
\end{table*}

\begin{table}[b]
\caption{The same sample, impurities and solver with and without the Hartree
term in the kernel. $N_{\rm eff}$ is the Bragg-ring weight defined in
Sec.~\ref{lll:sec:dis}; the contrast is
$(n_{\max}-n_{\min})/(n_{\max}+n_{\min})$. Energies in $\ee$.}
\label{lll:tab:noH}
\small

\begin{tabular}{@{}llrrrr@{}}
\toprule
 & $\nu$ & 0.50 & 0.32 & 0.18 & 0.10 \\
\midrule
full HF  & $E/N$ & $-0.449$ & $-0.391$ & $-0.325$ & $-0.279$ \\
         & $N_{\rm eff}$ & 9.4 & 10.4 & 20.5 & 38.5 \\
         & contrast & 0.749 & 0.965 & 1.000 & 1.000 \\
         & $\langle V\rangle_{\rm el}/V_{\rm rms}$ & $-0.075$ & $-0.107$ & $-0.231$ & $-0.514$ \\
         & $f_{\rm lake}/\nu$ & 1.10 & 1.12 & 1.23 & 1.50 \\
\midrule
no $U_H$ & $E/N$ & $-0.730$ & $-0.758$ & $-0.777$ & $-0.777$ \\
         & $N_{\rm eff}$ & 53.7 & 55.9 & 65.7 & 18.1 \\
         & contrast & 1.000 & 1.000 & 1.000 & 1.000 \\
         & $\langle V\rangle_{\rm el}/V_{\rm rms}$ & $-0.607$ & $-1.441$ & $-0.944$ & $+0.032$ \\
         & $f_{\rm lake}/\nu$ & 1.61 & 3.12 & 4.17 & 4.00 \\
\bottomrule
\end{tabular}
\end{table}

\section{Thomas--Fermi/IRM computational details}
\label{sec:SM_tf}

\subsection{Computational details (TF/IRM)}
\subsubsection{Active-set formulation of the local-TF obstacle problem}
\label{tf:app:active}

On a periodic mesh, let \(\bm n\) be the density vector at fixed mean
\(\bm 1^T\bm n=N_{\rm site}n\). In the local-TF calculation the discrete
functional has the form
\begin{equation}
 E(\bm n)=\frac{1}{2}\bm n^T(\pi I+V)\bm n+
 \bm V_{\rm imp}^T\bm n+\text{constant},
\label{tf:eq:discreteenergy}
\end{equation}
with \(\bm n\ge0\). The matrix \(V\) is the Fourier representation of the
Coulomb kernel with its zero mode deleted. On an occupied support \(S\),
\begin{equation}
 (\pi I+V)_S\bm n_S+\bm V_{{\rm imp},S}
 =\mu\bm 1_S.
\label{tf:eq:activeequation}
\end{equation}
Two conjugate-gradient solves give
\begin{equation}
 \bm n_S=\bm x_S+\mu\bm y_S,\quad
 A_S\bm x_S=-\bm V_{{\rm imp},S},\quad
 A_S\bm y_S=\bm1_S,
\label{tf:eq:activesolves}
\end{equation}
and the density constraint fixes
\begin{equation}
 \mu=\frac{N_{\rm site}n-\bm1_S^T\bm x_S}
 {\bm1_S^T\bm y_S}.
\label{tf:eq:activemu}
\end{equation}
The Karush-Kuhn-Tucker conditions require positive density on \(S\) and
\((\pi I+V)\bm n+\bm V_{\rm imp}\ge\mu\) outside it. Updating \(S\) until
these inequalities stabilize solves the nonlinear depletion problem without
smearing the density support. The classical check is obtained by replacing
\(\pi I\) by zero on a nontrivial support.

\subsubsection{Derivation of the transport quadrature and remote-scattering limit}
\label{tf:app:transport}

The radial \(k'\) integral in Eq.~\eqref{tf:eq:tau} gives the two-dimensional
density of final states. With \(t=\theta/2\), one has
\(q=2k_F\sin t\), \(1-\cos\theta=2\sin^2t\), and
\(d\theta=2dt\). Substitution of Eq.~\eqref{tf:eq:screenedpotential} then
gives Eq.~\eqref{tf:eq:rho}. The equality
\(\rho/(h/e^2)=1/(k_F\ell)\) follows from
\(\sigma=ne^2\tau_t/m=(e^2/h)k_F\ell\) for \(g_s=2\).

For \(k_Fd\gg1\), set \(\sin t\simeq t\) in the forward-scattering region.
The leading integral is
\begin{equation}
 \int_0^\infty dt\,t^2e^{-4k_Fdt}
 =\frac{1}{32k_F^3d^3},
\label{tf:eq:smallangle}
\end{equation}
which produces Eq.~\eqref{tf:eq:remoteasymptote}. This derivation explains why
the IRM scale can drop faster with \(d\) than the classical real-space
percolation scale.

\subsubsection{Short-range transport contribution}
\label{tf:app:short_range}

For the effective correlator in Eq.~\eqref{tf:eq:shortcorrelator}, the angular
average of \(1-\cos\theta\) is unity. Equation~\eqref{tf:eq:tau} therefore gives
\begin{equation}
 \frac{1}{\tau_{\rm sr}}=\frac{mW_{\rm sr}}{\hbar^3}.
\label{tf:eq:shorttau}
\end{equation}
Combining this result with \(\rho=m/(ne^2\tau_t)\) and
\(h=2\pi\hbar\) yields Eq.~\eqref{tf:eq:shortrho}. Thus
\(\gamma_{\rm sr}\) has a direct operational meaning: in the absence of
remote disorder, the white-noise component alone reaches
\(\rho=h/e^2\) at \(n a_{\rm B}^2=\gamma_{\rm sr}\). This is a transport
parameter, not an additional long-distance percolation parameter.

\subsubsection{Scope of the crossover function}
\label{tf:app:scope}

The numerical \(F_{1/2}(x)\) is a controlled output of a local Hartree-TF
functional with continuum Gaussian disorder. It does not include exchange
and correlation corrections to the compressibility, nonlocal density
response, finite quantum-well width, correlated impurities, quantum
tunneling between puddles, or interference localization. The large-\(x\)
agreement with EPB validates the correct classical limit of this model. The
finite-\(x\) values should instead be interpreted as a quantitatively useful
local-TF crossover benchmark. A microscopic small-\(d\) prediction requires
specifying the ultraviolet physics omitted here.